\documentclass[12pt,preprint]{aastex}
\newcommand{\be}{\begin{equation}}
\newcommand{\ee}{\end{equation}}
\newcommand{\bea}{\begin{eqnarray}}
\newcommand{\eea}{\end{eqnarray}}
\newcommand{\nn}{\nonumber}

\newcommand{\appleq}{\la}

\newcommand{\Bphi}{B_\phi}
\newcommand{\Bz}{B_z}
\newcommand{\Bzs}{B_{z,S}}
\newcommand{\Bphis}{B_{\phi,S}}

\newcommand{\Ps}{P_S}
\newcommand{\Pave}{\langle P\rangle}

\newcommand{\Rs}{{R_S}}

\newcommand{\Gz}{\Gamma_z}
\newcommand{\Gphi}{\Gamma_\phi}

\newcommand{\W}{{\cal W}}
\newcommand{\M}{{\cal M}}
\newcommand{\V}{{\cal V}}

\newcommand{\rhoc}{\rho_c}
\newcommand{\mvir}{m_{vir}}

\newcommand{\Stod}{Stod\'olkiewicz }

\shorttitle{A Genetic Algorithm-Based Exploration of Three Filament Models}
\shortauthors{Fiege et al.}

\begin{document}
% \title{Surveying Magnetic Models of an Infrared-Dark Cloud Using a Genetic Algorithm: A Case for Magnetic Support in the G11.11-0.12 Filament}
\title{A Genetic Algorithm-Based Exploration of Three Filament Models: A Case for the Magnetic Support of the G11.11-0.12 Infrared-Dark Cloud}

\author{Jason D. Fiege, Doug Johnstone, Russell O. Redman and Paul A. Feldman}
\affil{National Research Council of Canada, Herzberg Institute of Astrophysics, 5071 West Saanich Road, Victoria, British Columbia}

\begin{abstract}
The G11.11-0.12 infrared-dark cloud has a filamentary appearance, both in extinction against the diffuse infrared emission of the Galactic plane and in emission at 850$\micron$.  We use a novel computational technique based on an advanced genetic algorithm to explore thoroughly 3 different models of self-gravitating, pressure truncated filaments and to constrain their parameters.  Specifically, the models tested are the non-magnetic \citet{ostriker64} model, a generalized version of the magnetic \citet{stod63} model, and the magnetic \citet{FP1} model.  Previous results showed that G11.11-0.12 has a much steeper $\sim r^{-4}$ radial density profile than other filaments, where the density varies approximately as $r^{-2}$, and that this steep density profile is consistent with the \citet{ostriker64} model.  We present a more complete analysis that shows that the radial structure of G11.11-0.12 is consistent with regimes of each of these models.  All of the magnetic models that agree with the data are threaded by a dominant poloidal magnetic field, and most have dynamically significant fields.  Thus, G11.11-0.12 is an excellent candidate for radial support by a magnetic field that is predominantly poloidal.  We predict the polarization patterns expected for both magnetic models and show that the two magnetic models produce different polarization patterns that should be distingished by observations.
\end{abstract}


%% In the first two sections, notice the use of the natbib \citet

\section{Introduction}
\label{sec:intro}
Infrared-dark clouds (IRDCs) are a class of molecular clouds seen in extinction against the diffuse infrared emission of the Galactic plane.  About 2000 of these objects were discovered by the Midcourse Space Experiment (MSX) satellite in spectral bands from $7-25\micron$ \citep{egan98}.  They reside at distances of $2-8$ kpc typically \citep{egan98,carey98}, and most are located roughly toward the Galactic centre.  The IRDCs have unusual physical characteristics compared to nearby examples of molecular clouds.  \citet{egan98} found that IRDCs appear to have properties intermediate between massive molecular cloud cores ($n\approx 10^4-10^6$ {\rm cm}$^{-3}$; $T\approx 30-100$ K) and isolated dark clouds ($n\approx 10^2-10^4 {\rm cm}^{-3}$; $T\approx 8-15$ K), having densities typical of molecular cloud cores, but temperatures typical of isolated dark clouds.  Many IRDC filaments appear to be at a relatively early phase of evolution, with fewer signposts of star formation than nearby molecular clouds.

The IRDCs and nearby molecular clouds share the common feature that a large fraction of both populations are filamentary, but the IRDCs are often less fragmented.  Relatively ``pristine'' IRDC filaments have simpler density structures than fragmented, star-forming filaments, which makes them good candidates for detailed comparison with theoretical models.  One drawback is that IRDCs are more distant than many other filaments and are therefore less well resolved.  

The G11.11-0.12 cloud studied in this paper is a good example of a filamentary IRDC, which was observed by SCUBA in 850$\micron$ emission \citet{carey00}, following its discovery by the MSX survey.  \citet{johnstone03} (hereafter Paper I) fit a non-magnetic equilibrium filament model to the radial structure of two regions of G11.11-0.12.  This paper presents a more thorough analysis, which compares the SCUBA 850$\micron$ emission with 3 existing models of filamentary clouds in detail.  These models are the pressure-truncated non-magnetic Ostriker \citep{ostriker64} model, a generalized \citet{stod63} (hereafter GS) helical field model, and the Fiege \& Pudritz (hereafter FP) helical field model.  We discuss these models in the next section.  Note that the Ostriker model is the zero magnetic field limit of both the GS and FP models, but this limit is sufficiently important to warrant separate treatment.

This paper uses a novel genetic algorithm-based technique to explore the parameter space of each model completely and to fit solutions to the data.  This is the first astronomical research application for a new multi-object genetic algorithm (``Ferret'') developed by one of the authors (J.F.).  The technique is discussed in Section \ref{sec:GAs}.

\section{Theoretical Models of Filamentary Clouds}
\label{sec:theory}

The goal of this paper is a thorough comparison of the G11.11-0.12 IRDC with 3 different theoretical models of filamentary molecular clouds.  These models have different magnetic field configurations, or no magnetic field at all in the case of the Ostriker model.  However, they share the common features of self-gravity, external pressure truncation and isothermality, meaning that the velocity dispersion is constant everywhere within the cloud.  This last feature is somewhat restrictive because the velocity dispersion is well-known to increase with size scale in molecular clouds, and the equation of state does affect the radial density structure.  However, the effective equation of state of the gas is not known, and we restrict our analysis to existing, classical models of filaments. 

\subsection{Non-Magnetic Filaments: The Ostriker Model}
\label{sec:ostriker}
The simplest equilibrium model of a filamentary cloud is that of a self-gravitating, non-magnetic, isothermal filament.  \citet{ostriker64} studied such clouds, and found an exact solution for the density structure 
\footnote{
This solution is equivalent to the non-magnetic limit of the magnetic model discovered by \citet{stod63}, discussed in \S\ref{sec:stod}.
}
\be
\rho=\frac{\rhoc}{\left[1+r^2/(8 r_0^2)\right]^2},
\label{eq:ostriker}
\ee
where $r_0$ is the core radius of the filament.  The core radius is defined by
\be
r_0=\frac{\sigma}{\sqrt{4\pi G\rhoc}},
\label{eq:r0}
\ee
where $\sigma$ is the 1-dimensional velocity dispersion of the gas, assumed to be constant and possibly including both thermal and non-thermal contributions, $G$ is the gravitational constant, and $\rhoc$ is the central density along the axis of the filament.  Note that the density varies as $\sim r^{-4}$ at large radii $r\gg r_0$.

In the absence of external pressure, the Ostriker solution extends to infinite radius, and the mass per unit length, which we refer to as the line mass for brevity, is exactly 
\be
\mvir=\frac{2\sigma^2}{G}.
\label{eq:mvir}
\ee
This quantity is denoted $\mvir$, since it also represents the virial line mass in the virial equation \ref{eq:virial1} below.

An Ostriker filament is truncated at finite radius if it is bounded by an external pressure.  In this case, the filament obeys the virial equation
\be
\frac{m}{\mvir}=1-\frac{\Ps}{\Pave},
\label{eq:virial1}
\ee
where $m$ is the line mass, $\Ps$ is the external pressure, and $\Pave$ is the mean (volume weighted) internal pressure defined as follows:
\be
\Pave = \frac{\int_\V P d\V}{\V}.
\ee
\citep{FP1} (hereafter FP1).  
Generally, we use calligraphic characters to refer to quantities defined per unit length.  Thus, $\V \equiv \pi \Rs^2$ is the cross-sectional area of the filament (volume per unit length), in which $\Rs$ is the radius of the filament where it is truncated by the pressure of the external medium.

Thus, the line mass of the Ostriker filament is reduced if the filament is bounded by a finite external pressure.  Filaments with a line mass exceeding $\mvir$ cannot be represented by an hydrostatic equilibrium model and require magnetic support against self gravity.

We parameterize the pressure truncation radius $\Rs$ of the Ostriker model by the concentration parameter
\be
C\equiv \log_{10} \left( \frac{\Rs}{r_0} \right).
\label{eq:Cdef}
\ee
Thus, the Ostriker model is characterized by a single dimensionless parameter $C$, plus two parameters $\rhoc$ and $\sigma$, which determine the density scale and core of the solution via equation \ref{eq:r0}.

\subsection{The Generalized \Stod Model}
\label{sec:stod}
\citet{stod63} discovered an analytic solution for isothermal self-gravitating filamentary clouds threaded by a poloidal  ($\Bz$) magnetic field.  \Stod assumed that the magnetic pressure is proportional to the gas pressure, so that
\be
\beta \equiv \frac{P}{P_{mag}}=\frac{8\pi\sigma^2\rho}{B^2} = const
\label{eq:beta}
\ee
within the cloud.  

Magnetostatic models always have the freedom to assume how the mass is distributed relative to the poloidal magnetic flux surfaces.  The two most common assumptions are that $\beta$ is constant, or that the mass (per unit length for a filament) and magnetic flux enclosed by each poloidal magnetic flux surface are proportional.  The \Stod model is of the former type, while the FP model, discussed in \S\ref{sec:FP} below, is of the latter.

The \Stod model was generalized by Nakamura, Hanawa and Nakano \citep{nakamura93} to include a toroidal ($\Bphi$) magnetic field component, so that the field lines are helical in general.  The radial density structure is identical to that of the Ostriker model (equation \ref{eq:ostriker}), except that the core radius $r_0$ is replaced by $r_{0,Stod}$, which is modified by the magnetic field:
\be
r_{0,Stod}^2=\frac{\sigma^2}{4\pi G\rhoc}\left[1+\frac{1+\cos^2\theta}{2\beta}\right]
\label{eq:r0stod}
\ee
where $\theta$ is the pitch angle of the magnetic field, relative to the axis, at radial infinity.  The poloidal and toroidal components of the magnetic field are given respectively as follows:
\bea
\Bz &=& \Bz(0)\sqrt{ 1+\cos^2\theta\left(\frac{r^2}{8 r_{0,Stod}^2}\right) } \left(1+\frac{r^2}{8 r_{0,Stod}^2}\right)^{-3/2} \\
\Bphi &=& \Bz(0)\frac{r}{2\sqrt{2}r_{0,Stod}}\sin\theta \left(1+\frac{r^2}{8 r_{0,Stod}^2}\right)^{-3/2},
\label{eq:Bstod}
\eea
where $\Bz(0)$ is the magnetic field at the radial centre of the filament, given by
\be
\Bz(0)=\sqrt{\frac{8\pi\sigma^2\rho}{\beta}}.
\ee

The pressure-truncated GS model is characterized by 3 dimensionless parameters $C, \beta$ and $\theta$, where $C$ is the same concentration parameter defined by equation \ref{eq:Cdef}, plus plus two parameters $\rhoc$ and $\sigma$, which determine the density scale and core radius exactly as in the Ostriker solution.  Note that there exists an interesting and exact degeneracy in the density structure of the \citet{stod63} model.  All solutions characterized by 
\be
(1+\cos^2\theta)/2\beta=\kappa
\label{eq:kappaDef} 
\ee
form a 1-parameter set having exactly the same core radius and density structure, controlled by the constant $\kappa$, although they may have different magnetic field strength and geometry, which are controlled by $\theta$ and $\beta$.  This paper fits models to the surface density structure of G11.11-0.12 only.  Thus, we consider only $\kappa$ in our analysis.

The density structure of the GS model tends to that of the Ostriker model in the non-magnetic limit $\beta \rightarrow \infty$.  Note that $r_{0,Stod}>r_0$ for any choice of $\beta$ and $\theta$, which indicates that the magnetic field always offers net magnetic support.  Thus, the GS helical field model is always dominated by the poloidal magnetic field.  This is true even for the special case $\theta=\pi/2$, where the magnetic field tends to a purely toroidal field as $r \rightarrow \infty$.  Even in this limit, the poloidal field dominates globally over the toroidal field and provides net support.  The GS model differs in this regard from the FP model, which may be dominated by either the poloidal or toroidal field.

\subsection{The Fiege \& Pudritz Model}
\label{sec:FP}

The primary difference between the GS model and the FP model (Fiege \& Pudritz 2000a) is that they make different assumptions regarding the magnetic flux loading.  The FP model assumes uniform flux to mass loading, whereas the GS model assumes that the plasma $\beta$ parameter is constant (see (\S\ref{sec:stod}).  The assumption of a constant magnetic flux to mass ratio in the FP model implies that that all magnetic field lines are loaded with equal mass.  Thus all field lines are equivalent, and there is no possibility of exciting magnetic interchange instabilities.  Also, any filament with this property could be assembled from a cloud of uniform density and magnetic field by radially compressing and twisting flux tubes under the condition of magnetic flux freezing.  

The magnetic flux to mass ratios $\Gz$ and $\Gphi$  are defined as follows:
\bea
\Gz &=& \sqrt{\frac{\rho_c}{\sigma_c^2}} \left(\frac{\Bz}{\rho}\right) \label{eq:Gz} \nn\\
\Gphi &=& \frac{1}{\sqrt{4\pi G}} \left(\frac{\Bphi}{r\rho}\right). \label{eq:Gphi},
\eea
which determine the flux to mass loading of the poloidal and toroidal field components respectively.  We refer the reader to FP1 for an illustration of $\Gz$ and $\Gphi$.  FP1 found that the virial properties of the filaments that they studied could be described by the isothermal FP model with 
\bea
5 \appleq \Gphi &\appleq& 25 \nn\\
\Gz &\appleq& 8.
\label{eq:GammaLimits}
\eea

In contrast to the Ostriker and GS solutions, the density and magnetic field of the FP model cannot be written down in closed form.  One must integrate numerically the equations of self-gravitating magnetostatic equilibrium, which are given in FP1.  However, the magnetic generalization of equation \ref{eq:virial1} is only slightly more complex:
\be
\frac{\Ps}{\Pave}=1-\frac{m}{\mvir}\left(1-\frac{\M}{|\W|}\right),
\label{eq:virial}
\ee
where $\Ps$, $\Pave$, $m$, and $\mvir$ are defined as above, and the remaining terms are
defined as follows. The gravitational binding energy per unit length is given by
\be
\W=-m^2 G,
\label{eq:Wdef}
\ee
and the total magnetic energy per unit length is given by
\be
\M=\frac{1}{4\pi}\int B_z^2 d\V - \left( \frac{\Bzs^2+\Bphis^2}{4\pi} \right) \V.
\label{eq:MDef}
\ee
In the equations above, subscript $S$ refers to quantities evaluated at the cylindrical surface of the cloud.  

The assumption of constant magnetic flux to mass ratio makes the FP model somewhat more general than GS model, in the sense that the magnetic energy term $\M$ in equation \ref{eq:virial} may be either positive, negative or zero, corresponding to net support, net magnetic confinement or magnetically neutral solutions, respectively.  The magnetic energy term in the GS model is always positive, or zero in the special case where the solution tends to the non-magnetic Ostriker limit discussed in \S\ref{sec:ostriker}. 

The sign of $\M$ in equation \ref{eq:virial} determines whether the magnetic field provides net magnetic support against self-gravity ($\M>0$) or confinement ($\M<0$).  The former case ($\M>0$) is the result of a dominant poloidal magnetic field, for which the field lines provide a radially expansive pressure.  The latter case ($\M<0$) results from a dominant toroidal magnetic field, which provides a net binding energy.  The FP model applies equally well to either of these regimes.  FP1 tabulated virial data that allowed them to estimate $\Ps/\Pave$ and $m/\mvir$ for a number of filamentary clouds, which provided an estimate the magnetic energy term $\M/|\W|$.  They found that $\M/|\W| < 0$ for most of their sample, which is only possible for filaments dominated by the toroidal field.  

Solutions with flux to mass parameters in the ranges given by equation \ref{eq:GammaLimits} may be dominated by either the poloidal or the toroidal field component.  Toroidal-field dominated filaments are characterized by density profiles that are more centrally peaked than the Ostriker model, but fall off more gradually with radius.  FP1 found numerically that the density falls off as $r^{-1.9\pm 0.1}$ at large radius, which is in agreement with the observed density profiles of several filamentary clouds (Alves et al. 1998, Lada et al 1999, Johnstone \& Bally 1999).  \citet{fiege03} showed analytically that this shallow density profile is due to the dominance of the toroidal field at large radii ($r\gg r_0$).  Steeper density profiles are possible in the regime dominated by the poloidal field in the FP model.  FP1 studied this regime, but they focused on the toroidal-field dominated regime of the model, which was more consistent with observational data compiled from the literature.

Poloidal field and toroidal-field dominated regimes of the FP model are separated by a magnetically neutral regime of the model, defined by $\M=0$, where the expansive pressure of the poloidal field is exactly balanced by the radial confinement due to the toroidal field.  The virial equation \ref{eq:virial} cannot distinguish solutions in this neutral set from the non-magnetic truncated Ostriker solution.  This illustrates a fundamental degeneracy of the FP model. It is generally true that any choice of $\M/|\W|$ and surface pressure corresponds to a one parameter set of models, where $\Gz$ and $\Gphi$ are critically balanced to yield the same $\M/|\W|$.  This degeneracy occurs because $\Bz$ and $\Bphi$ enter the magnetic energy term of the virial equation (equation \ref{eq:MDef}) with opposite sign, so that poloidal and toroidal terms can balance each other.  In general, FP models with indistinguishable virial properties can still have observable differences in their internal density structure.  This study is based on detailed models of the internal density structure, which might distinguish between such differences.

To summarize, the FP model requires 3 dimensionless parameters, namely the concentration parameter $C$ and two magnetic flux to mass ratios $\Gz$ and $\Gphi$, which correspond respectively the flux to mass ratios of the poloidal and toroidal magnetic field components.  These dimensionless parameters determine the shape of the density, pressure, and magnetic profiles.  Two dimensional parameters $\rhoc$ and $\sigma$ are also required, which determine the density scale and core radius as in the other models.  $C$, $\rhoc$ and $\sigma$ are as defined for the Ostriker and GS models.  Both the GS and FP models assume that the field may be helical in general, although a pure poloidal field is allowed by both models.  

\section{Analysis of the G11.11-0.12 Filament}
\label{sec:analysis}
\begin{figure}
\includegraphics[width=.49\textwidth]{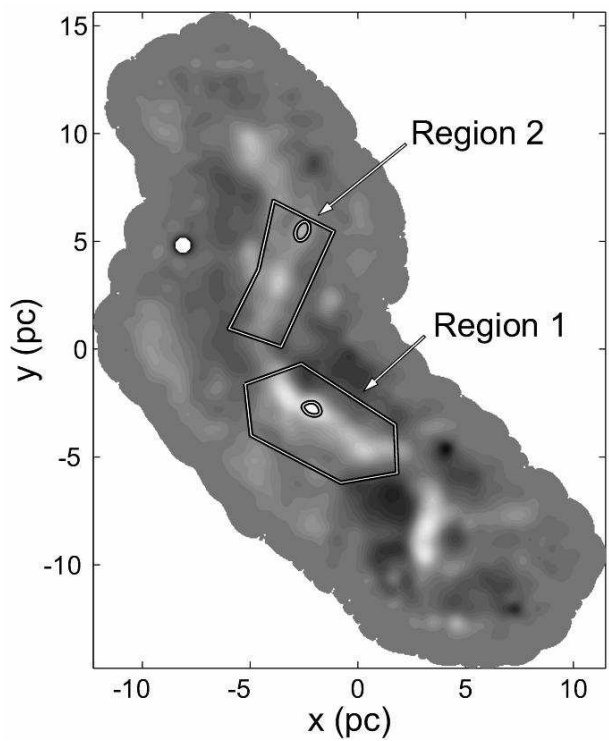}
\caption{The G11.11-0.12 filamentary infrared dark cloud at 850 $\micron$.  The labeled polygons are the regions chosen for analysis.  Each of these regions contains one bright core (shown) that we removed to emphasize the filamentary structure.}
\label{fig:map}
\end{figure}

Figure \ref{fig:map} shows an $850\micron$ map of the G11.11-0.12 filamentary 
infrared-dark cloud.  The observational details and data reduction are described 
in Paper I.  At an estimated distance of $3.6$ kpc 
\citep{carey98}, the G11.11-0.12 filament is $\sim 22$ pc long and has a width of 
$\sim 1$ pc.  The filament is curved and fragmented into several cores, but none of 
these cores appears to be forming stars.  Thus, the fragmentation and subsequent 
evolution is probably in a less advanced stage than in other well-studied filaments 
such as the Orion ``integral-shaped'' filament \citep{dutrey93,johnstone99}.

The first step of our analysis was to locate accurately the central ``spine'' of the 
filament.  We began with an initial guess by graphically locating 10-20 points along 
the ridge covering almost the entire length of the filament.  We smoothed this curve 
slightly using a low-order windowed polynomial fit, and constructed orthogonal lines 
at 50 points equally spaced along the curve.  We interpolated the intensity onto each 
of these lines, and located the peak emission.  We used this sequence of points as an 
improved guess for the location of the spine, and iterated the procedure until no point 
along the spine moved by more than 0.1 pixel (1 pixel $ = 3''$).  

Our analysis ``straightened'' the filament by taking the radius to any point as the 
nearest perpendicular distance to the curve defining the spine.  This procedure can 
be problematic for points on the concave side of a curved spine and more distant than 
its radius of curvature, which can, in principle, intersect with more than one orthogonal.  
However, G11.11-0.12 is sufficiently linear that this was not a problem.  Note that one of 
us used this procedure previously in an analysis of the Orion integral-shaped filament 
\citep{MWF}.

Two regions selected for analysis are shown in Figure \ref{fig:map}.  The brightest peaks 
were clipped from the data as shown, since the contours within these regions are quite 
round and filament models do not apply.  We binned the data to reduce the apparent scatter, 
and estimated the uncertainty from the standard deviation about the mean within each bin. 
We show scatter plots of the binned density as a function of radius for each region, along 
with fits to the data, in $\S$\ref{sec:OstrikerResults}-\ref{sec:FPResults}.

In Paper I, we fit a non-pressure-truncated Ostriker model to each of two regions of 
G11.11-0.12.  We determined that the southern segment, corresponding 
roughly to our region 1, could be fit by an Ostriker model with 
$n_c=(3.3\pm 0.7)\times 10^4 {\rm cm}^{-3}$ and $r_0=0.10\pm 0.02\,$pc.  
We found that the northern segment, corresponding to our region 2 roughly, has a central 
number density of $n_c=(1.1\pm 0.3)\times 10^4 {\rm cm}^{-3}$ and a core radius of 
$r_0=0.12\pm 0.02$ pc, assuming a dust temperature of 13 K and a dust emissivity 
of $\kappa_{850}=0.02 ~{\rm cm}^2 ~{\rm g}^{-1}$.  
At the $3.6$ kpc distance of G11.11-0.12 the filament is only marginally
resolved on these scales and thus care was taken to properly convolve the model
to fit the data. The full width half maximum of an Ostriker filament is several 
$r_0$, however, and thus the core radius fits are reasonably well determined.

\section{Models and Observational Constraints}
\label{sec:models}
Section \ref{sec:FP} discussed how the poloidal and toroidal field components can compensate for each other to produce a 1-parameter set of models with approximately the same density (or surface density) structure.  The degeneracy in $\Gz$ and $\Gphi$ is approximate for the FP model, but the corresponding degeneracy in $\beta$ and $\theta$ is exact for the GS model.  The goal of our analysis is to explore completely the parameter space of each model and to find the largest possible class of solutions that is consistent with the data.  This determines the structure of the parameter space in detail, which would illuminate any other less obvious model degeneracies that might exist.

It is useful to enumerate the parameters of each of the 3 models discussed in $\S$\ref{sec:theory} before discussing the results of our parameter space search.  There is a single dimensionless theoretical parameter $C$ for the truncated Ostriker model, and 3 dimensionless theoretical parameters for each of the magnetic models: ${C,\beta,\theta}$ for the GS models and ${C,\Gz,\Gphi}$ for the FP model.  All models require 5 additional parameters that reflect various observational complexities and unknown quantities.  Two dimensional parameters $\rhoc$ and $\sigma$ determine the density scale and core radius.  One parameter $i$ encodes the unknown inclination of the filament relative to the plane of the sky.  Finally, 2 additional dimensional parameters are required to convert the surface density $\Sigma$ calculated from the models to a predicted $850\micron$ flux density $f$ that can be compared with the observations.  These are the flux conversion factor FCF, defined by equation \ref{eq:FCF} below, and a DC flux level $f_{DC}$ that accounts for the unknown zero point of the $850\micron$ map.  The latter parameter is necessary because the JCMT measures only a differential signal between chop positions and is insensitive therefore to the absolute zero point of the flux in the map.  The total number of parameters is 6 for the Ostriker model and 8 for the magnetic models, but all 5 of the dimensional parameters are constrained by independent observations or reasonable assumptions.  If one is concerned with the density structure only, and not with the magnetic field geometry, then the degenerate parameters $\beta$ and $\theta$ of the GS model can be replaced by a single independent parameter $\kappa$ (see discussion following equation \ref{eq:Bstod}), thereby decreasing the parameter space of the GS model to 7 dimensions.

We determined the velocity dispersion within the filament from C$^{18}$O~(2--1) and $^{13}$CO~(2--1) maps of the southern half of the G11.11-0.12, including all of region~1.  These observations were obtained with RxA3 on the JCMT during several observing sessions in 2000 July.  Typical line widths were around 0.9~{\rm km~s}$^{-1}$ for C$^{18}$O a 1.2~{\rm km~s}$^{-1}$ for $^{13}$CO.  Since the C$^{18}$O data were often quite noisy, we loosened the constraint on the 1-dimensional velocity dispersion of the gas to be :
\be
0.8~{\rm km~s}^{-1} \appleq \sigma \appleq 1.2~{\rm km~s}^{-1}.
\label{eq:sigmaConstraint}
\ee
We assumed that the central density along the spine of the filament is in the range 
\be
10^{-21}~g~{\rm cm}^{-3} \appleq \rhoc \appleq 10^{-17}~\rm{g~cm}^{-3},
\ee
which corresponds to the range $2.6\times (10^2-10^6)~{\rm cm}^{-3}$ in number density.  The central density actually determined by our modeling procedure is well inside of these bounds.  

The flux conversion factor FCF is given in terms of the 850$\micron$ extinction coefficient $\kappa_{850}$ and dust temperature $T_d$ as follows:
\be
FCF\equiv \left(\frac{\Sigma}{ {\rm g~cm}^{-2}}\right)\left(\frac{f}{ {\rm Jy~beam}^{-1}}\right)^{-1}=0.26\left(\frac{\kappa_{850}}{0.01 {\rm cm}^2~{\rm g}^{-1}}\right)^{-1}\left[\exp\left(17{\rm K}/T_d\right)-1\right]~{\rm g~cm}^{-2}~{\rm Jy}^{-1}~{\rm beam}
\label{eq:FCF}
\ee
(see Paper I).  We combine the observational constraints on $T_d$ and $\kappa_{850}$ into a single constraint on FCF, since the temperature and extinction coefficient enter into our models only through FCF.  The bounds on the $T_d$ and $\kappa_{850}$ are given by
\bea
8  {\rm K} &\appleq& T_d \appleq 15 {\rm K} \nn\\
0.01~{\rm cm}^2~{\rm g}^{-1} &\appleq& \kappa_{850} \appleq 0.02~{\rm cm}^2~{\rm g}^{-1},
\label{eq:TKConstraints} 
\eea
which together imply the constraint
\be
0.27~{\rm g~cm}^{-2}~{\rm Jy}^{-1}~{\rm beam} \appleq FCF \appleq 1.92~{\rm g~cm}^{-2}~{\rm Jy}^{-1}~{\rm beam}.
\label{eq:FCFConstraints}
\ee
The assumptions that lead to these constraints are discussed in Paper I.  Note that the Ostriker solutions presented in Paper I required a $\kappa_{850}$ value of $0.02~{\rm cm}^2~{\rm g}^{-1}$ and a temperature of 13 K.  These values are at or near the upper bounds given by equation \ref{eq:TKConstraints}.

The DC flux level $f_{DC}$ was assumed to be small compared to the flux level over most of the map:
\be
|f_{DC}| \le 0.25~{\rm Jy~beam}^{-1}.
\ee
We verified that the actual $f_{DC}$ values obtained from the model fitting were well within this range.  Thus, we are confident that we have not been too restrictive in setting this constraint.

Finally, we assumed for all models that the axis of the filament lies with 45 degrees of the plane of the sky:
\be
-\frac{\pi}{2} \le i \le \frac{\pi}{2},
\label{eq:inc}
\ee
Note, however, that this quantity is unconstrained by observations, and it is conceivable that G11.11-0.12 might be even more severely inclined.

\section{Parameter Space Exploration and Data Modeling}
\label{sec:GAs}

\subsection{Genetic Algorithm Basics}
\label{sec:GABasics}
Biological evolution can be thought of as a powerful and robust search and optimization algorithm.  Evolution drives individuals within a population toward better adapted forms characterized by greater fitness and improved survivability.  Genetic algorithms are a class of search and optimization algorithms that are based on biological evolution, in which the parameters of a model are encoded on the ``genome'' of an individual, and the goodness of the resulting solution translates into the fitness of the individual, and hence determines the likelihood that it will survive.  Genetic algorithms are often associated with computational intelligence research in computer science, since they learn by experimenting with their computational environment, and are capable of finding innovative solutions to complex optimization and design problems with little human intervention.

Traditional genetic algorithms encode parameters on ``genes'' composed of low-cardinality (often binary) integer strings and process information via crossover, mutation and selection operators.  The biological counterparts of these genetic operators motivate their respective roles.  Crossover combines the parameters represented by two ``parent'' individuals to produce ``offspring'' with intermediate characteristics.  Mutation provides occasional random perturbations to a small subset of the population.  The selection operator serves as an information filter that destroys poor solutions, while allowing good solutions to survive, often by a tournament-based ranking scheme.  These processes operate on a population of solutions iteratively, resulting in a global improvement in the quality of the solutions with each successive generation.  \citet{goldberg89} is the standard introduction to the subject and provides code for a basic genetic algorithm. 

Genetic algorithms are guided stochastic search algorithms, which are much more efficient than random search techniques.  It is the non-random aspects of evolution, namely natural selection and crossover, which make evolution feasible and powerful as an algorithm for search and optimization.  In general, genetic algorithms are extremely robust, and perform well in the presence of multiple local optima, discontinuities, noise and functions with deceptive gradients.  They are suitable for mathematically difficult problems, high-dimensional parameter space searches, problems that aim to find multiple optima or entire classes of solutions, and multi-objective problems.  However, they are not efficient for easy problems that can be handled by straightforward calculus-based techniques.

One of us (J.F.) has developed a multi-objective genetic algorithm called ``Ferret'' with features that are optimized for parameter exploration and optimization problems encountered in the physical sciences.  The algorithm has some similarities to the multi-objective algorithm discussed by \citet{fonseca93}, but we have incorporated several new features to improve the flexibility and performance of the code.  A detailed description of the algorithm is beyond the scope of this paper, but the most significant challenges involved in developing the code are discussed below.

\subsection{Solution Class-Finding \& Multi-Objective Genetic Algorithms}
\label{sec:multiobj}

The data-fitting problem discussed in this paper is an example of a solution ``class-finding'' problem, where the goal is to find a class of acceptable solutions to a problem that obeys a user-defined acceptance criterion.  For the G11.11-0.12 problem, we define a $\chi^2$ that is representative of a good solution, and search for the class of all solutions with $\chi^2$ less than this value. 

Ferret is designed with multi-objective capabilities that will be useful for future polarization modeling work, which would fit $850\micron$ continuum, the polarization position angle, and the polarization percentage simultaneously.  It is important to note that most of the computational machinery needed for a multi-objective code is also required for solution class-finding.  Ferret's multi-objective capabilities and the details of multi-objective optimization are discussed in \citet{europa}, where the code is being used model the internal structure of the Jovian moon Europa, using observational constraints obtained from the Galileo mission to Jupiter.

Simple genetic algorithms suffer generally from a property known as ``genetic drift'', which causes solutions to converge toward a single arbitrary solution within the optimal set.  Genetic drift is analogous to an undesirable loss of diversity within a biological population.  This is problematic because it artificially reduces the extent of the solution class discovered by the algorithm.  This problem must be treated carefully because it is very difficult for most genetic algorithms to recover, once diversity is lost in the solution class.  Sophisticated countermeasures are required to combat genetic drift, which make class-finding and multi-objective codes much more complex than a simple genetic algorithm.  Ferret uses a niching technique similar to that used by \citet{fonseca93}, in addition to several other new techniques that are unique to our code.  Ferret is the only multi-objective genetic algorithm that we are aware of that can recover completely from an extreme (artificially induced) loss of solution diversity within a few generations.

Genetic algorithms have not been used widely in astrophysics, even though they are well-suited for many problems of astrophysical interest.  \citet{charbonneau95} provided a basic single-objective genetic algorithm PIKAIA and used it to solve several illustrative problems including modeling the rotation curves of galaxies, time series analysis of Doppler velocities in spectral lines of $\delta$  Scuti stars, and modeling spherically symmetric magnetized winds from solar-type stars.  This paper and the code provided therein provide a good alternate introduction to genetic algorithms.

\subsection{Code Testing and Verification}
\label{sec:tests}
We tested Ferret on a variety of multi-parameter multi-objective test problems of varying complexity.  These tests included the following:
\begin{enumerate}
\item Simple functions with multiple minima
\item Noise tolerance tests
\item Functions with deceptive gradients, which often fool calculus-based techniques
\item Multi-dimensional solution class-finding problems, using niching and other diversity-preserving strategies
\item Class-finding problems with discontinuities and complex geometries
\item Multi-objective problems, using niching and other diversity-preserving strategies
\item Discrete and semi-discrete problems where some of the variables take only discrete values
\item Critical-parameter detection problems, where the solutions reside in a low-dimensional subspace of the parameter space.  The code improves performance by determining the relative importance of each parameter and focusing on the parameters that are significant to the problem.  This allows solutions to be found in test problems designed with up to 500 parameters.
\end{enumerate}
The code performed well and located the solution set for each of these problems.

We show the results for a test problem that is similar in spirit to the parameter search conducted in this paper.  We consider an N-dimensional function $f=f_1 ~ f_2$, where $f_1$ and $f_2$ are defined as follows:
\bea
f_1 &=& sign\left[\cos\left(\frac{2\pi r_0}{\lambda_1}\right)-\cos\left(\theta+\frac{2\pi r}{\lambda_1}\right)\right] \nn\\
f_2 &=& \sum_{i=3}^N 2 ~round\left(\frac{|x_i-x_0|}{\lambda_2}\right)-floor\left(\frac{|x_i-x_0|}{\lambda_2}\right)
\label{eq:f1f2}
\eea
where $r$ are $\theta$ are cyclindrical coordinates in the $(x_1, x_2)$ subspace defined by
\bea
r=\sqrt{(x_1^2+x_2^2)} \nn\\
\theta = tan^{-1} (x_2/x_1),
\eea
and $r_0$ is a constant.  The functions ``sign(x)'', ``round(x)'' and ``floor(x)'' are the familiar discontinuous functions that are available in most programming languages.  We designed this rather complicated function as a non-trivial test of several features of the code that are critical to its success on data modeling problems.  Specifically, we test the ability of the code to work efficiently in very high-dimensional parameter spaces, to avoid being trapped by numerous local minima, and to function as a solution class-finder by distributing solutions over a finite-volumed optimal region.  Furthermore, our test problem has a simple solution, but cannot be solved by any calculus-based numerical approach because the gradient is zero everywhere, except at discontinuities.  

\begin{figure}
\includegraphics[width=.49\textwidth]{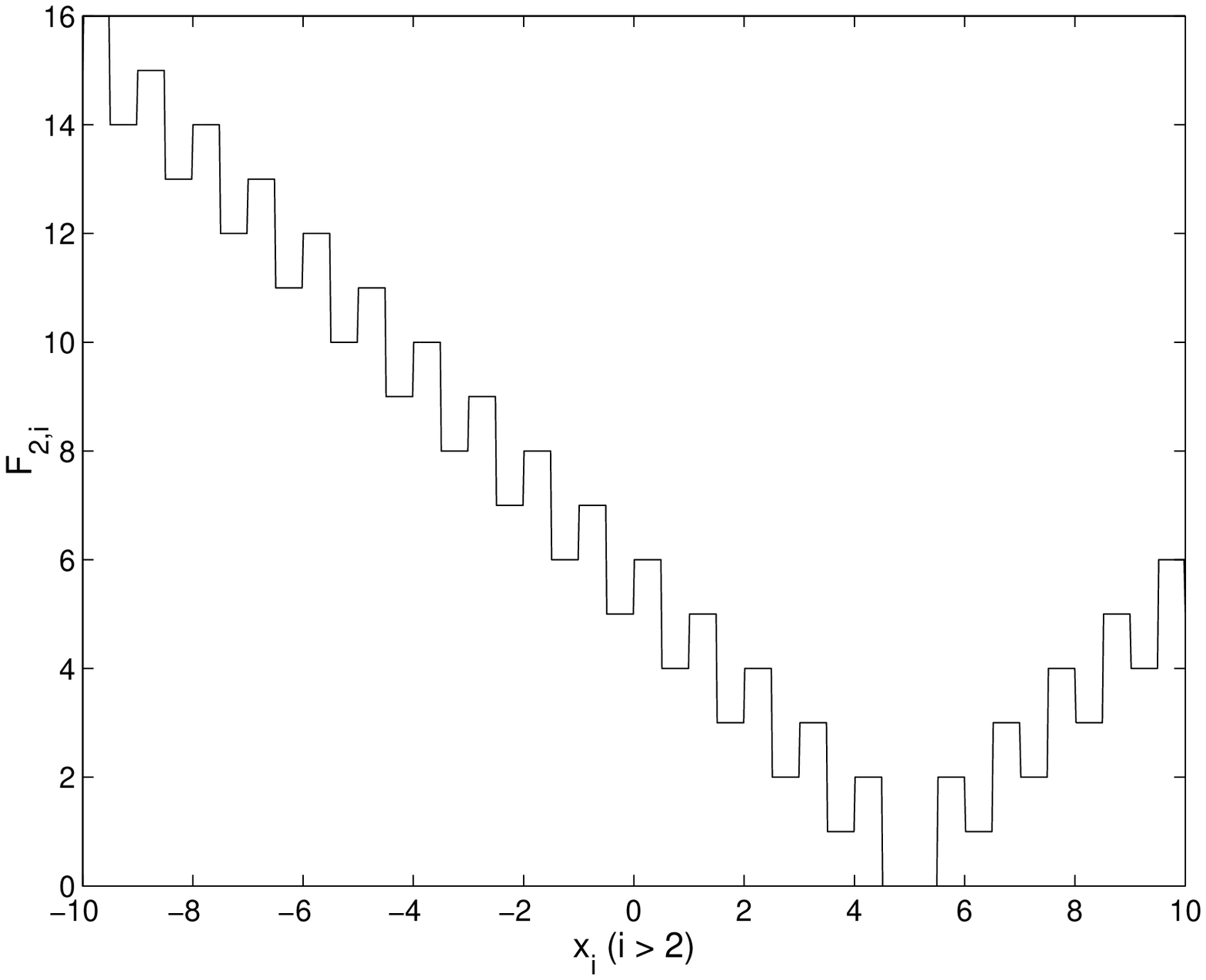}
\includegraphics[width=.49\textwidth]{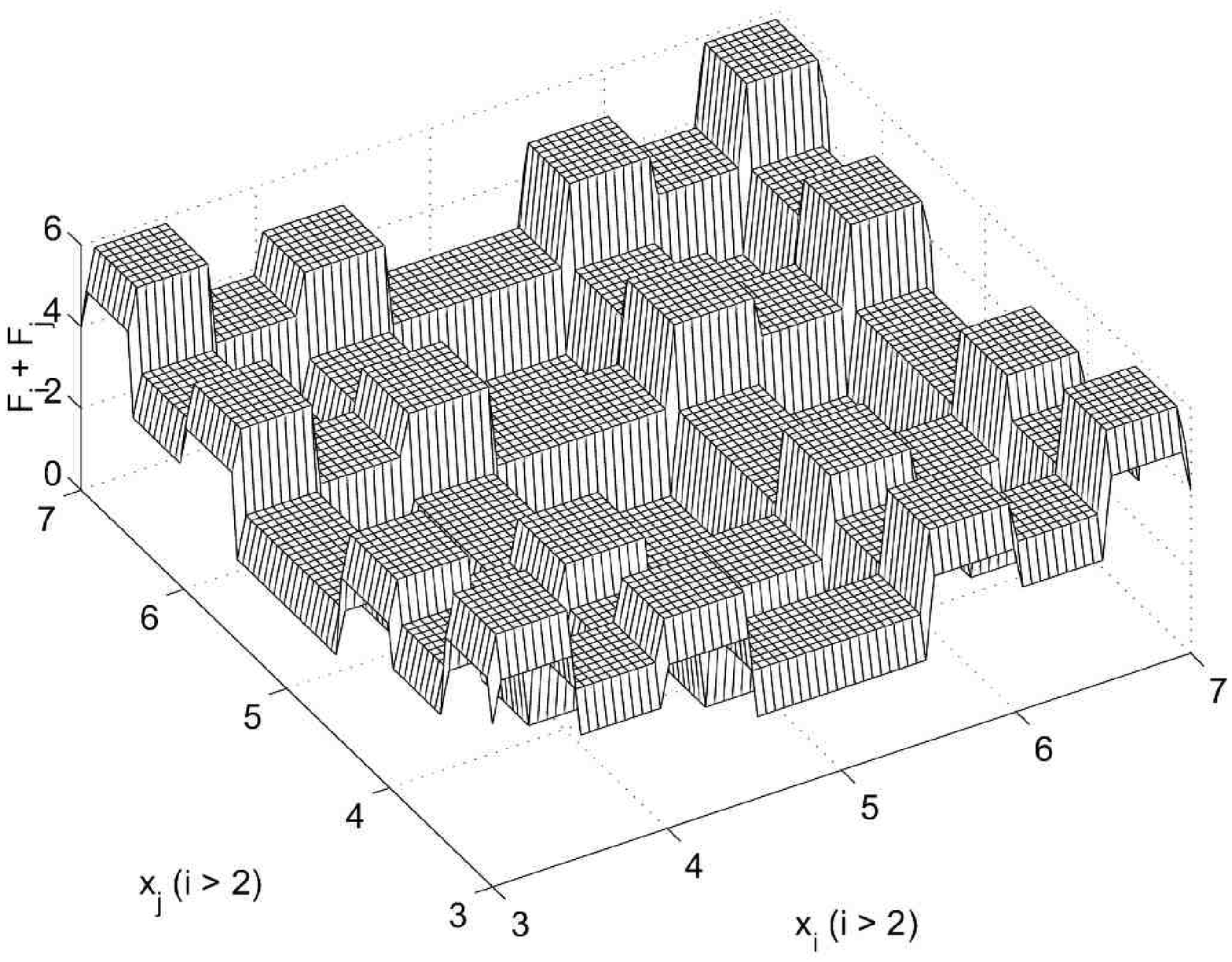}

\includegraphics[width=.49\textwidth]{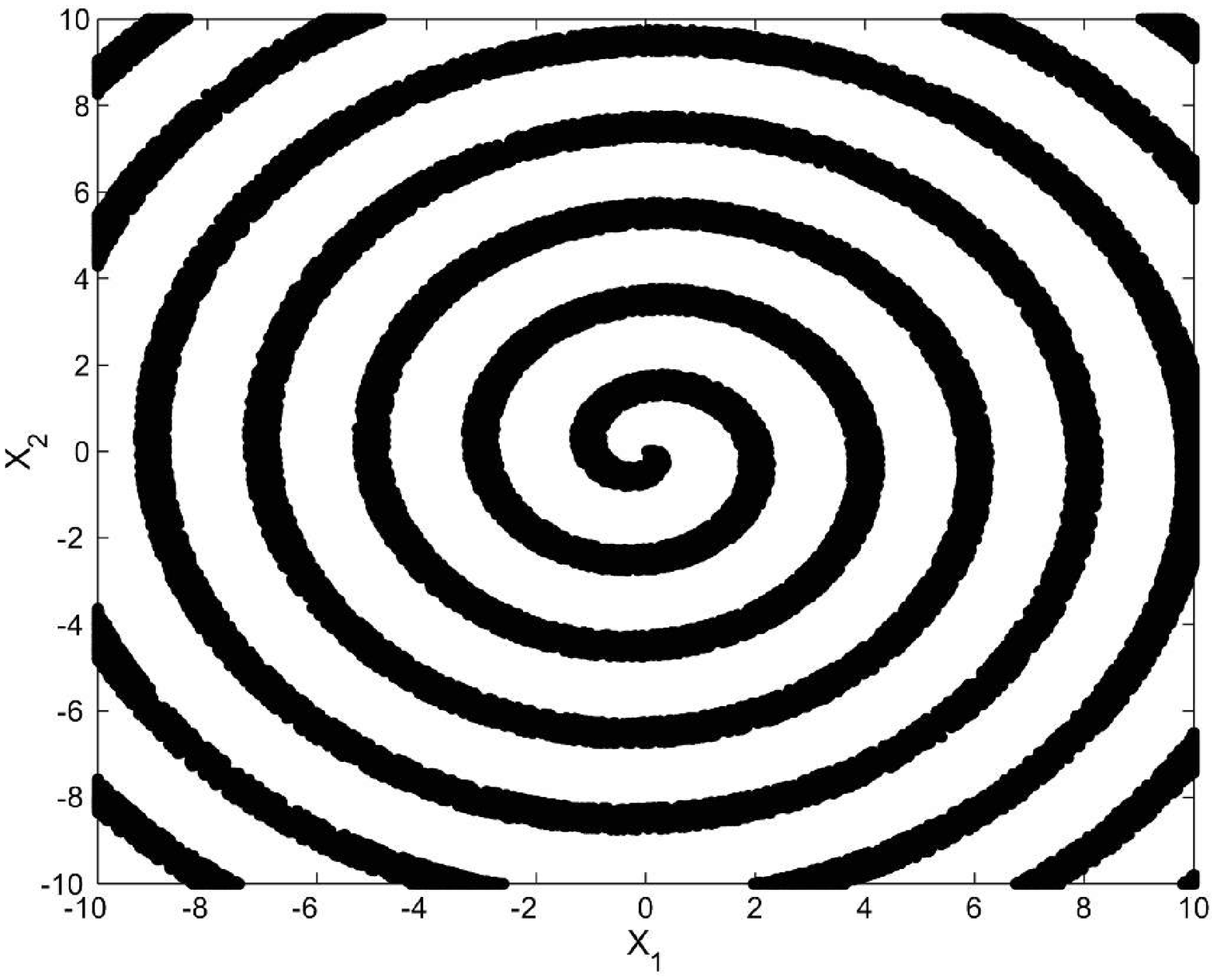}
\includegraphics[width=.49\textwidth]{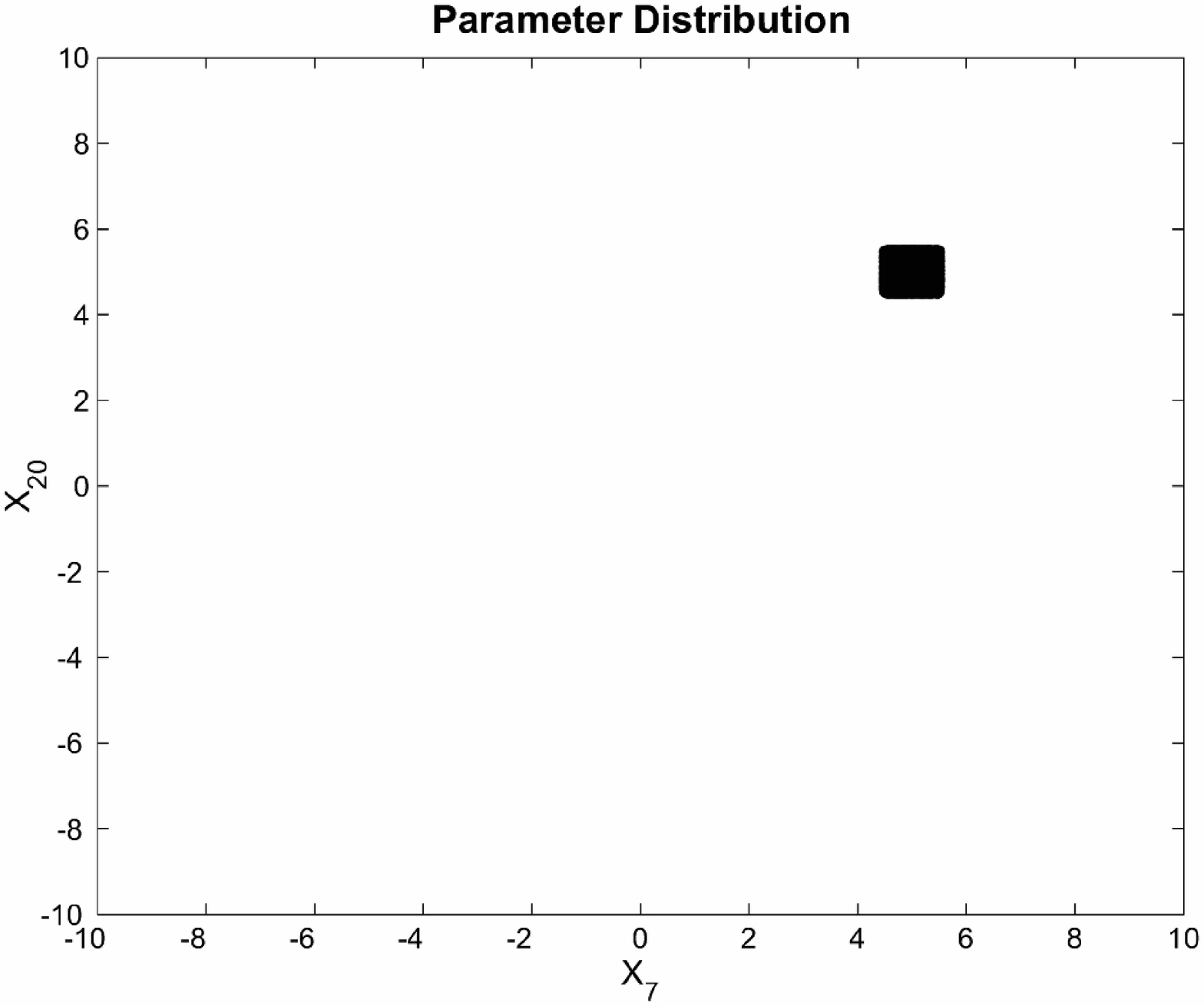}
\caption{High dimensional test problem for Ferret genetic algorithm with many ($\sim 9.4 \times 10^{23}$) local minima.  Panel a (top left): One term in the function $f_2$ given by equation \ref{eq:f1f2}.  Panel b (top right): Two terms of function $f_2$, zooming in on the global minimum at (5,5).  Panel c (lower left): Projection of the solution set in the $(x_1, x_2)$ subspace showing the flat-bottomed spiral minimum given by function $f_1$.  Panel d (lower right): Projection of the solution set in the $(x_{7},x_{20})$ subspace.  The solution set looks like this figure in all subspaces that do not include $x_1$ or $x_2$.}
\label{fig:test}
\end{figure}

The function $f_1$ is a discontinuous function with a long, flat-bottomed spiral whose minimum has thickness $2*r_0$ and repeats with radial wavelength $\lambda_1$ in the $(x_1, x_2)$ subspace.  In this finite domain, $f_1$ breaks up into 9 distinct minima of equal depth.  The function $f_2$ is another discontinuous function designed to generate many local minima, spaced by wavelength $\lambda_2$, and a single global minimum of width $\lambda_2$ centred at $x_0$.  We chose $N=20$,  $x_0=5$, and a range of $-10 \le x_i \le 10$ for all parameters.  Figure \ref{fig:test}a shows one term of $f_2$, where $\lambda_2=1$ and, over the same range as above.  Figure \ref{fig:test}b shows the sum over any two terms, where we have zoomed in to the region near the global minimum at (5,5).  

On our chosen range, each term of $f_2$ generates 19 local minima, including a single global minimum.  The summation over $N-2$ terms results in $19^{N-2}$ minima.  Multiplying by the 9 minima in $f_1$, we find that the combined function $f$ has $9\times 19^{N-2} = 9.4 \times 10^{23}$ isolated, finite-volumed minima, of which only 9 are global minima.  Our goal is to map out the 9 global minima while evading all other local minima, which should reveal the spiral structure of $f_1$ in the $(x_1, x_2)$ projection, and look like a $1 \times 1$ square box centred on (5,5) in any projection that does not include $x_1$ or $x_2$.

We ran the code using a population size of 1000 for 500 generations, and performed several trials.  In each trial, all global minima were found and covered with points in less than 100 generations.  Subsequent generations increased the density of points and improved the resolution of the boundaries.  Figure \ref{fig:test}c shows the projection of the solution set in the $(x_1, x_2)$ subspace, and Figure \ref{fig:test}d shows an example of another projection.  All subspaces that do not include $x_1$ or $x_2$ look like Figure \ref{fig:test}d.

This problem is more difficult that the parameter searches required to model the G11.11-0.12 cloud because our astrophysical models have fewer (6-8) parameters, and few local minima, other than the global minimum.  We conclude that our code succeeded on this test problem, and should be adequate for the modeling conducted later in this paper.

\subsection{The Objective Function for G11.11-0.12 models}
\label{sec:objfun}
In Paper I, we fit two \citet{ostriker64} solutions to the $850\micron$ continuum data for G11.11-0.12.  We extend this analysis by exploring and constraining the parameter spaces for the 3 models discussed in $\S$\ref{sec:theory}.  To accomplish this goal, we need to locate the class of solutions for each model, which has a reduced $\chi^2<\chi^2_{min}$ where $\chi^2_{min} > 1$ but still of order unity.  We also need to rule out solutions or particular regimes of each model (i.e. poloidal-field dominated models, toroidal-field dominated models, etc.) where this condition is not met.

We verified the success of the code by checking the values of $\chi^2$ {\em a posteriori}.  We ran the genetic algorithm typically for 500 generations, although a stable population of models satisfying our $\chi^2$ criterion usually developed after only $\sim 30-40$ generations.  Subsequent generations served to fill out the optimal class of solutions and to improve the resolution of the edges of the optimal region of parameter space.  In practice, we save all solutions that are in the optimal set of each generation, and filter through all generations at the end of the run to select all models from all previous generations that satisfied the $\chi^2$ criterion.  Our results show that the best solutions discovered are usually surrounded by a sheath of poorer solutions, as one might expect.  We ran the code several times while varying the population size and other genetic algorithm control parameters, to ensure that that the solution set was repeatable.

We normally display the output from our genetic algorithm as a scatter plot that represents a 2-dimensional projection of the region of parameter space that agrees with the data, according to our $\chi^2$ criterion.  In practice, a well-designed multi-objective or class-finding genetic algorithm distributes solutions as evenly and completely as possible throughout the optimal region.  However, there is no strict mathematical guarantee that this will occur.  The safest interpretation of our scatter plots is simply that the locus of points covers the multi-dimensional region of parameter space allowed by the observations, without attempting any further inference regarding the uniformity of the points or the probability distribution of the solutions.

As noted in \S\ref{sec:multiobj}, multi-objective problems require nearly the same computational machinery as the single-objective class-finding problem that is the focus of this paper.  Future work will focus on 2-dimensional polarization modeling, which will fit the intensity map, the polarization position angle and the polarization percentage simultaneously across the map. 

\section{Results}
\subsection{The Truncated Ostriker Model}
\label{sec:OstrikerResults}

\begin{figure}
\includegraphics[width=.49\textwidth]{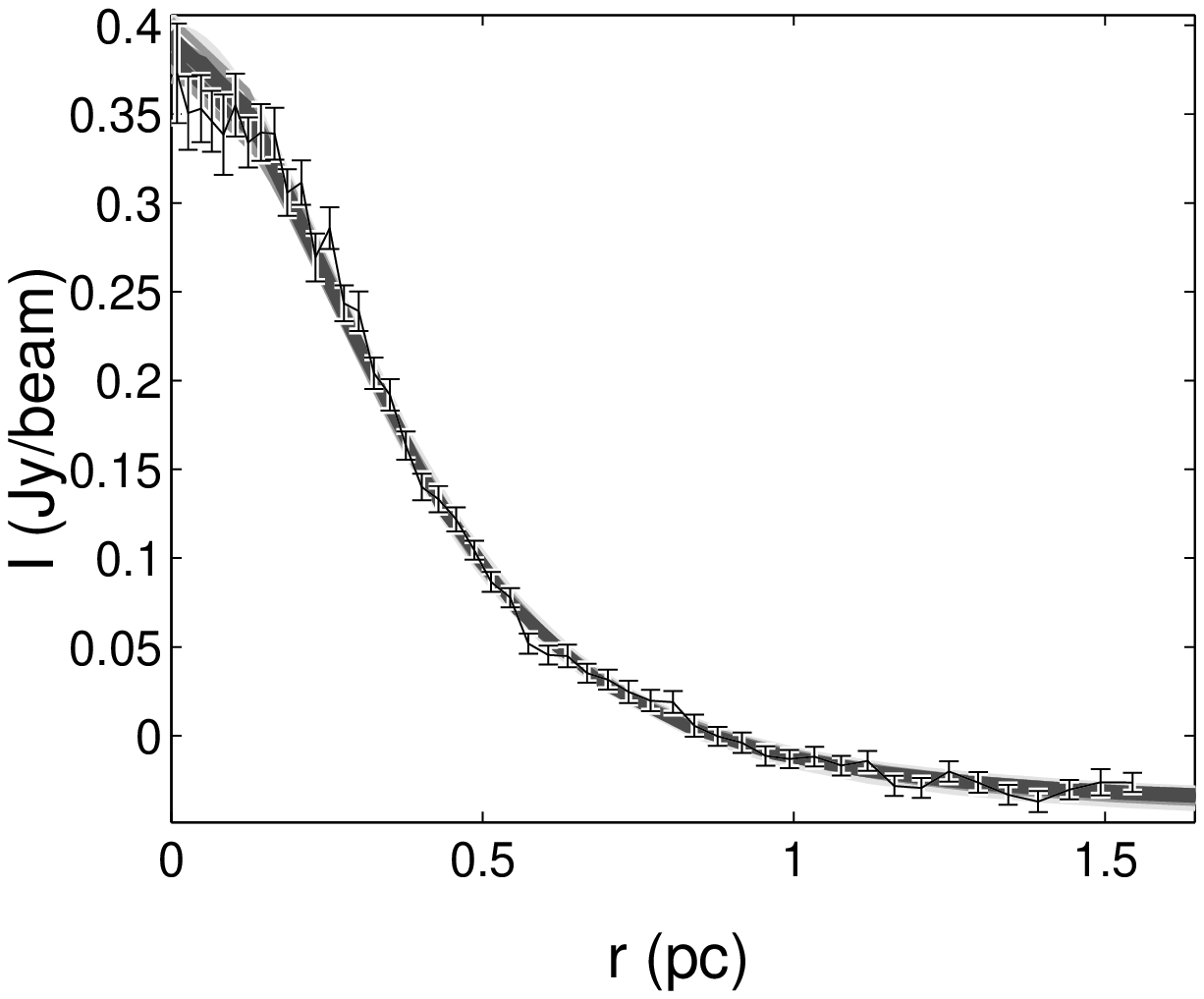}
\includegraphics[width=.49\textwidth]{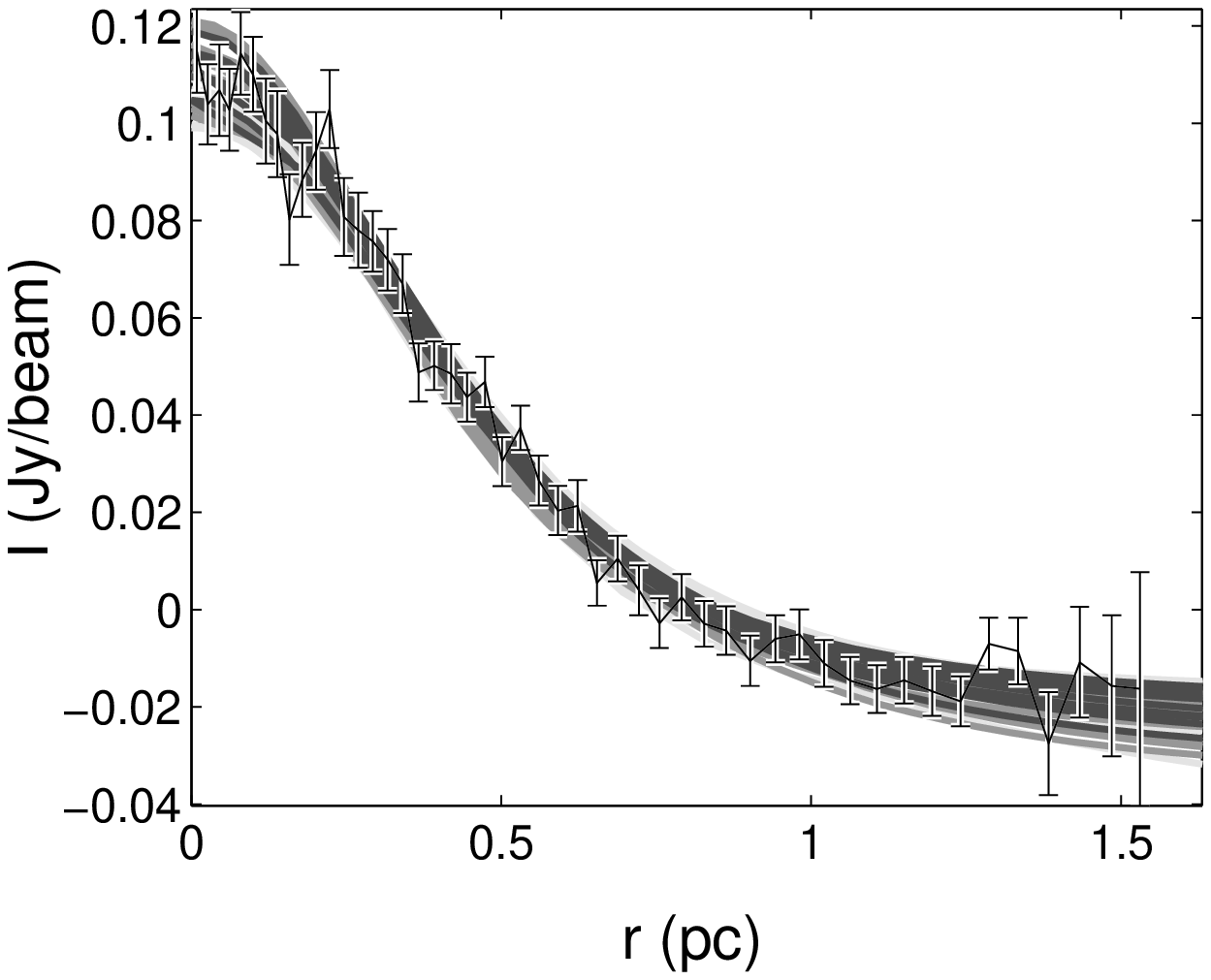}
\caption{Sample Ostriker model fits to the data.  Region 1 is on the left and region 2 is on the right.  The grayscale indicates the quality of the fit: black indicates solutions with $\chi^2 \le 1.167$, medium gray indicates solutions with $1.167< \chi^2 \le 1.333$, and light gray indicates solutions with $1.333 < \chi^2 \le 1.5$.  We use this convention in all of our scatter plots, except for those of \S\ref{sec:polarimetry}.  We always plot points in order of decreasing range of $\chi^2$, so that darker points, representing solutions of higher quality, are plotted after lighter points and may obscure them in some cases.}
\label{fig:OstFit}
\end{figure}

Figure \ref{fig:OstFit} shows examples of truncated Ostriker model fits to regions 1 (left panel) and 2 (right panel).  The radial profile of region 1 is smoother and less noisy than that of region 2, which results in better constrained fits.  The grayscale indicates the quality of the solutions: black designates solutions with $\chi^2 \le 1.167$, medium gray designates solutions with $1.167<\chi^2 \le 1.333$, and light gray is for solutions with $1.333< \chi^2 \le 1.5$.  We discuss the distribution of these solutions within the parameter space of the Ostriker model below.

\subsubsection{Observationally Constrained Parameters}
\label{sec:OstrikerResultsObs}

% NEW JUNE 2004 RUNS.
%
% Ost_F1
% ======
% nc: 0.98269, 2.3942 x 10^4 cm^-3
% sigma: 0.864592, 1.2 km/s
% C: 0.875279, 2
% FCF: 0.2738, 0.52934

% Ost_F2
% ======
% nc: 0.354422, 1.9904 x 10^4 cm^-3
% sigma: 0.8, 1.2 km/s
% C: 0.8, 2
% FCF: 0.30996, 1.5779

\begin{figure}
\includegraphics[width=.49\textwidth]{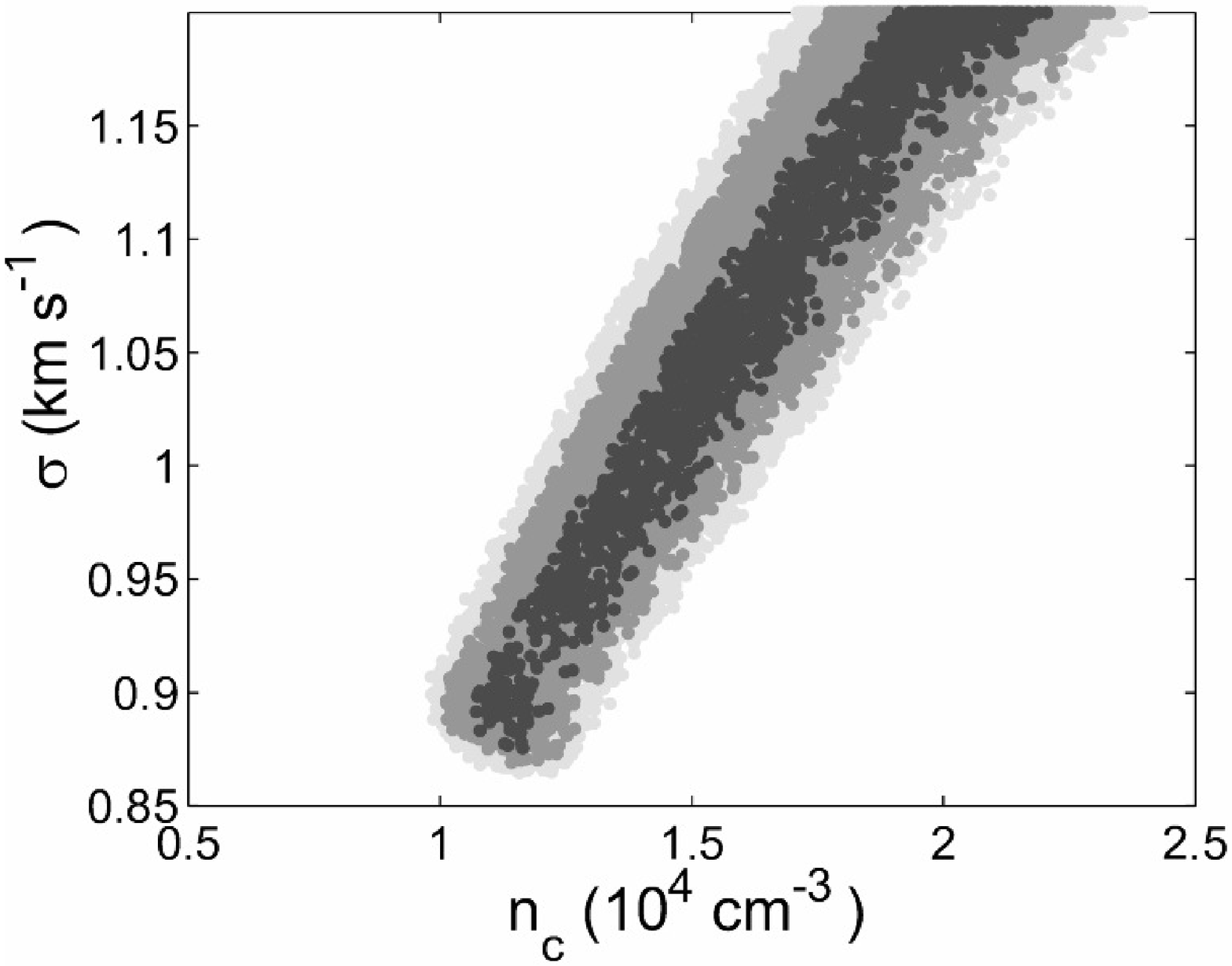}
\includegraphics[width=.49\textwidth]{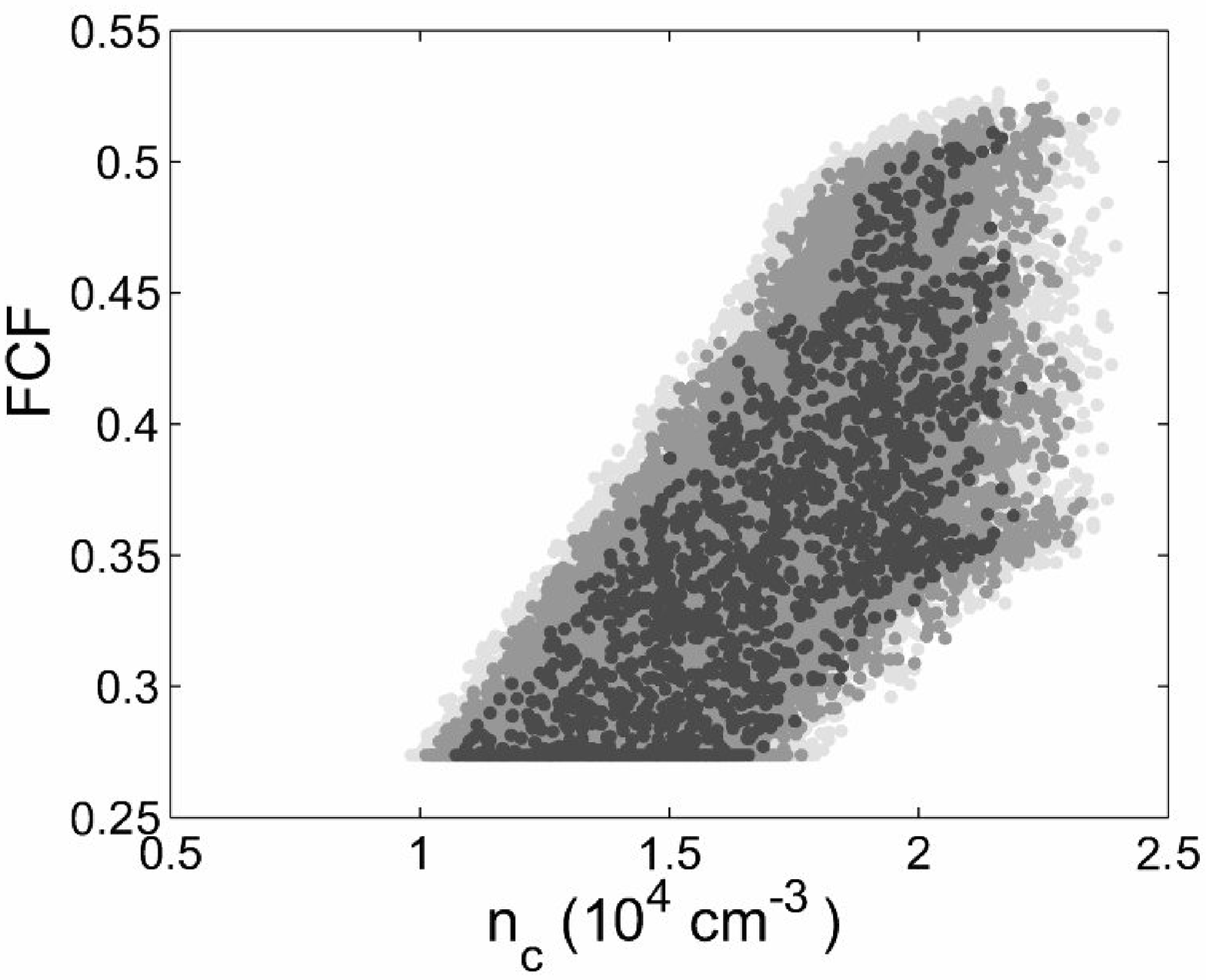}
\caption{Two views of the observationally constrained part of the parameter space $(\rhoc, \sigma, FCF)$ of the truncated Ostriker models for region 1, with $\chi^2\le 1.5$.  The grayscale indicates the quality of the solutions, according to the convention of Figure \ref{fig:OstFit}.  The best solutions form a core surrounded by an envelope of poorer solutions.}
\label{fig:Ost_F1_obs}
\end{figure}

Figure \ref{fig:Ost_F1_obs} shows the set of truncated Ostriker solutions in the observationally constrained part of the parameter space $(\rhoc, \sigma, FCF)$ for region 1.  We found solutions with central number densities in the range $0.98\times 10^4~{\rm cm}^{-3} \appleq n_c \appleq 2.4\times 10^4~{\rm cm}^{-3}$ and velocity dispersion constrained to the range $0.86~{\rm km~s}^{-1} \appleq \sigma \appleq 1.2~{\rm km~s}^{-1}$.  The flux conversion factor is constrained to values near the lower limit of the allowed range given by equation \ref{eq:FCFConstraints}: $0.27~{\rm g~cm}^{-2}~{\rm Jy}^{-1}~{\rm beam} \appleq FCF \appleq 0.53~{\rm g~cm}^{-2}~{\rm Jy}^{-1}~{\rm beam}$.

Our analysis found a central number density that is somewhat lower than the value given for the southern region by Paper I, which gave $n_c=(3.4\pm 0.6)\times 10^4~{\rm cm}^{-3}$ for region 1.  The grayscale indicates the quality of the solutions, following the same convention outlined in the caption of Figure \ref{fig:OstFit}.  There are no apparent gaps in the solution set, and the best solutions are surrounded by a sheath of poorer solutions, as expected.

We obtained similar results for region 2, but the solutions are much less well-constrained than in region 1.  The central number density of region 2 is constrained to the range $3.5 \times 10^3~{\rm cm}^{-3} \appleq n_c \appleq 2.0\times 10^4~{\rm cm}^{-3}$.  The velocity dispersion spanned the entire range of allowed values from $0.8~{\rm km~s}^{-1} \appleq \sigma \appleq 1.2~{\rm km~s}^{-1}$.  Allowed values of the flux conversion factor ranged from $0.31~{\rm g~cm}^{-2}~{\rm Jy}^{-1}~{\rm beam} \appleq FCF \appleq 1.6~{\rm g~cm}^{-2}~{\rm Jy}^{-1}~{\rm beam}$, which covers a significant fraction of the allowed range.  The central number density is consistent with the value of $(1.1\pm 0.3)\times 10^4~{\rm cm}^{-3}$ obtained in Paper I, but the range of densities that we found indicates that the uncertainty in their estimate was probably underestimated.

\begin{figure}
\includegraphics[width=.49\textwidth]{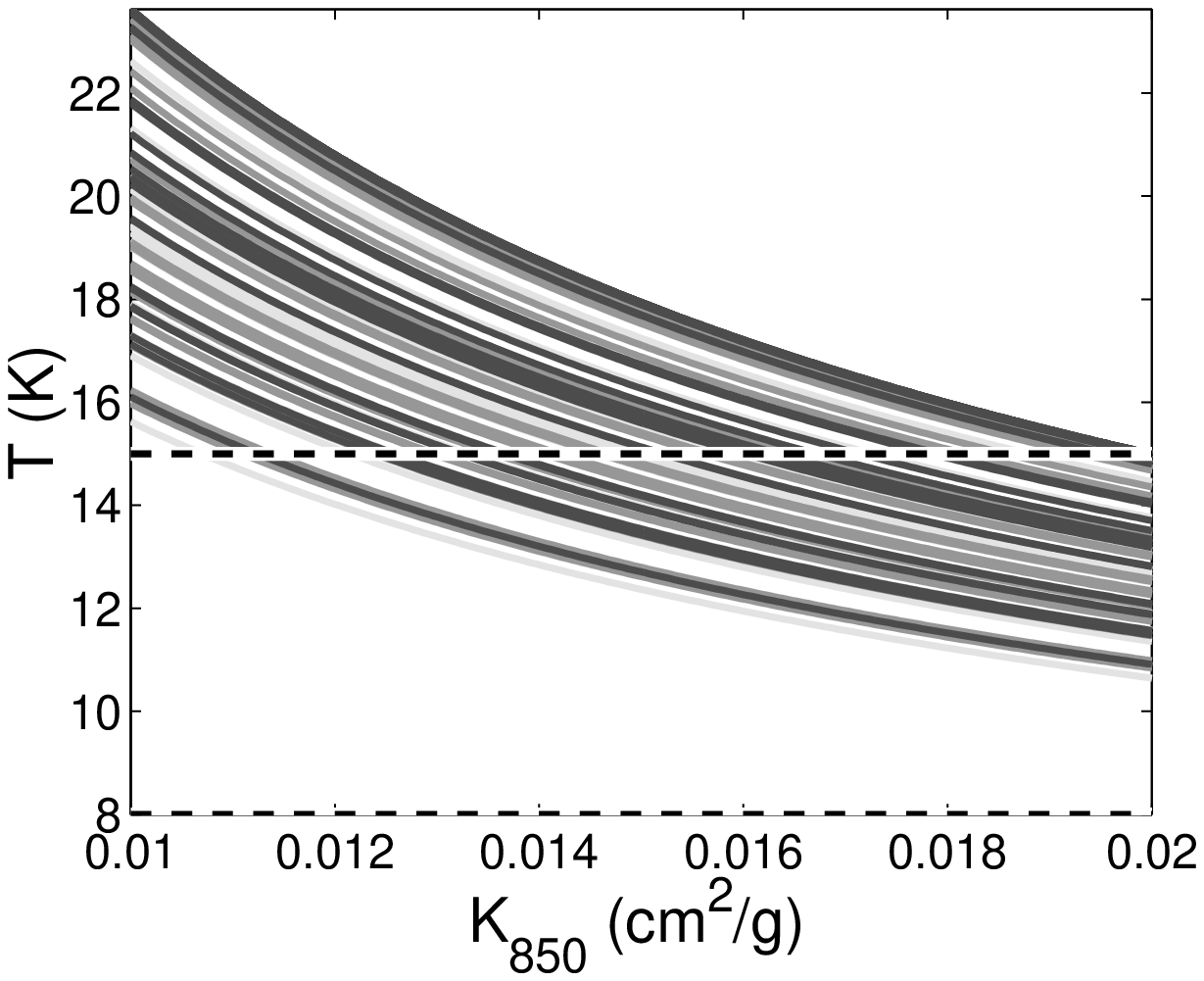}
\includegraphics[width=.49\textwidth]{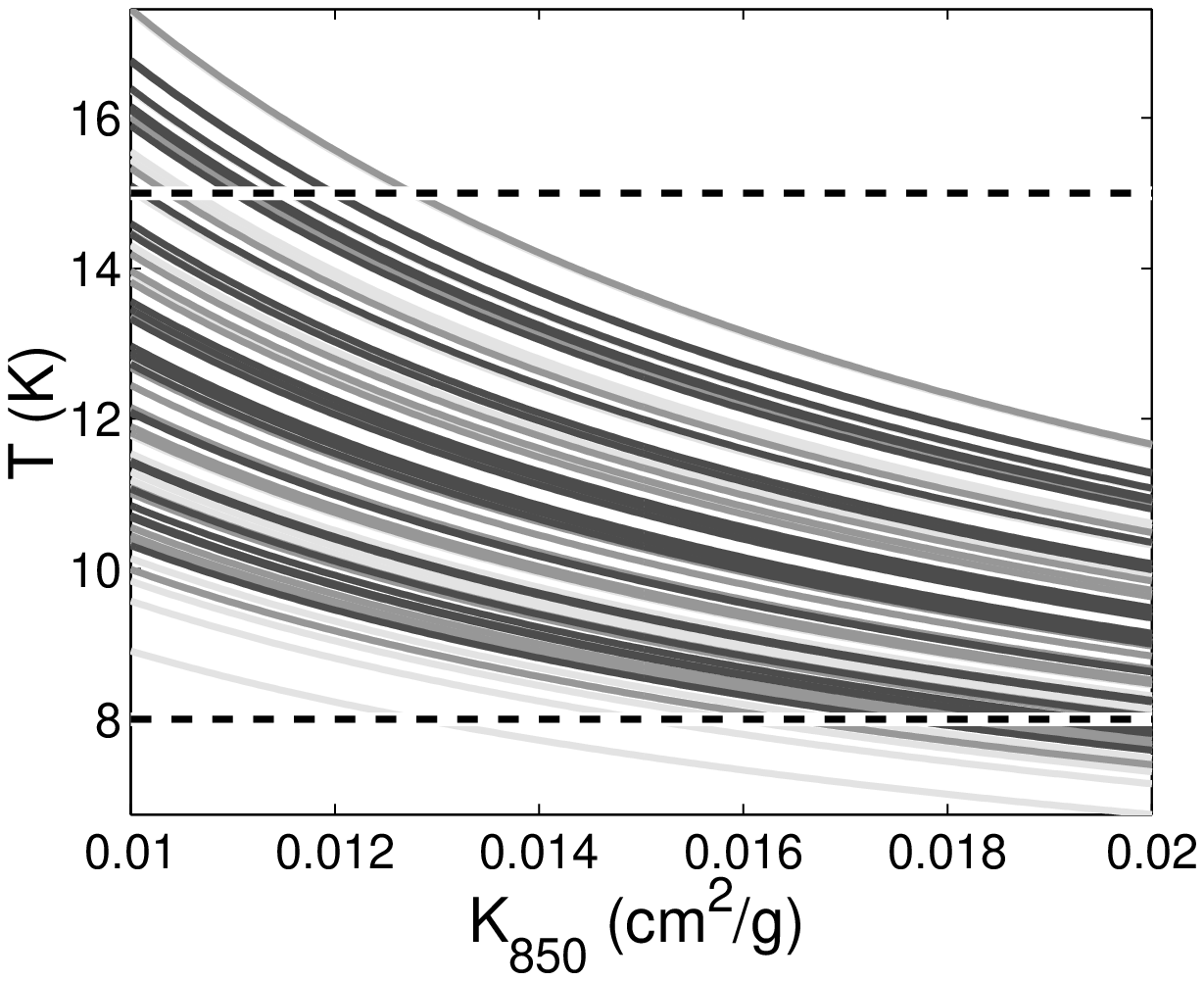}
\caption{Constraints on the temperature and 850$\micron$ extinction coefficient of the truncated Ostriker model.  We show a random sample of solutions for region 1 (left) and region 2 (right).  Each curve corresponds to a single value of the FCF for a single model, as defined by equation \ref{eq:FCF}.  The grayscale follows the convention of Figure \ref{fig:OstFit}.}
\label{fig:OstTK}
\end{figure}

As we discussed in \S\ref{sec:models}, our modeling technique combines the temperature (T) and 850 $\micron$ emissivity ($\kappa_{850}$) into a single flux conversion factor (FCF).  Each value of the FCF corresponds to a one parameter set of (T, $\kappa_{850}$) values, according to equation \ref{eq:FCF}.  Figure \ref{fig:OstTK} plots a random sample of these curves to show the possible ranges of these parameters for region 1 (left panel) and region 2 (right panel).  The dashed lines in this figure indicate the observational constraints on the temperature.  The results for region 1 require values of T and $\kappa_{850}$ that are in the upper range of the values given by equation \ref{eq:TKConstraints}.  This is consistent with the model presented in Paper I, which used a temperature of 13 K and an emissivity of $\kappa_{850}=0.02~{\rm cm~g}^{-1}$.  Region 2 is less well-constrained, but it is consistent with a broad range of T and $\kappa_{850}$ covering most of the allowed range.

The solution sets for both regions span the entire range of inclination angles allowed by equation \ref{eq:inc}.  We did not find any interesting correlations between the inclination angle and any other quantity.  Obviously, the inclination angle must be correlated with other parameters, but the underlying relationship is not evident in any 2-dimensional projection of the parameter space that can be displayed here.

\subsubsection{Theoretical Parameters}
\label{sec:ostrikerTheoryParameters}
\begin{figure}
\includegraphics[width=.49\textwidth]{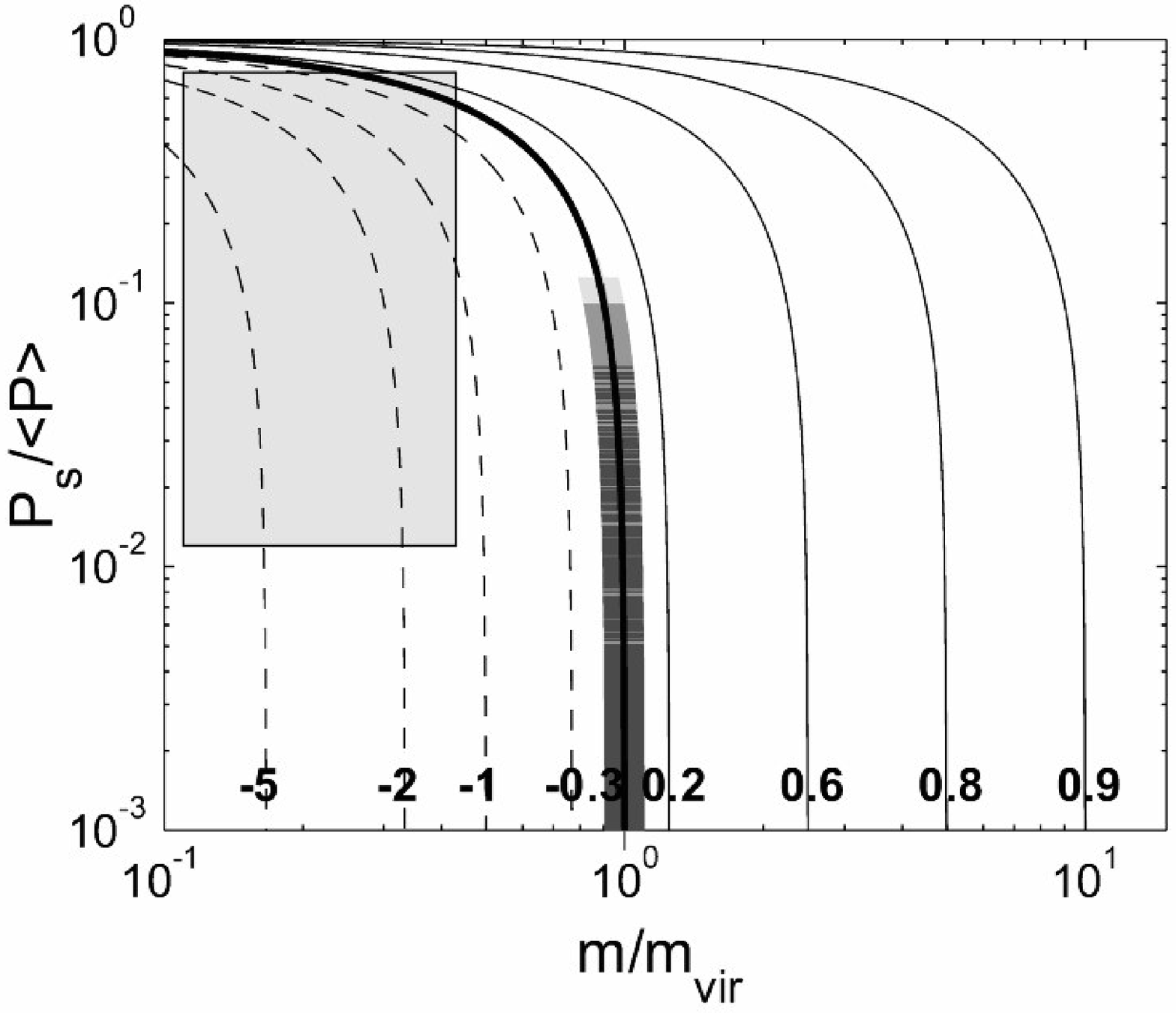}
\includegraphics[width=.49\textwidth]{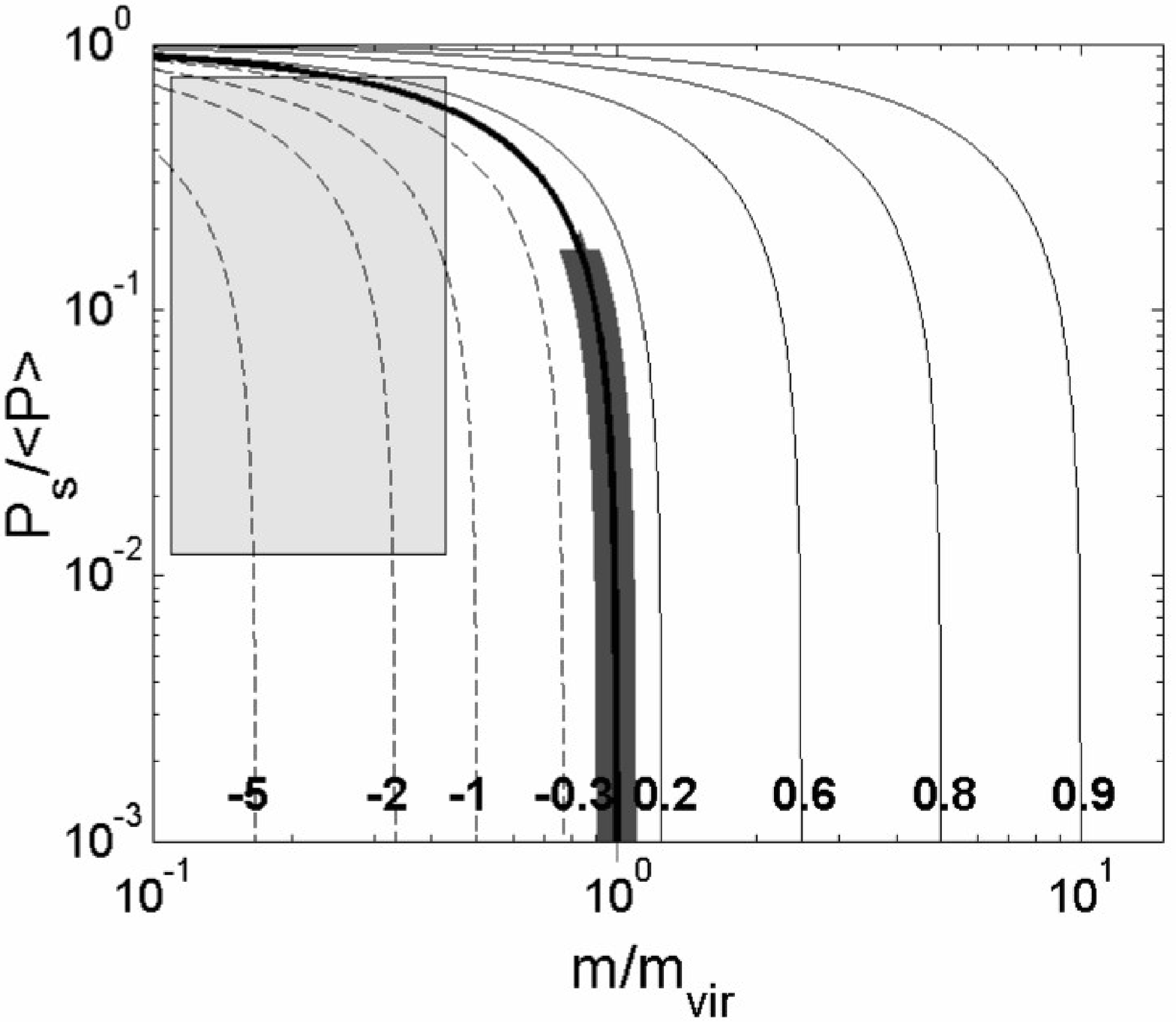}
\caption{The virial parameters of the truncated Ostriker model solutions.  Results for region 1 are shown on the left, and region 2 on the right.  The quantities $m/\mvir$ and $\Ps/<P>$ obey equation \ref{eq:virial1} for the truncated Ostriker model, and equation \ref{eq:virial} for more general magnetic models.  Magnetic models dominated by a toroidal field would lie to the left of this curve, and poloidal-field dominated models would lie to the right. The numbers labeling the curves indicate values of the magnetic parameter $\M/|\W|$ (see equation \ref{eq:virial}).  The gray box indicates the location of the toroidal-field dominated filamentary clouds studied by FP1.  This box is almost entirely on the $\M/|\W| < 0$ side of the diagram.  This region is inaccessible to the non-magnetic truncated Ostriker solutions, which are always constrained to lie on the magnetically neutral line (heavy line).  The grayscale follows the convention of Figure \ref{fig:OstFit}.}
\label{fig:OstVirial}
\end{figure}

Figure \ref{fig:OstVirial} shows the virial parameters of the truncated Ostriker model solutions.  The quantities $m/\mvir$ and $\Ps/<P>$ obey equation \ref{eq:virial1} for the truncated Ostriker model, and equation \ref{eq:virial} for more general magnetic models.  FP1 showed that these virial quantities provide an important constraint on the geometry of the magnetic field, since equation \ref{eq:virial} allows one to deduce the value of the magnetic parameter $\M/|\W|$ if $m/\mvir$ and $\Ps/<P>$ can be estimated.  Truncated Ostriker models always lie on the magnetic neutral line (dotted line) of this diagram.  It is notable that G11.11-0.12 appears to have a lower value of $\Ps/<P>$ and a much higher value of $m/\mvir$ than the filaments that were studied by FP1.  The fact that good fits were found to the Ostriker model indicates that the line mass of G11.11-0.12 is high enough for self-gravity to hold the filament together without additional binding energy due to a toroidally dominated magnetic field.

\subsection{The Generalized \Stod Model}
\label{sec:StodResults}

% NEW JUNE 2004 RUNS
%
% Stod_X1
% ======
% nc: 0.98372, 12.5724 x 10^4 cm^-3
% sigma: 0.8, 1.2 km/s
% C: 0.883544, 2
% FCF: 0.2738, 1.917
% logMagPar: -1, 1.0697

% Stod_X2
% ======
% nc: 0.396512, 3.5432 x 10^4 cm^-3
% sigma: 0.8, 1.2 km/s
% C: 0.75, 2
% FCF: 0.352347, 1.917
% logMagPar: -1, 0.70614

\begin{figure}
\includegraphics[width=.49\textwidth]{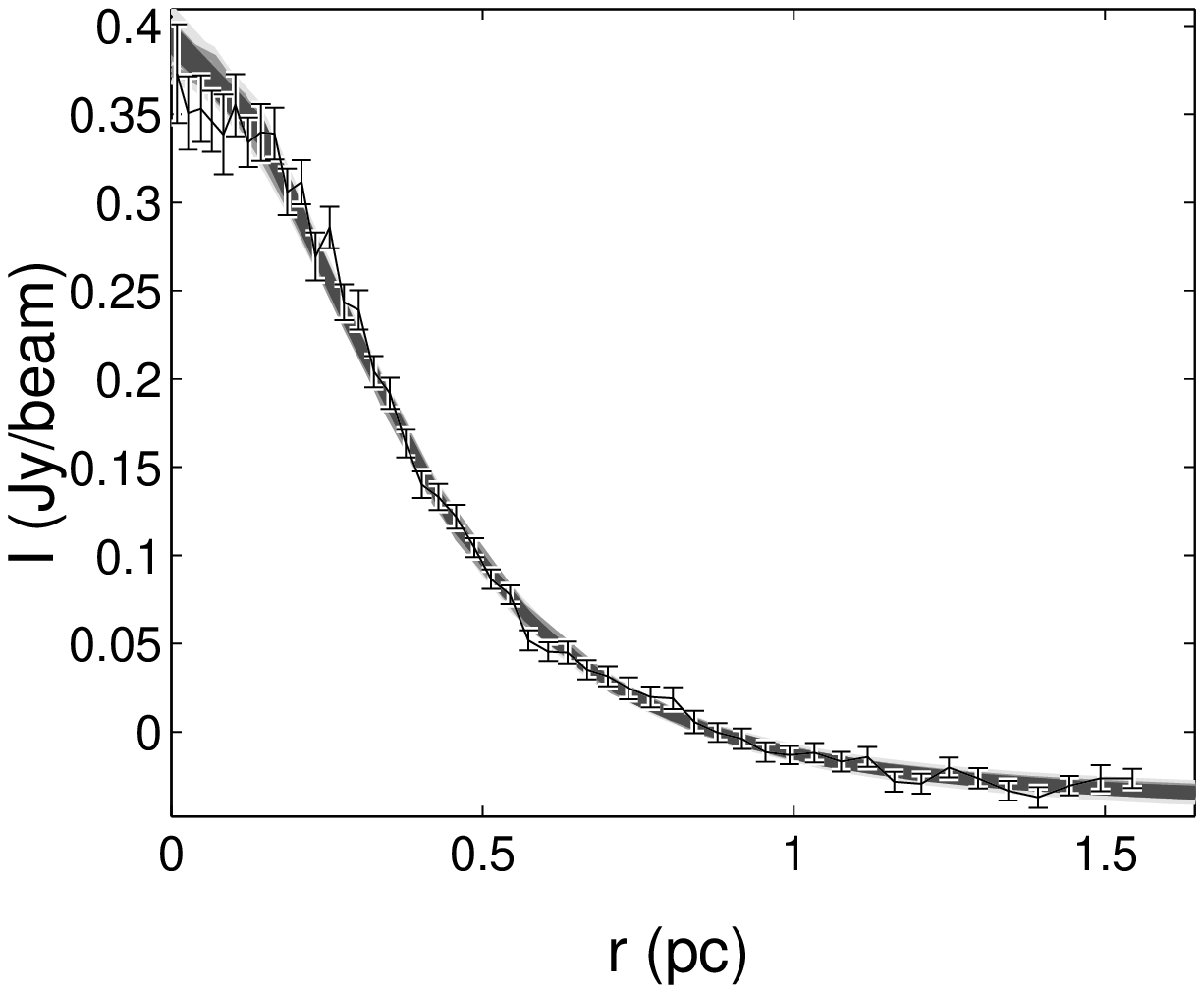}
\includegraphics[width=.49\textwidth]{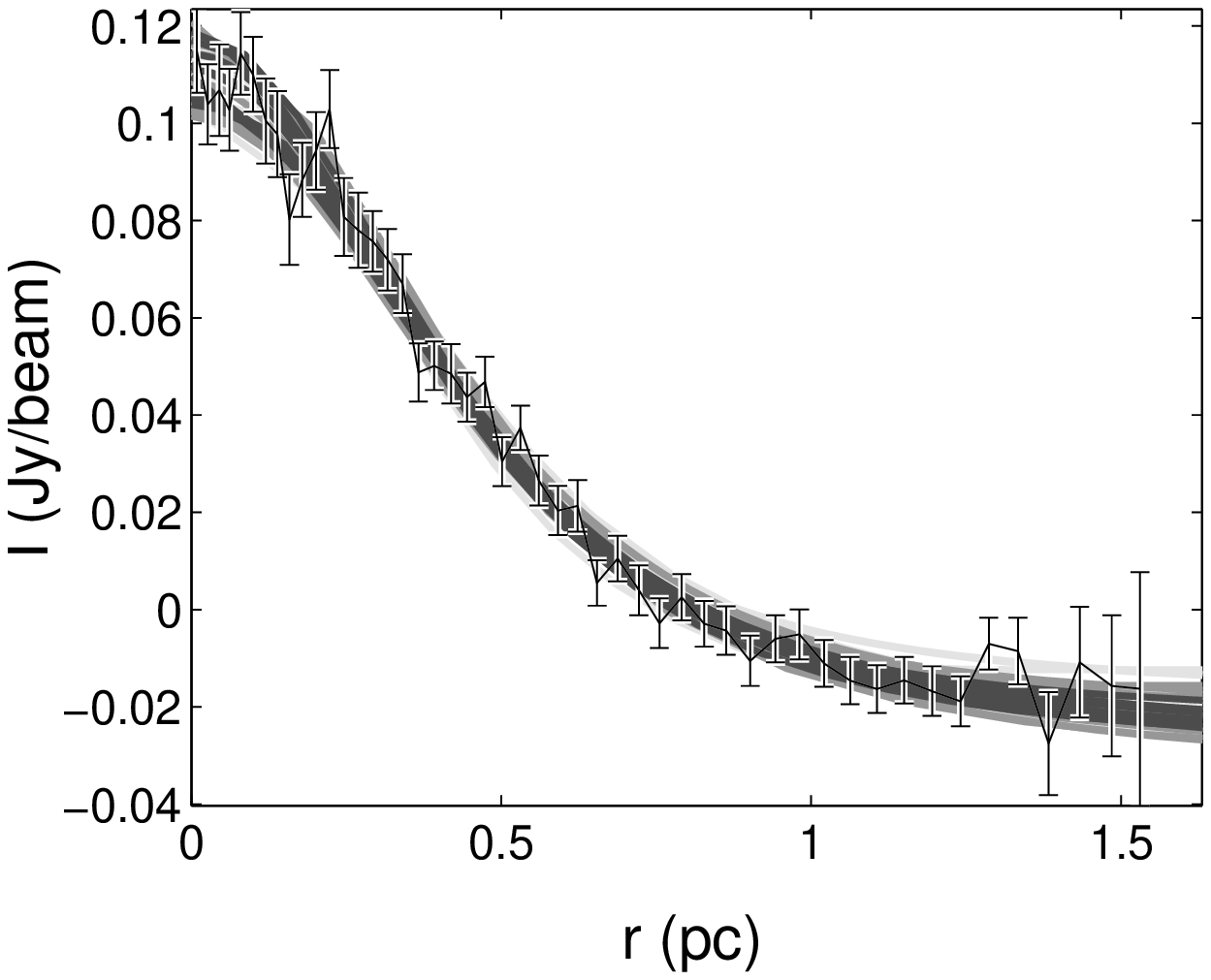}
\caption{Sample GS model fits to the 850 $\micron$ intensity data.  Region 1 is on the left and region 2 is on the right.  The grayscale indicates the goodness of fit, following the convention of Figure \ref{fig:OstFit}.}
\label{fig:StodFit}
\end{figure}

Figure \ref{fig:StodFit} shows examples of GS model fits to our data.  Both of the regions that were analyzed are well-fit by the model, but the solutions are not as well-constrained in region 2.

\subsubsection{Observationally Constrained Parameters}
\label{sec:StodResultsObs}
\begin{figure}
\includegraphics[width=.49\textwidth]{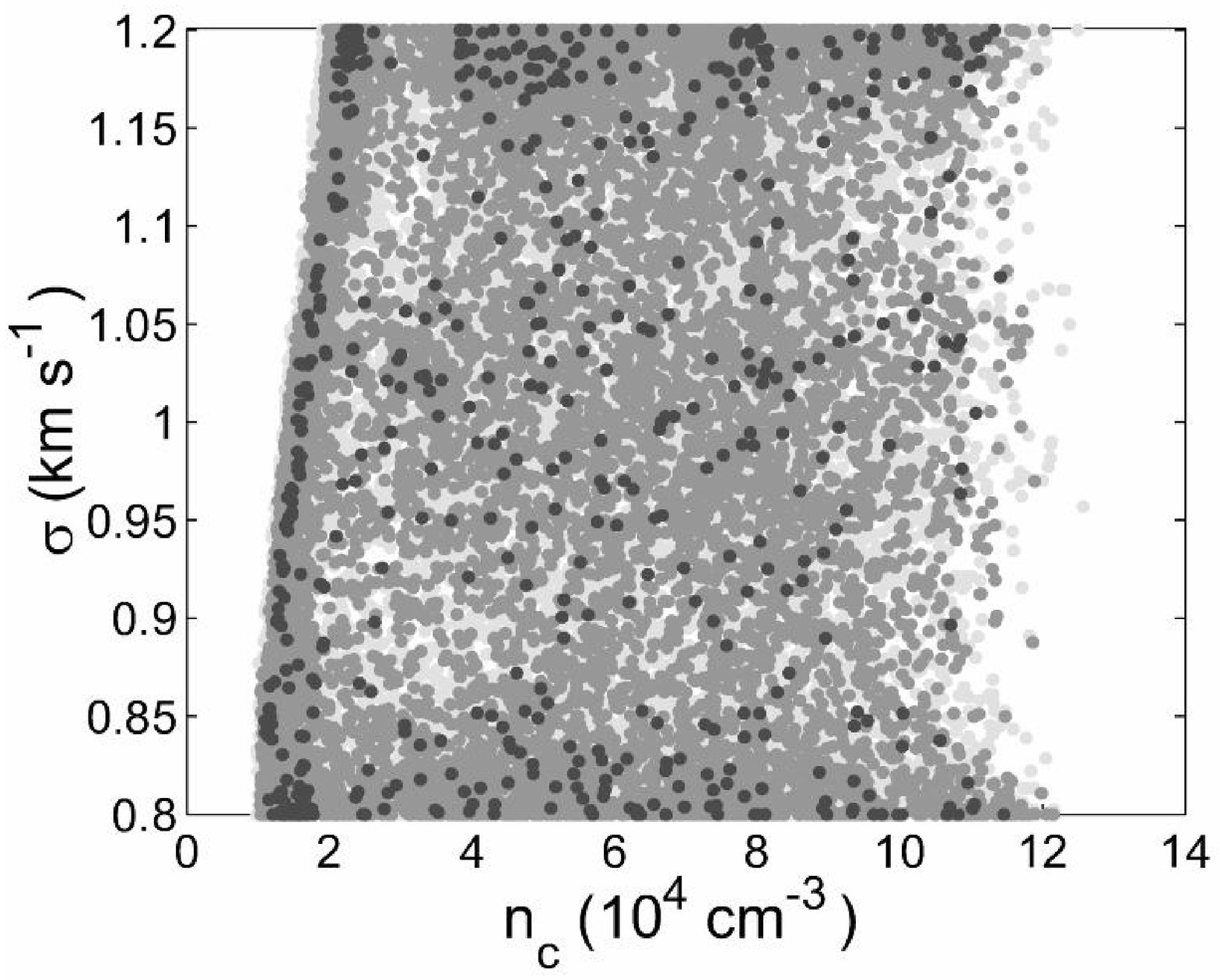}
\includegraphics[width=.49\textwidth]{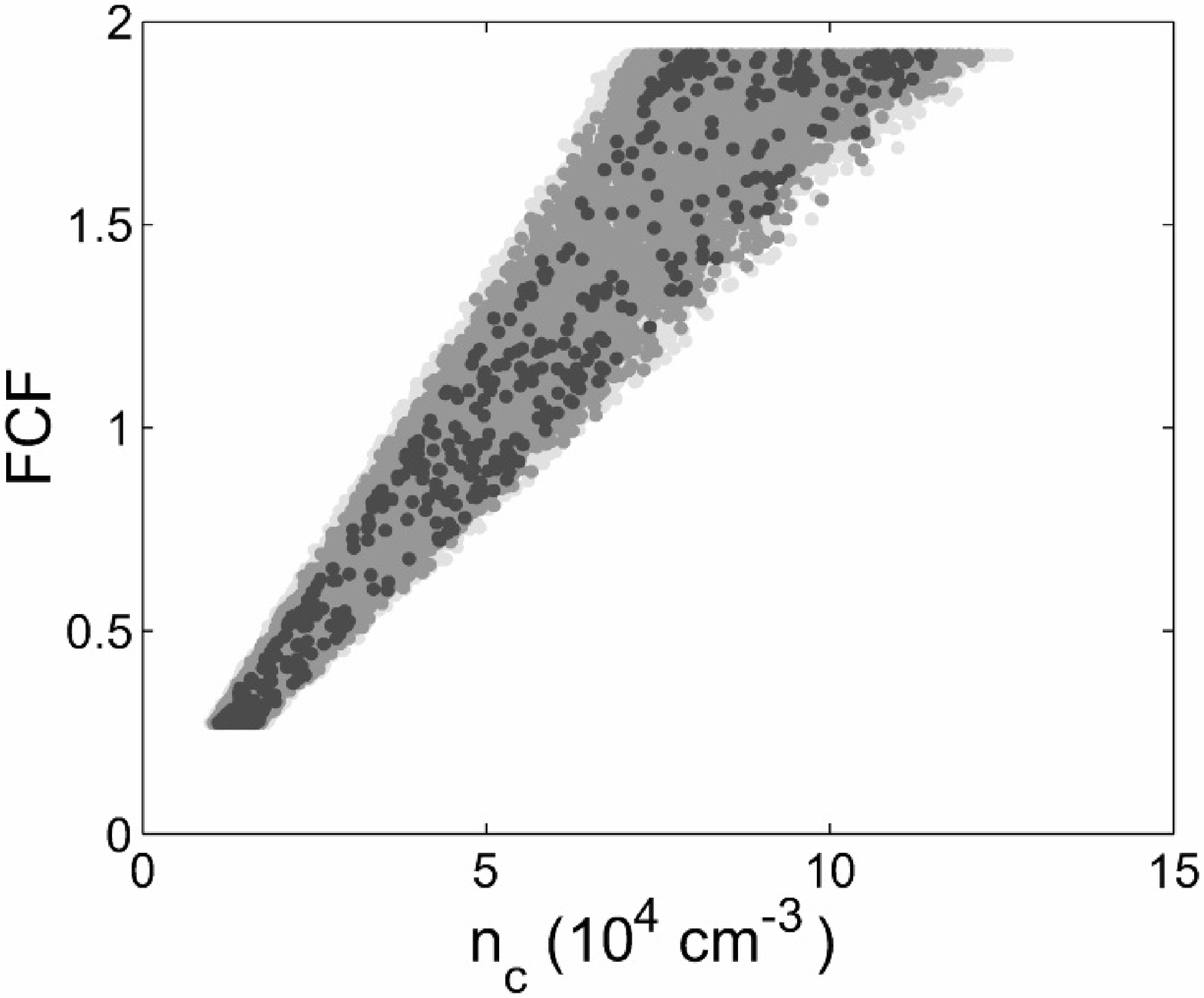}
\caption{Two views of the observationally constrained part of the parameter space $(\rhoc, \sigma, FCF)$ of the GS model for region 1, with $\chi^2\le 1.5$.  The allowed models reside approximately within a curved 2-dimensional sheet in this subspace.  The grayscale indicates the quality of the solutions, according to the convention of Figure \ref{fig:OstFit}.}
\label{fig:Stod_F1_obs}
\end{figure}

Figure \ref{fig:Stod_F1_obs} shows the distribution of GS models for region 1, in the observationally constrained part of the parameter space given by $(\rhoc, \sigma, FCF)$.  The set of models that fit the data are constrained approximately to a curved 2-dimensional sheet in this subspace.  Good fits are available for the entire range of $\sigma$ and $FCF$ explored.  The GS model is consistent with a range of central number density such that $9.8\times 10^3~{\rm cm}^{-3} \appleq n_c \appleq 1.3\times 10^5~{\rm cm}^{-3}$.  

Models were less well-constrained for region 2, and the corresponding figures are omitted for brevity.  Region 2 is consistent with central number densities of $4.0\times 10^3~{\rm cm}^{-3} \appleq n_c \appleq 3.5\times 10^4~{\rm cm}^{-3}$, which are a factor of $~2.5-3.5$ lower than region 1.  The central number density range is consistent with the values that we obtained for the truncated Ostriker model, but the GS model has a higher upper limit.

The solution set for the \Stod model spans the entire range of inclination angles allowed by equation \ref{eq:inc}.  There are no obvious correlations between the inclination angle and any other quantity, for reasons that are discussed at the bottom of $\S$\ref{sec:OstrikerResultsObs}.

\subsubsection{Theoretical Parameters}
\label{sec:StodResultsTheory}

The theoretical parameters for the GS model are the concentration parameter $C$ (equation \ref{eq:Cdef}) and the magnetic parameter $\kappa$, discussed earlier in this section.  There were no interesting correlations between these quantities.  For region 1, solutions filled the rectangular region of parameter space decribed by $0.89\le C \le 2.0$ and $ 0.1\le  \kappa \le 12$.For region 2, solutions filled the rectangular region of parameter space decribed by $0.75 \le C \le 2.0$ and $ 0.1 \le \kappa \le 5.1$.

\begin{figure}
\includegraphics[width=.49\textwidth]{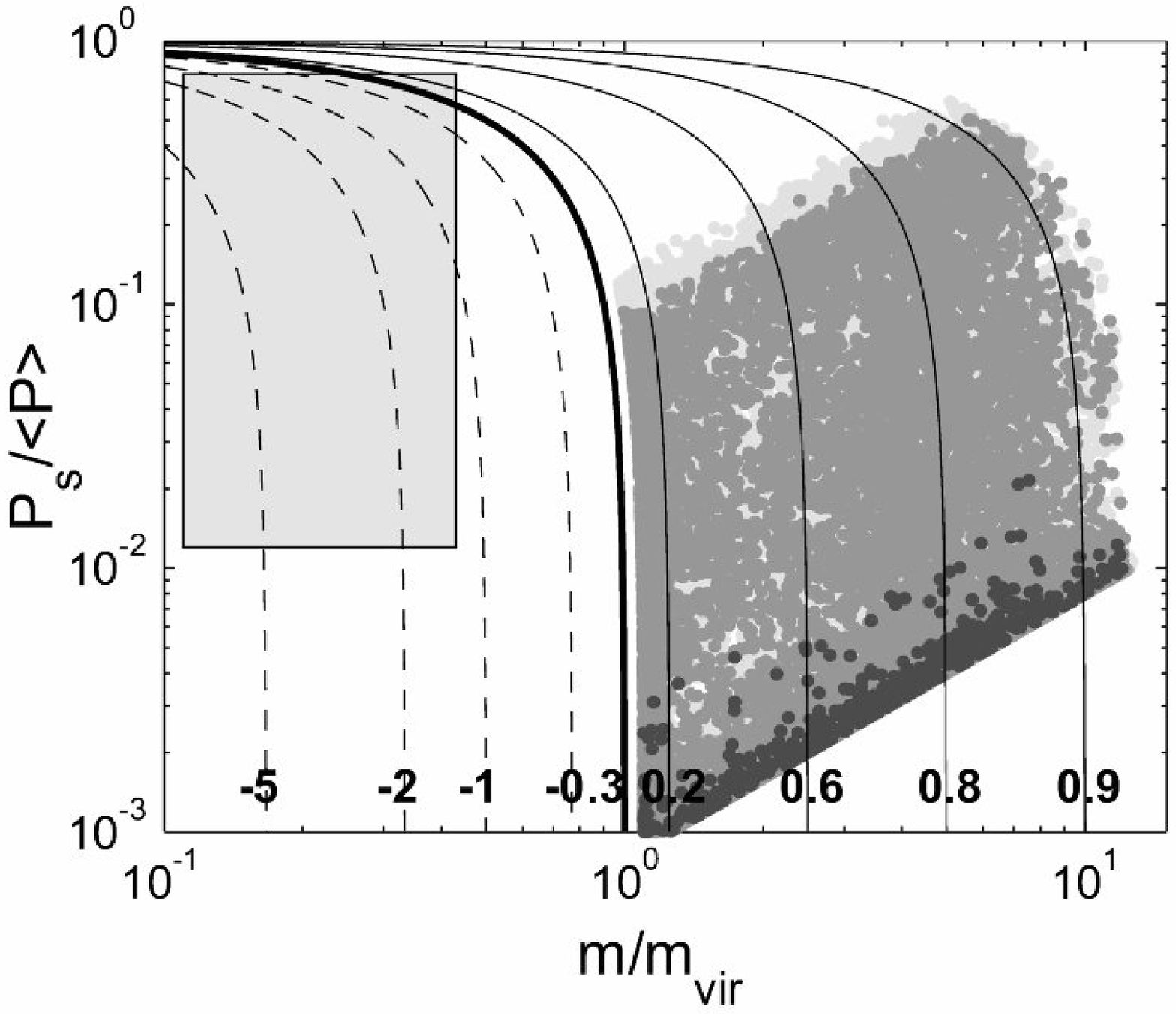}
\includegraphics[width=.49\textwidth]{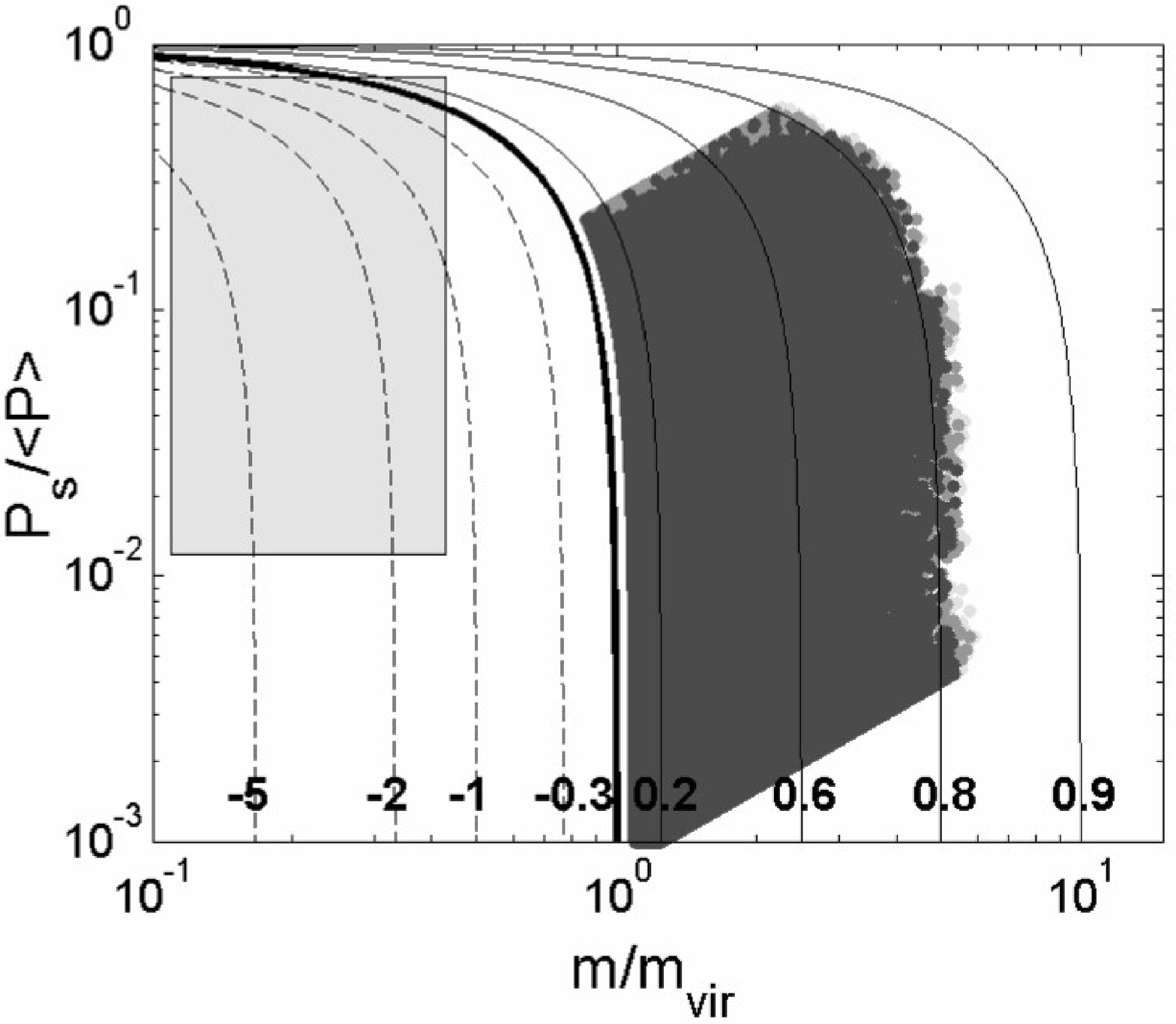}
\caption{The virial diagram for the GS model.  Region 1 is shown on the left and region 2 on the right.  The diagram is explained by the caption of Figure \ref{fig:OstVirial}, and the grayscale follows the convention of Figure \ref{fig:OstFit}.  Note that the magnetic field of the GS model always contributes to net magnetic support.  Thus the $\M/|\W| < 0$ (magnetic confinement) side of the diagram, where the filaments studied by FP mostly reside, is inaccessible to this model.}
\label{fig:StodVirial}
\end{figure}

Figure \ref{fig:StodVirial} shows the virial diagrams for the allowed GS solutions that are consistent with each of the two regions that were studied.  Both regions are consistent with a poloidally dominant magnetic field with $\M/|\W| \appleq 1$.  Note, however, that toroidal-field dominated solutions are not possible within the context of the GS model.

\begin{figure}
\includegraphics[width=.49\textwidth]{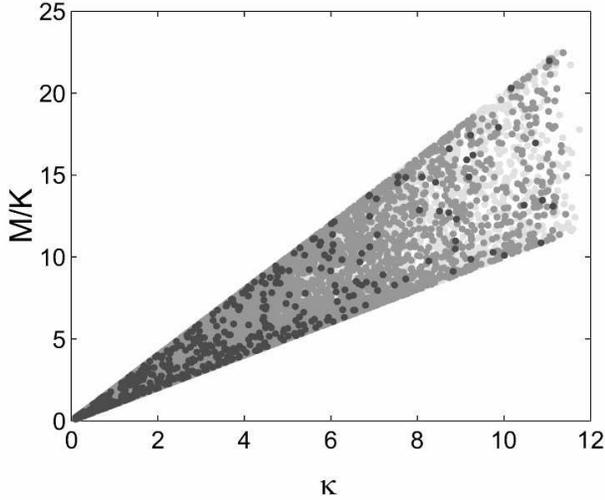}
\caption{The magnetic energy ratio $M/K$ of the GS model as a function of the magnetic parameter $\kappa=(1+\cos^2\theta)/2\beta$.  Only solutions that are consistent with region 1 are shown.}
\label{fig:StodMagEnergy}
\end{figure}

Figure \ref{fig:StodMagEnergy} shows how the magnetic energy ratio $M/K$ scales with the magnetic parameter $(1+\cos^2\theta)/2\beta$ within the allowed range of solutions for region 1.  The plasma $\beta$ parameter is equal to the inverse of the magnetic energy ratio, and the scatter in this figure is due to the range of magnetic pitch angles $\theta$ allowed by the GS model.

\subsection{The Fiege \& Pudritz Model}

% NEW JUNE 2004 RUNS
%
% FP_X1
% =====
% nc: 0.907477, 11.0505 x 10^4 cm^-3
% sigma: 0.8, 1.2 km/s
% C: 0.888889, 2
% FCF: 0.2738, 1.917
% Gz: 0, 15
% Gphi: 0, 3.7437
% magRatioBound0.2: Gz = 3.6278+/-0.25601
% magRatioBound0.5: Gz = 5.5745+/-0.15853
% magRatioBound1: Gz = 7.7057+/-0.12926
% magRatioBound2: Gz = 10.8015+/-0.11142

% FP_X2
% =====
% nc: 0.266063, 4.0309 x 10^4 cm^-3
% sigma: 0.8, 1.2 km/s
% C: 0.8, 2
% FCF: 0.2738, 1.917
% Gz: 0, 15
% Gphi: 0, 6
% magRatioBound0.2: Gz = 1.6814+/-1.314
% magRatioBound0.5: Gz = 3.1244+/-2.0648
% magRatioBound1: Gz = 6.7682+/-0.95407
% magRatioBound2: Gz = 9.9224+/-0.70532
% magRatioBound5: Gz = 14.9172+/-0.092529

\begin{figure}
\includegraphics[width=.49\textwidth]{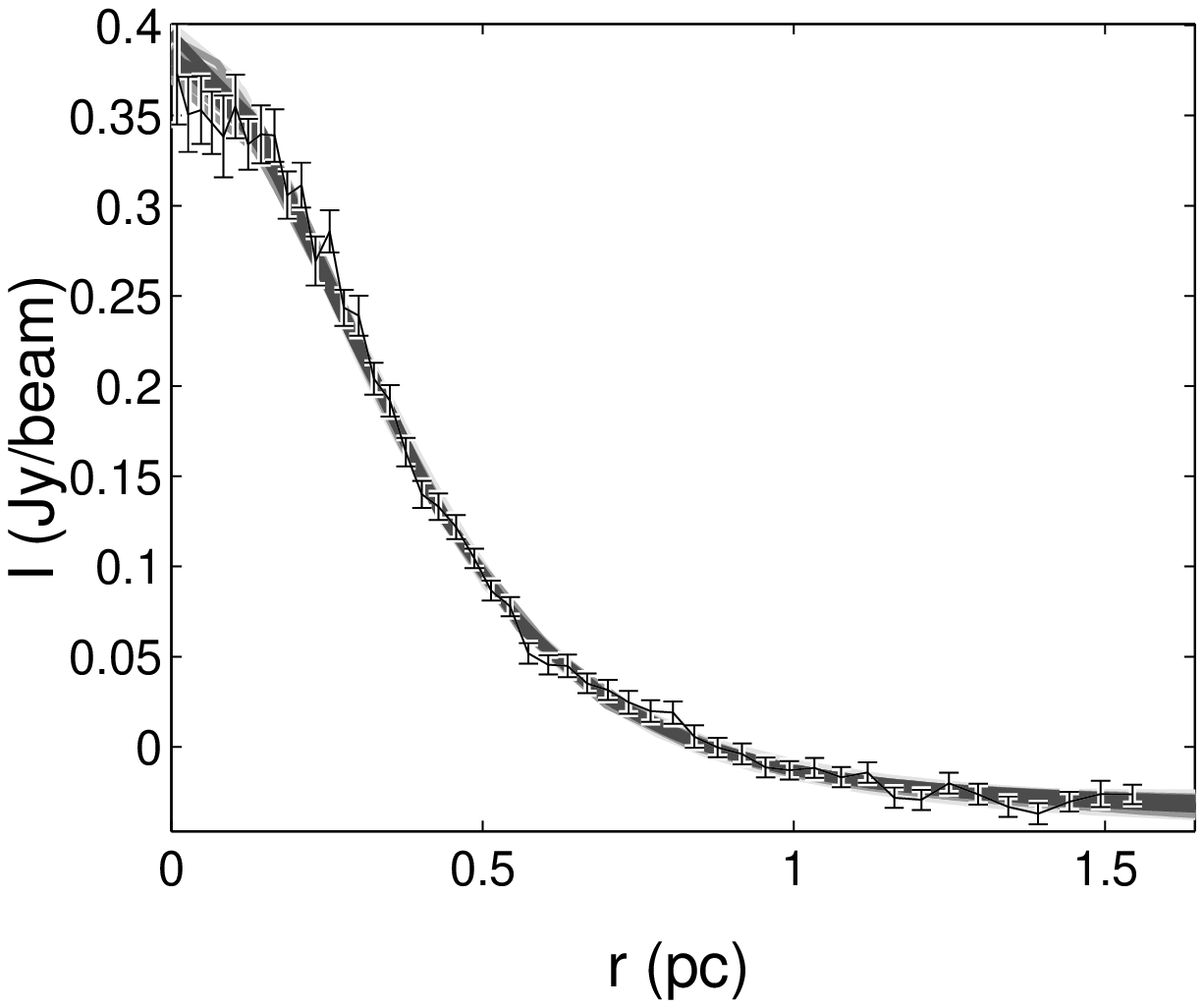}
\includegraphics[width=.49\textwidth]{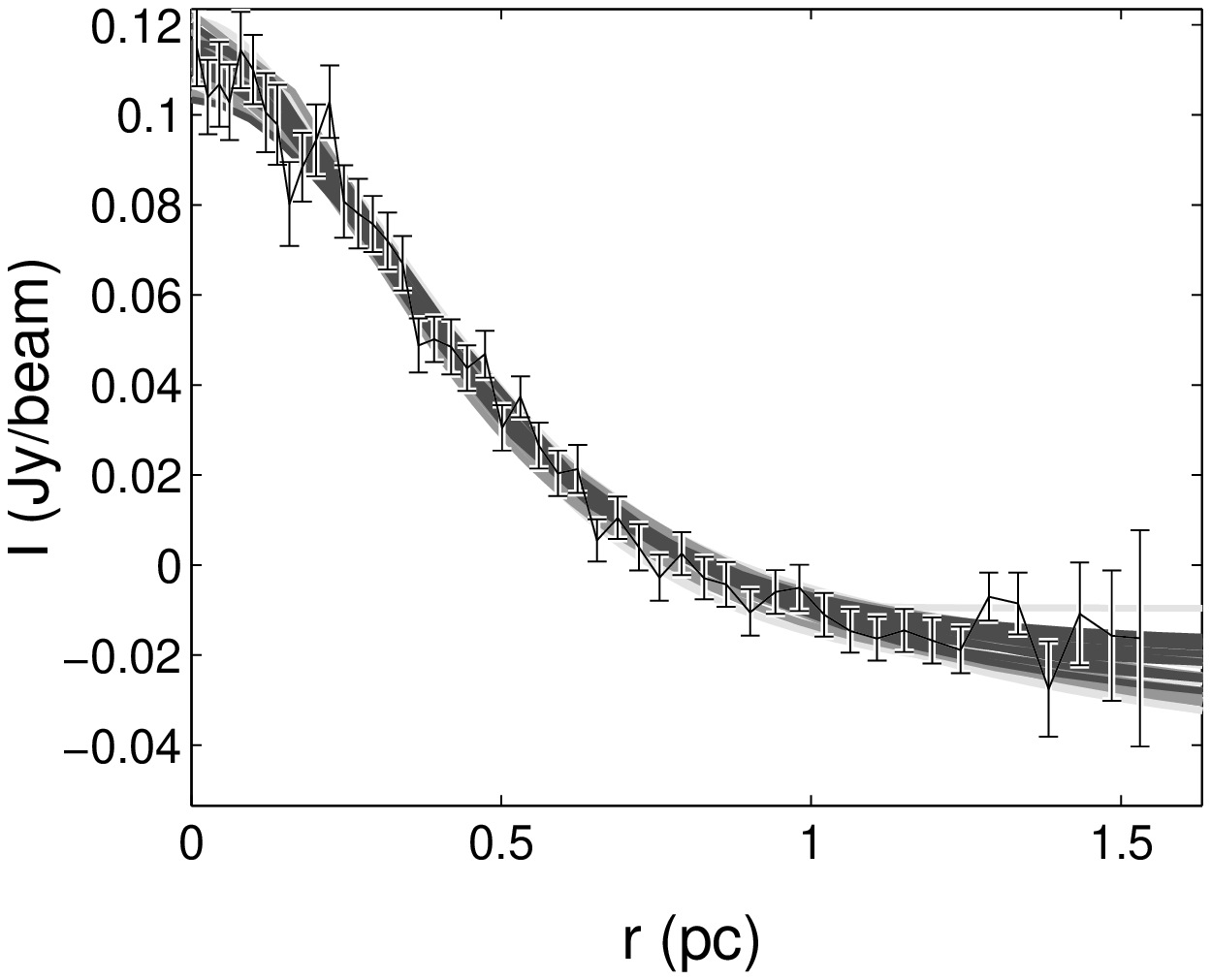}
\caption{Sample Fiege \& Pudritz model fits to the 850 $\micron$ intensity data.  Region 1 is on the left and region 2 is on the right.  The grayscale indicates the goodness of fit, following the convention of Figure \ref{fig:OstFit}.}
\label{fig:FPFit}
\end{figure}

\label{sec:FPResults}
Figure \ref{fig:FPFit} shows examples of FP model fits to both regions of G11.11-0.12 that were analyzed.  Both regions are well-fit by the model, but the solutions are less well-constrained in region 2.  This section discusses the distribution of these solutions in the parameter space of the FP model.

\subsubsection{Observationally Constrained Parameters}
\label{sec:FPResultsObs}
\begin{figure}
\includegraphics[width=.49\textwidth]{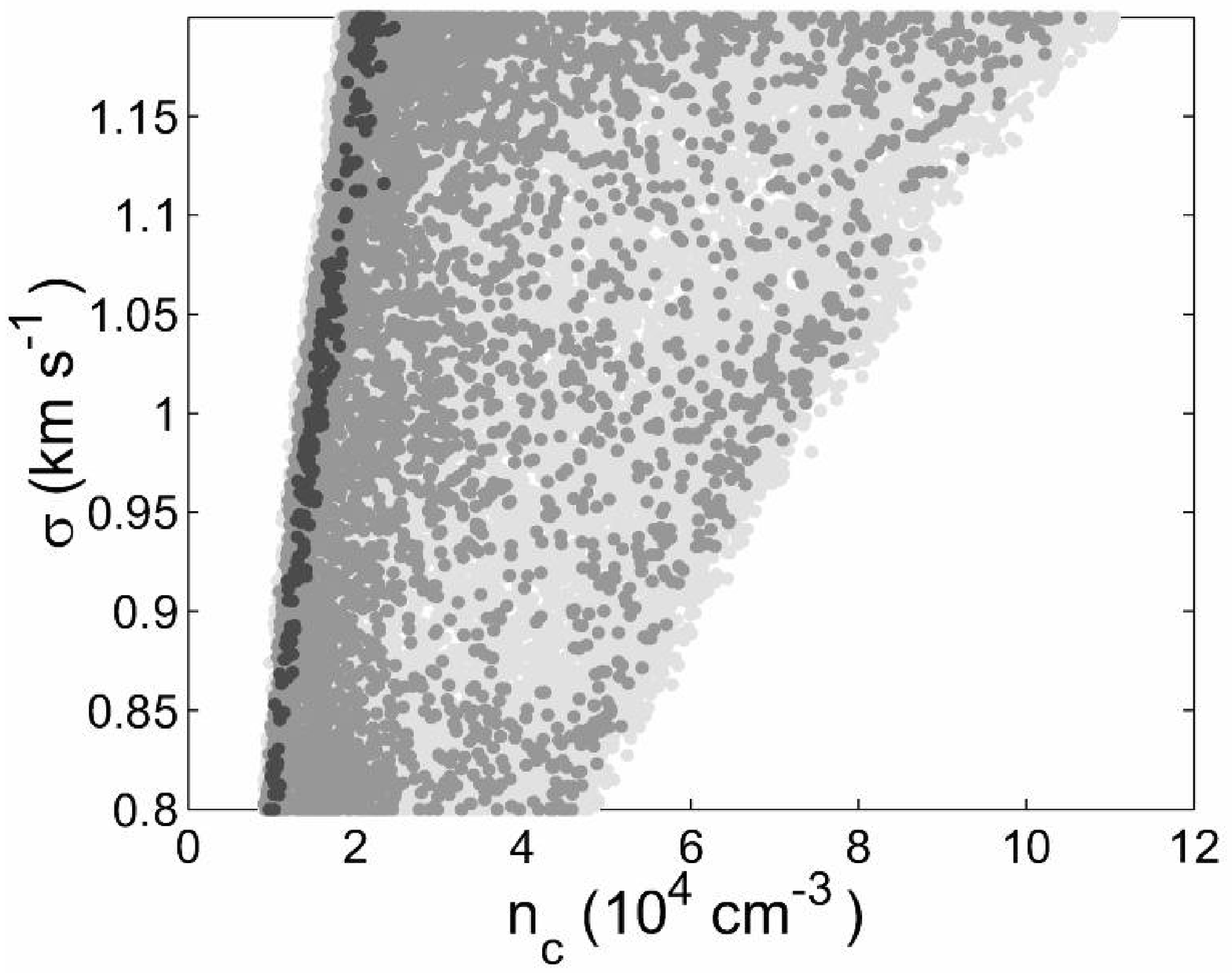}
\includegraphics[width=.49\textwidth]{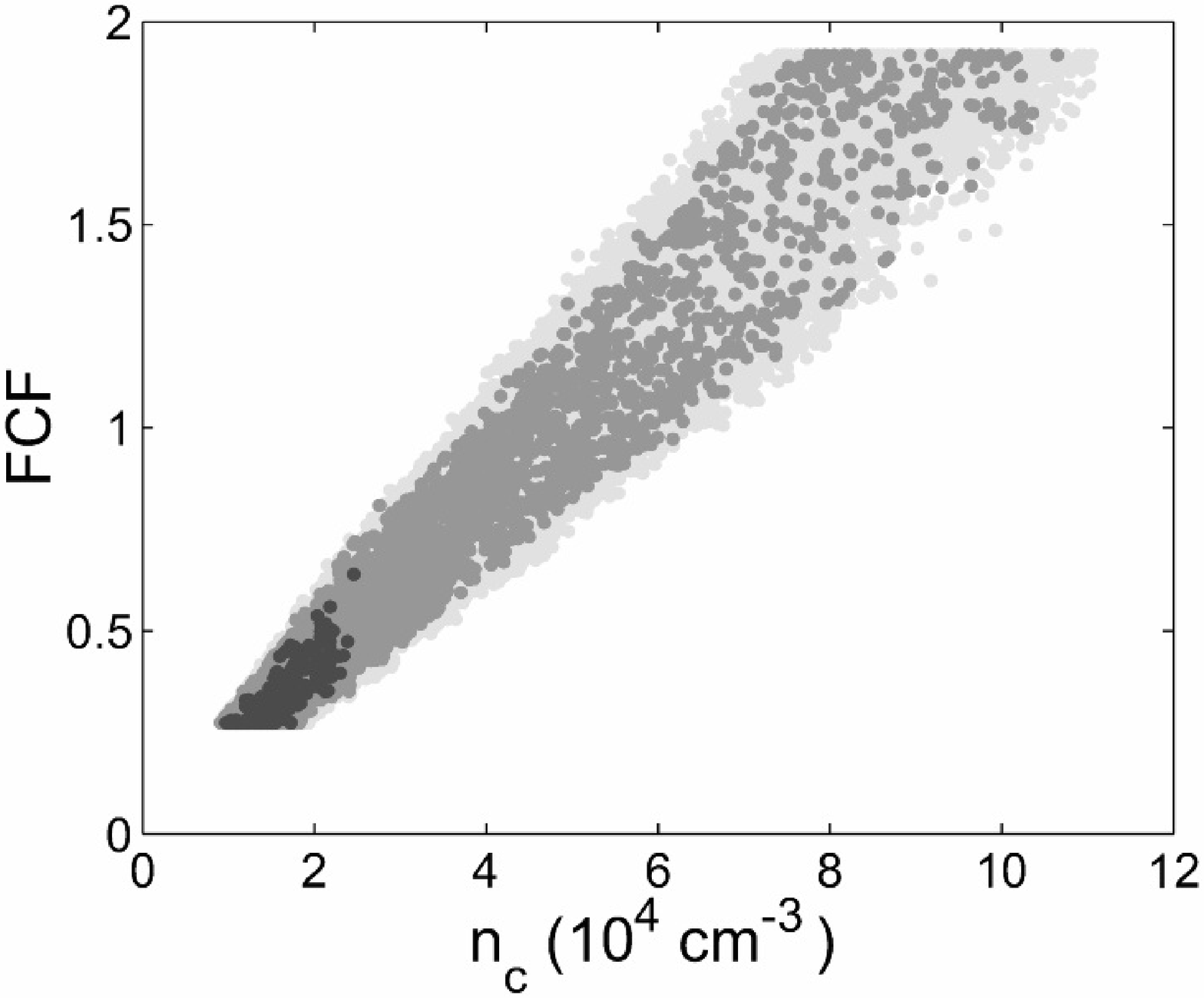}
\caption{Two views of the observationally constrained part of the parameter space $(\rhoc, \sigma, FCF)$ of the FP model for region 1, with $\chi^2\le 1.5$.  The grayscale indicates the quality of the solutions, according to the convention of Figure \ref{fig:OstFit}.}
\label{fig:FP_F1_obs}
\end{figure}

Figure \ref{fig:FP_F1_obs} shows two different viewpoints of the distribution of FP models for region 1, in the observationally constrained part of the parameter space given by $(\rhoc, \sigma, FCF)$.  The set of models that fit the data occupy an approximately planar region in this subspace.  We obtained good fits for essentially the entire range of $\sigma$ and $FCF$ explored.  The FP model is consistent with central number densities in the range $9.1\times 10^3~{\rm cm}^{-3} \appleq n_c \appleq 1.1\times 10^5~{\rm cm}^{-3}$.  The highest number density that we obtained for an Ostriker model was $2.4\times 10^4~{\rm cm}^{-3}$; thus, the FP model is consistent with central densities of up to $\sim 5$ times higher.  This is significant because higher densities may be more consistent with the critical densities of molecules such as H$_2$CO (Carey et al. 1998), HCO$^+$ and CS (Redman et al. 2000), and N$_2$H$^+$ (Feldman et al. 2003).  Higher densities are also preferred by current ideas on dust coagulation, which assume densities of $\sim 10^6~{\rm cm}^{-3}$, as discussed in Paper I.  The velocity dispersion and FCF spanned the full range allowed by the search.

We found similar results for region 2 within the following parameter ranges: $2.7\times 10^3~{\rm cm}^{-3} \appleq n_c \appleq 4.0\times 10^4~{\rm cm}^{-3}$.  The number density range for both regions is similar to our results for the GS model, but much higher than values obtained for the Ostriker model.

\begin{figure}
\includegraphics[width=.49\textwidth]{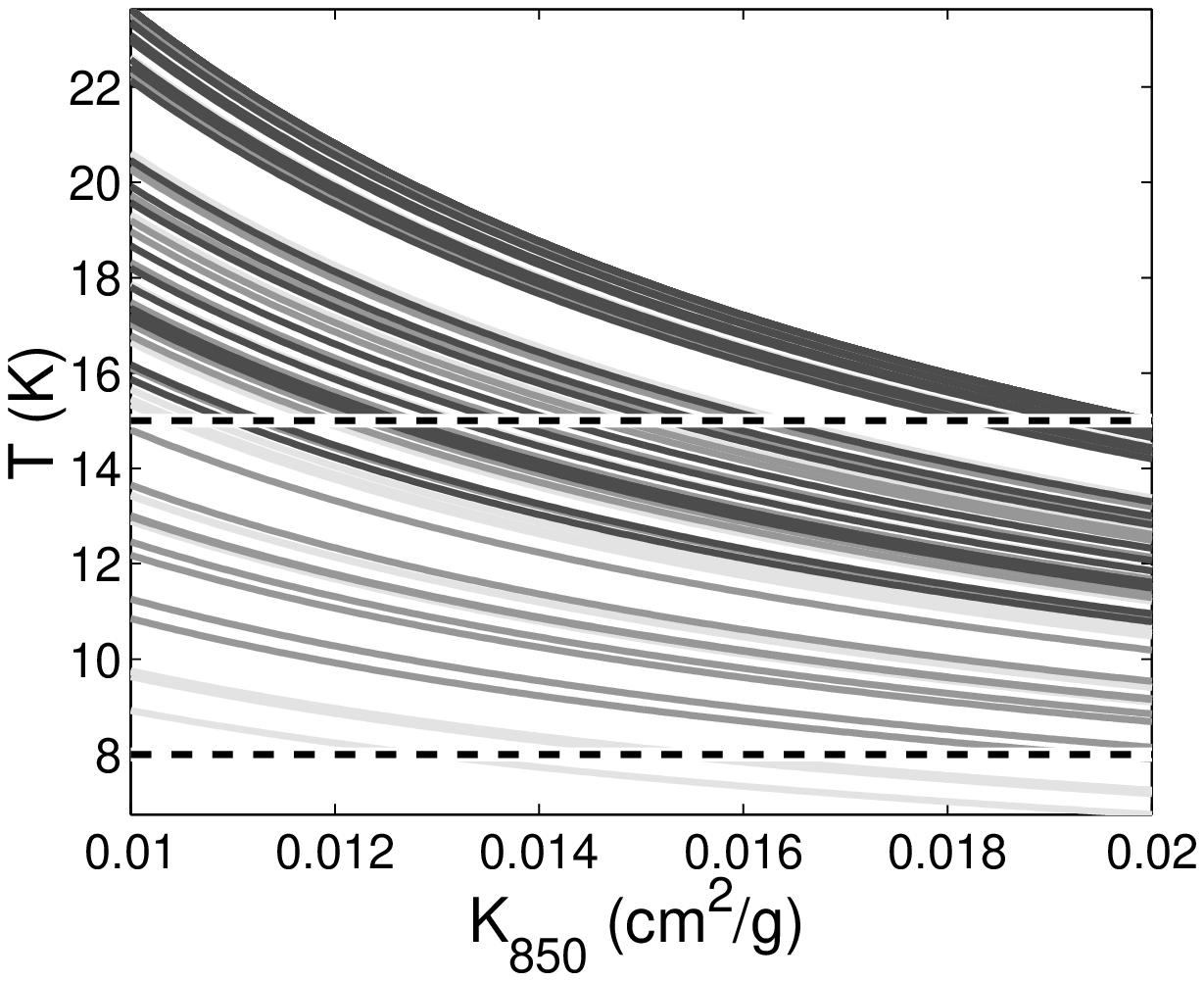}
\includegraphics[width=.49\textwidth]{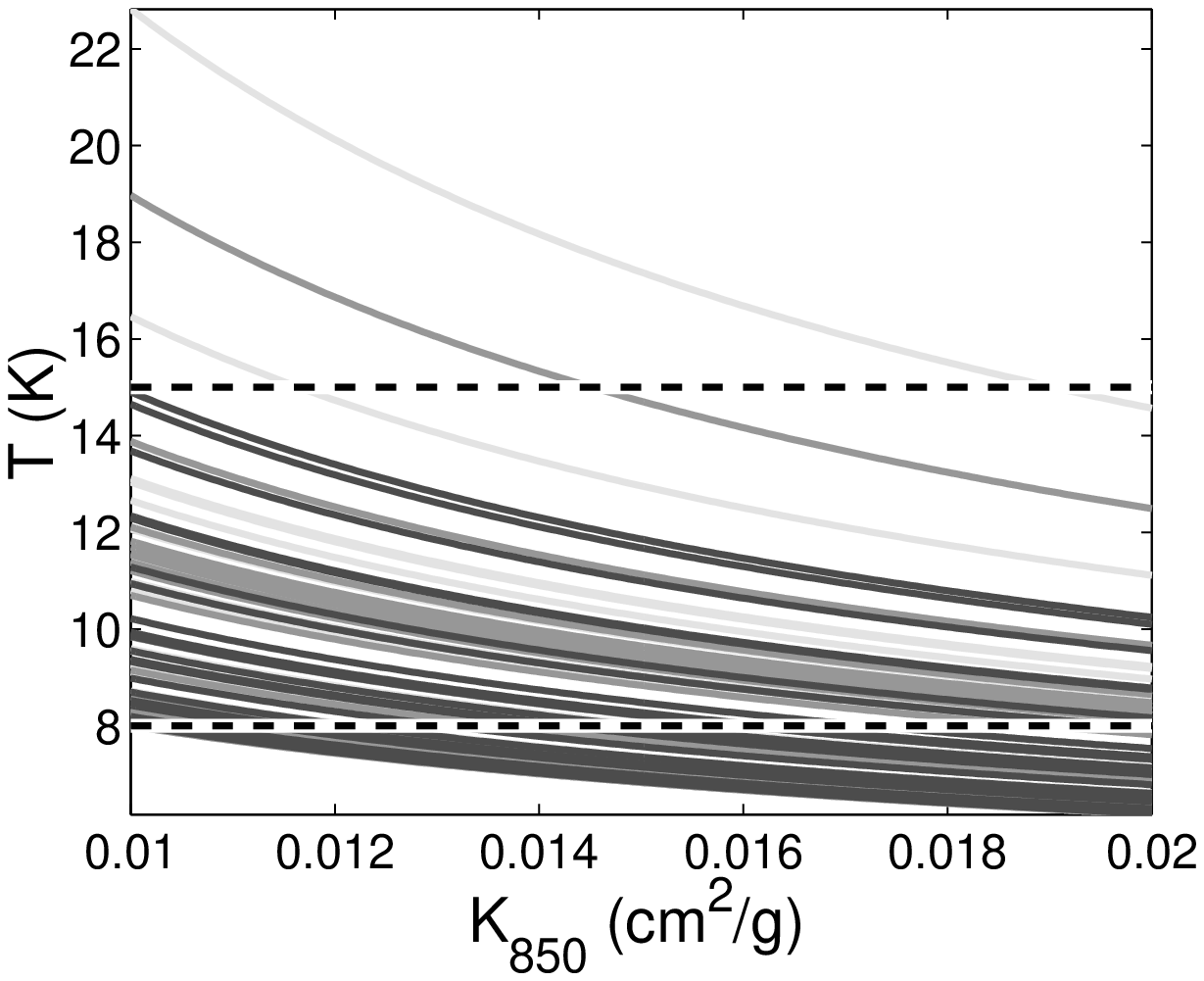}
\caption{Constraints on the temperature and 850$\micron$ extinction coefficient of the FP model.  We show a random sample of solutions for region 1 (left) and region 2 (right).  Each curve corresponds to a single value of the FCF for a single model, as defined by equation \ref{eq:FCF}.  The grayscale follows the convention of Figure \ref{fig:OstFit}.}
\label{fig:FPTK}
\end{figure}

Figure \ref{fig:FPTK} shows the allowed values of T and $\kappa_{850}$ for a sample of FP models for region 1 (left) and region 2 (right).  FP solutions exist for nearly any value within the range allowed by equation \ref{eq:TKConstraints}.

The solution sets for both regions span the entire range of inclination angles allowed by equation \ref{eq:inc}.  We did not find any obvious correlations between the inclination angle and any other quantity for the FP model.

\subsubsection{Theoretical Parameters}
\label{sec:FPResultsTheory}% >> analyze('Z1')

\begin{figure}
\includegraphics[width=.49\textwidth]{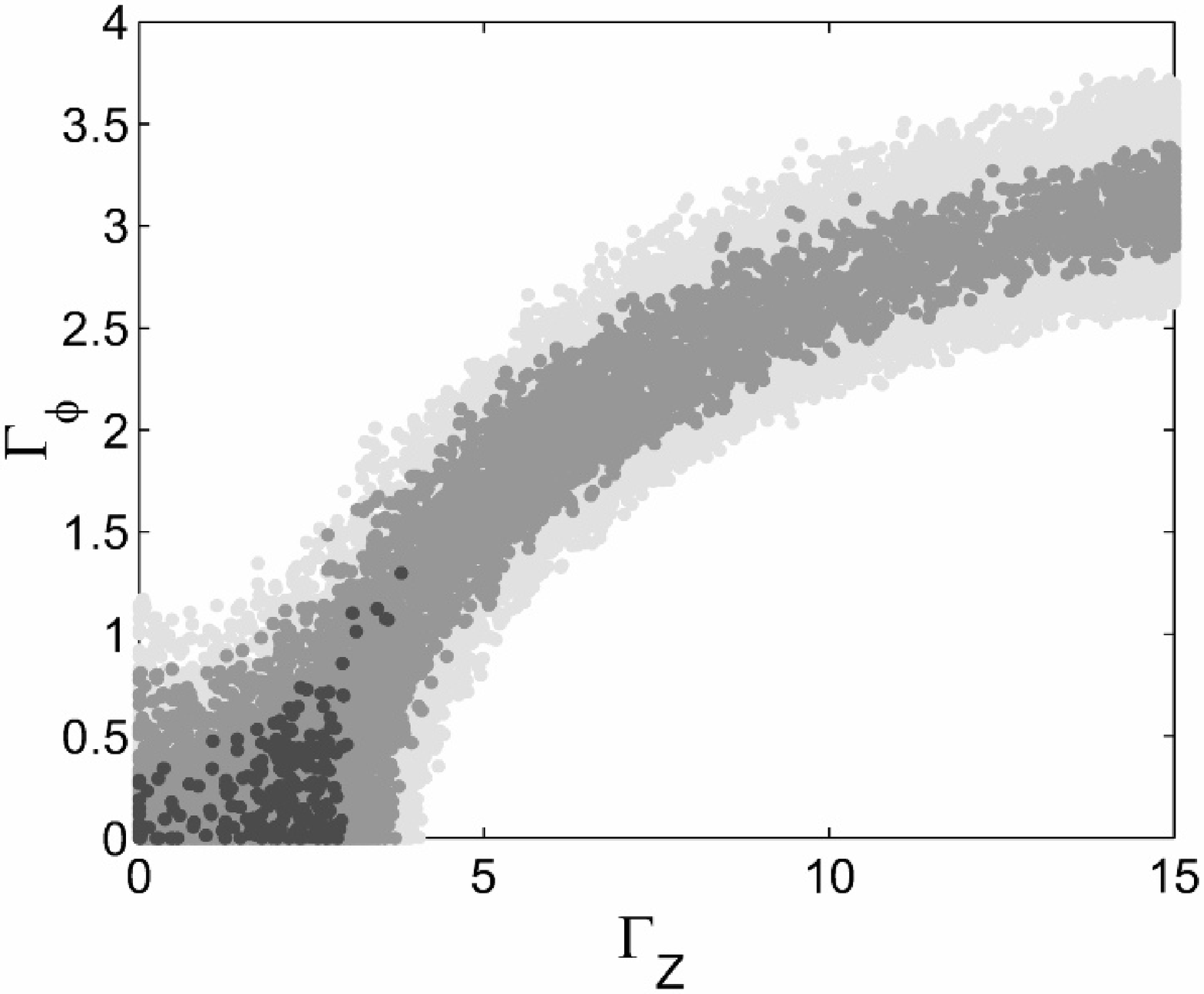}
\includegraphics[width=.49\textwidth]{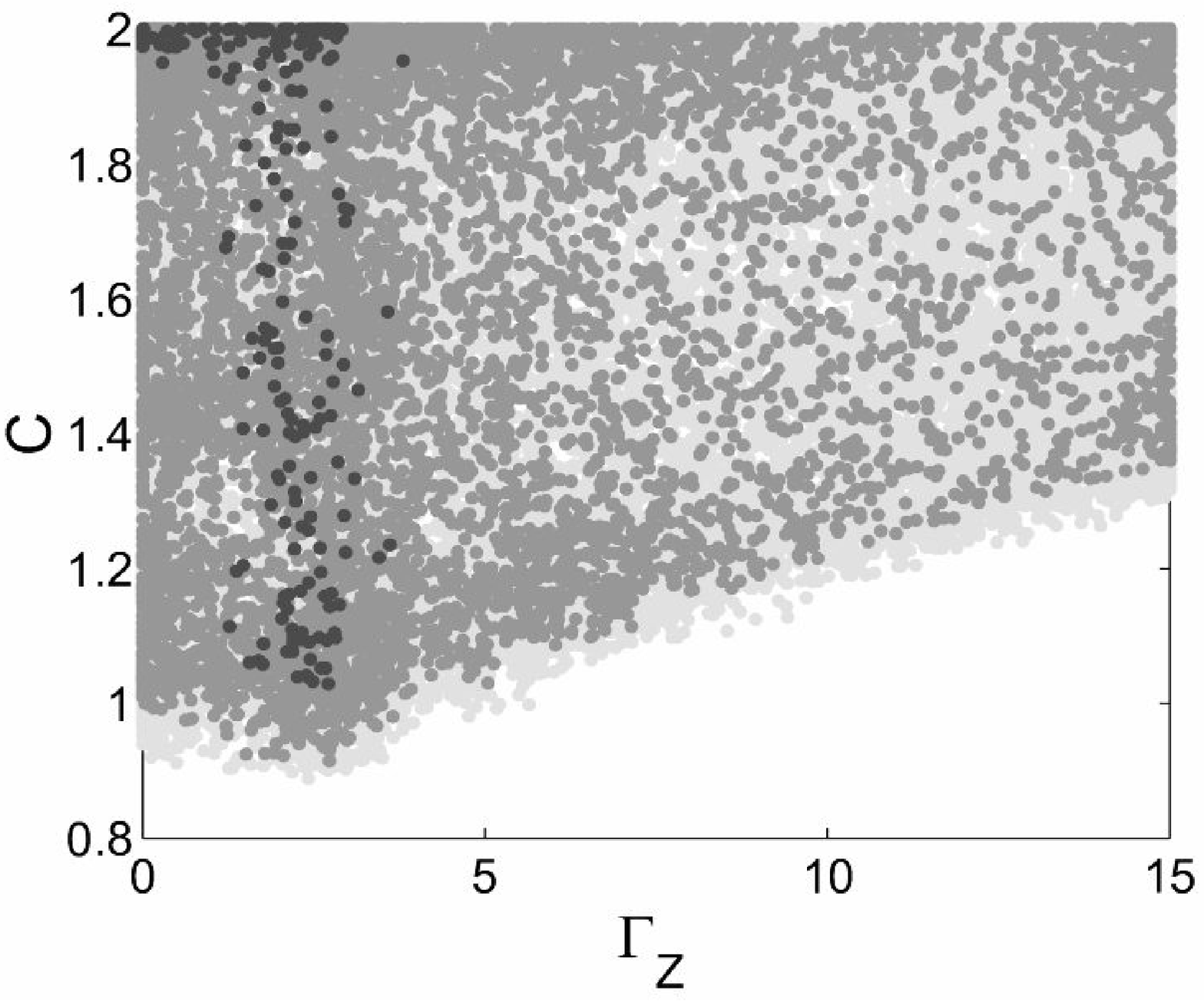}
\caption{Distribution of the theoretical parameters of the FP model ($\Gz$, $\Gphi$, $C$) for region 1.  (left) $\Gz$ and $\Gphi$ are tightly correlated and lie approximately on a curve in the ($\Gz$, $\Gphi$) subspace.   (right) The lower bound of the concentration parameter $C$ tends to increase slightly as $\Gz$ increases.  
}
\label{fig:FP_F1_Theory}
\end{figure}

Figure \ref{fig:FP_F1_Theory} shows the distribution of FP model solutions in the theoretical part of the parameter space ($\Gz$, $\Gphi$, $C$) for region 1.  This region is consistent with values of $\Gz\appleq 15$ and $\Gphi\appleq 3.7$.  There is a very clear trade-off between flux to mass ratios $\Gz$ and $\Gphi$, which lie approximately on a curve in the ($\Gz$, $\Gphi$) subspace, such that $\Gphi$ increases with increasing $\Gz$.  This trend occurs because $\Gz$ and $\Gphi$ have an opposite effect on the radial structure of the cloud.  Equation \ref{eq:MDef} and the discussion of \S\ref{sec:FP} show that any value of the magnetic virial parameter $\M/|\W|$ corresponds to a one parameter set of $\Gz$ and $\Gphi$.  Figure \ref{fig:FP_F1_Theory} demonstrates that this exact degeneracy in the virial balance is also manifest as an approximate degeneracy in the radial density structure of the cloud.  This degeneracy is expected from the results obtained by FP1, who showed that poloidal and toroidal fields have opposite effects on the shape of the density profile.  Dominant toroidal fields result in centrally peaked density profiles with a slow decrease outside of the core region, while dominant poloidal fields result in flat central density profiles with a sharp decrease in the outer regions.  Our results are an indication that these two effects can be critically balanced against each other to explain a given density profile for a significant range of $\Gz$ and $\Gphi$.  

% NEW JUNE 2004 RUNS
%
% FP_X1
% =====
% nc: 0.907477, 11.0505 x 10^4 cm^-3
% sigma: 0.8, 1.2 km/s
% C: 0.888889, 2
% FCF: 0.2738, 1.917
% Gz: 0, 15
% Gphi: 0, 3.7437
% magRatioBound0.2: Gz = 3.6278+/-0.25601
% magRatioBound0.5: Gz = 5.5745+/-0.15853
% magRatioBound1: Gz = 7.7057+/-0.12926
% magRatioBound2: Gz = 10.8015+/-0.11142

% FP_X2
% =====
% nc: 0.266063, 4.0309 x 10^4 cm^-3
% sigma: 0.8, 1.2 km/s
% C: 0.8, 2
% FCF: 0.2738, 1.917
% Gz: 0, 15
% Gphi: 0, 6
% magRatioBound0.2: Gz = 1.6814+/-1.314
% magRatioBound0.5: Gz = 3.1244+/-2.0648
% magRatioBound1: Gz = 6.7682+/-0.95407
% magRatioBound2: Gz = 9.9224+/-0.70532
% magRatioBound5: Gz = 14.9172+/-0.092529

It is very interesting that the swath of solutions shown in Figure \ref{fig:FP_F1_Theory} just grazes the non-magnetic limit ($\Gz=0$, $\Gphi=0$), which is equivalent to a truncated Ostriker model.  The Ostriker model is consistent with the data, but it can be viewed as an outlying model near the boundary of a more general class of magnetic solutions.

Region 1 is consistent with values of the concentration parameter $C$ in the range $0.89 \appleq C \appleq 2$, which corresponds to a truncation range of $7.8 \appleq \Rs/r_0 \appleq 100$.  We found that $\Rs$ is poorly determined within this range.  This is not surprising because the density is very low at these large radii and has little effect on the surface density structure.  The concentration parameter is slightly correlated with $\Gz$ and tends to slowly decrease as $\Gz$ increases.  This is explained by the opposite roles of these parameters in the surface density structure.  Solutions with large values of $\Gz$ have large, flat central density profiles with more emitting material near the center than low $\Gz$ solutions.  Decreasing $C$ compensates for high $\Gz$ by decreasing the column depth of the surrounding gas.

\begin{figure}
\includegraphics[width=.49\textwidth]{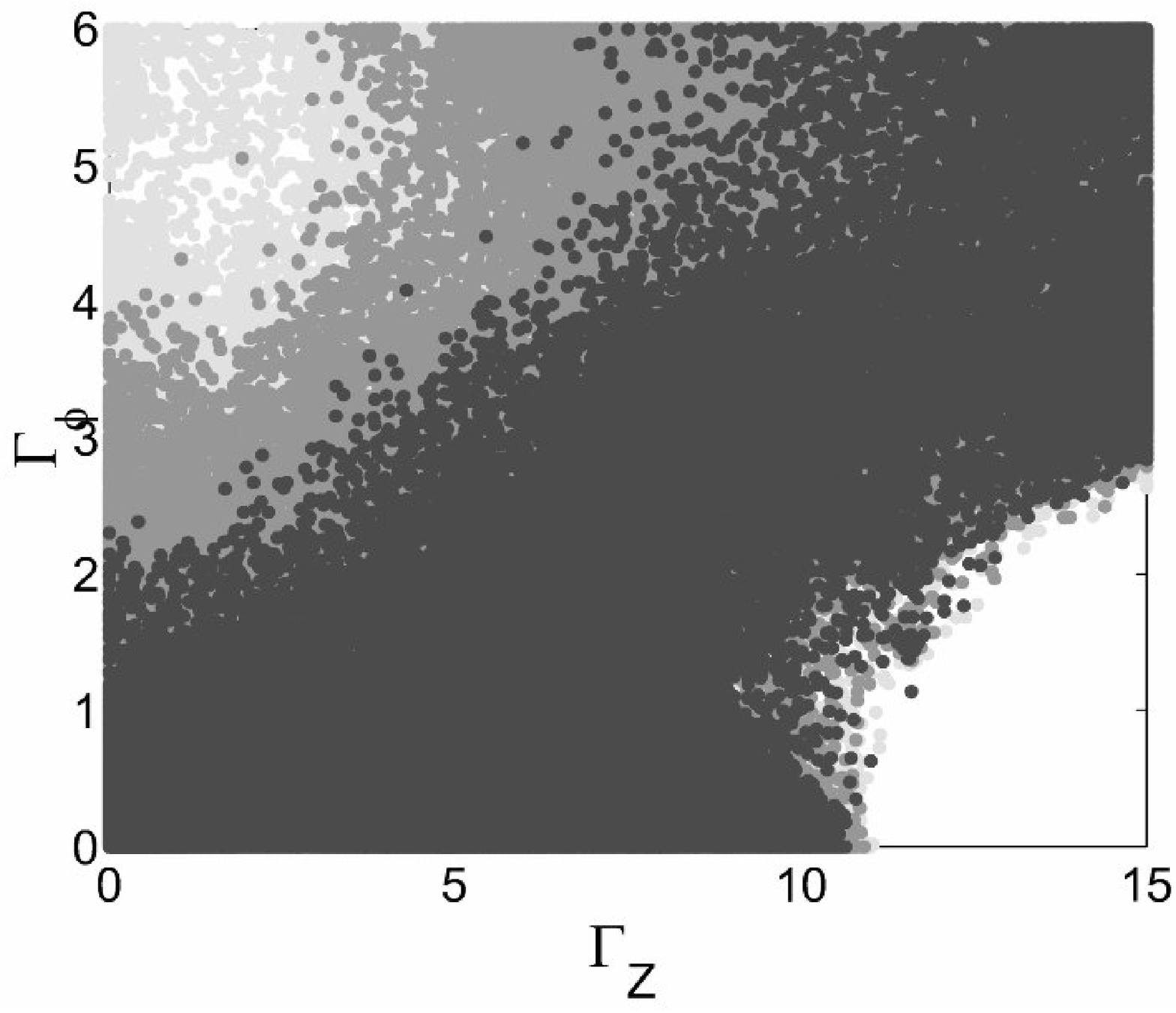}
\includegraphics[width=.49\textwidth]{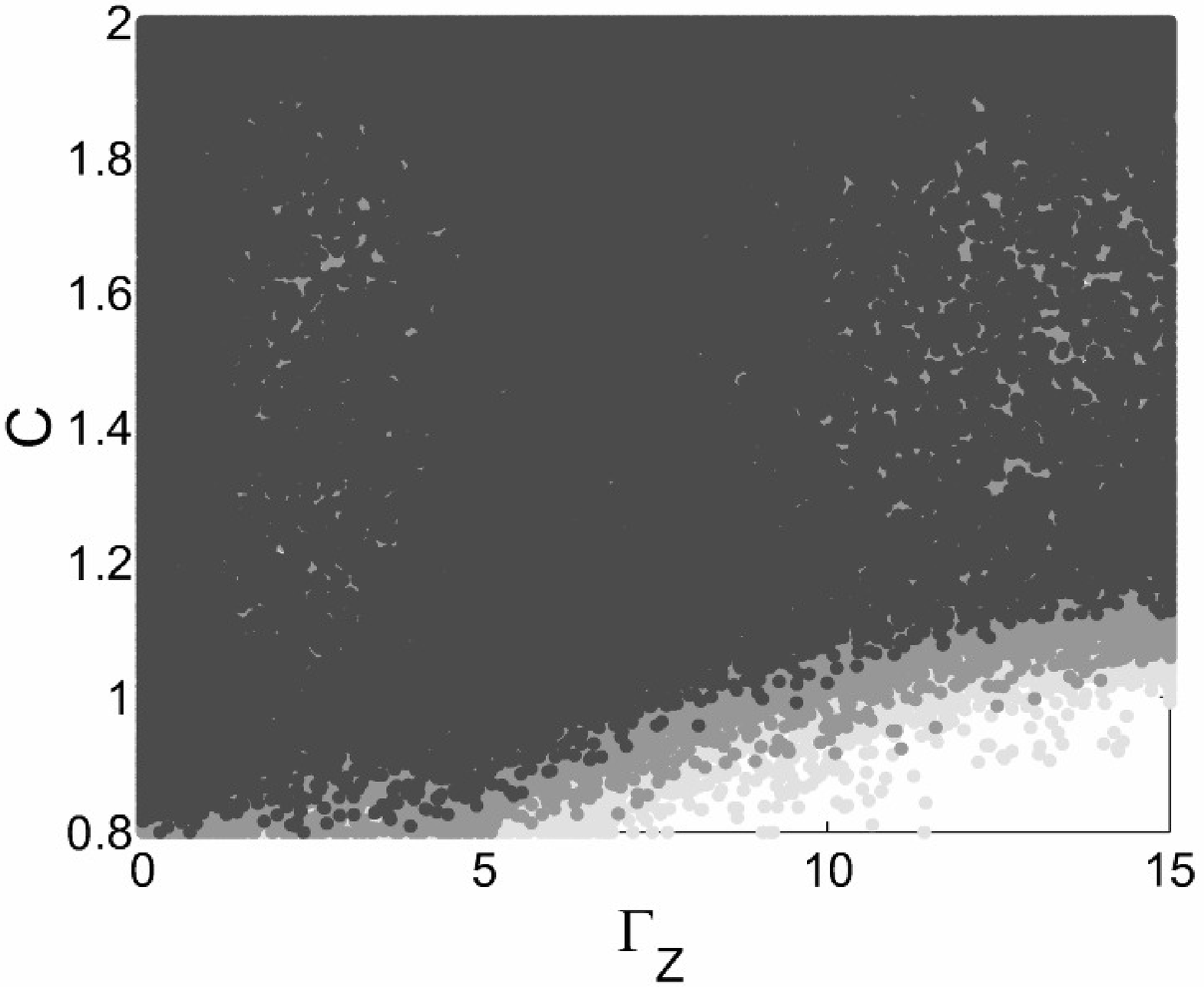}
\caption{Distribution of the theoretical parameters of the FP model ($\Gz$, $\Gphi$, $C$) for region 2.  (left) $\Gz$ and $\Gphi$ are correlated and lie approximately on a curve in the ($\Gz$, $\Gphi$) subspace.  (right) The concentration parameter $C$.
}
\label{fig:FP_F2_Theory}
\end{figure}

The corresponding figures are shown in Figure \ref{fig:FP_F2_Theory} for region 2.  Region 2 is consistent with values of $\Gz\appleq 15$ and $\Gphi\appleq 6$.  $\Gphi$ and $\Gz$ are correlated, as in region 1, but less tightly.  The solutions have a similar range of concentrations as in region 1: $0.80 \appleq C \appleq 2.0$.

\begin{figure}
\includegraphics[width=.49\textwidth]{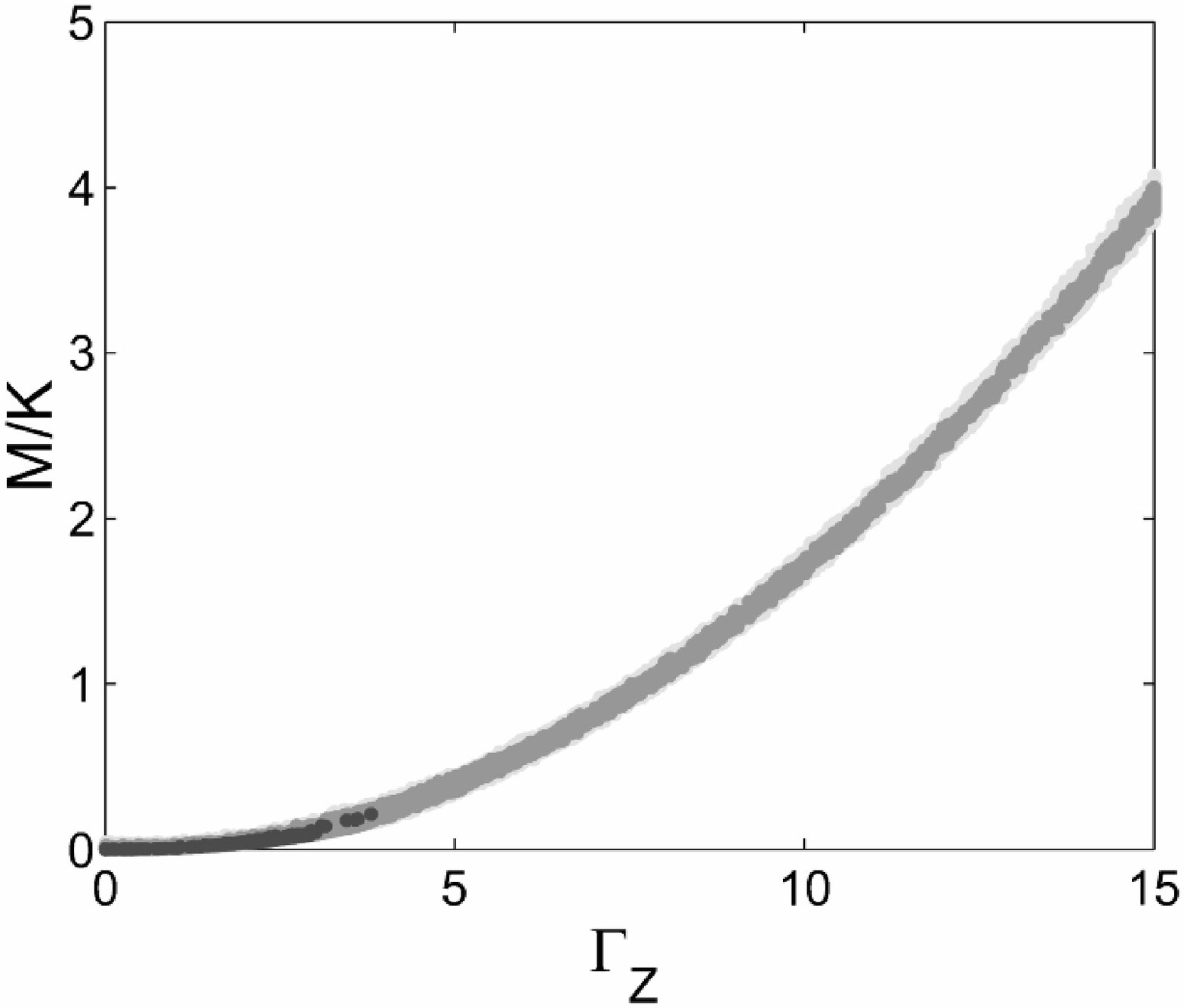}
\includegraphics[width=.49\textwidth]{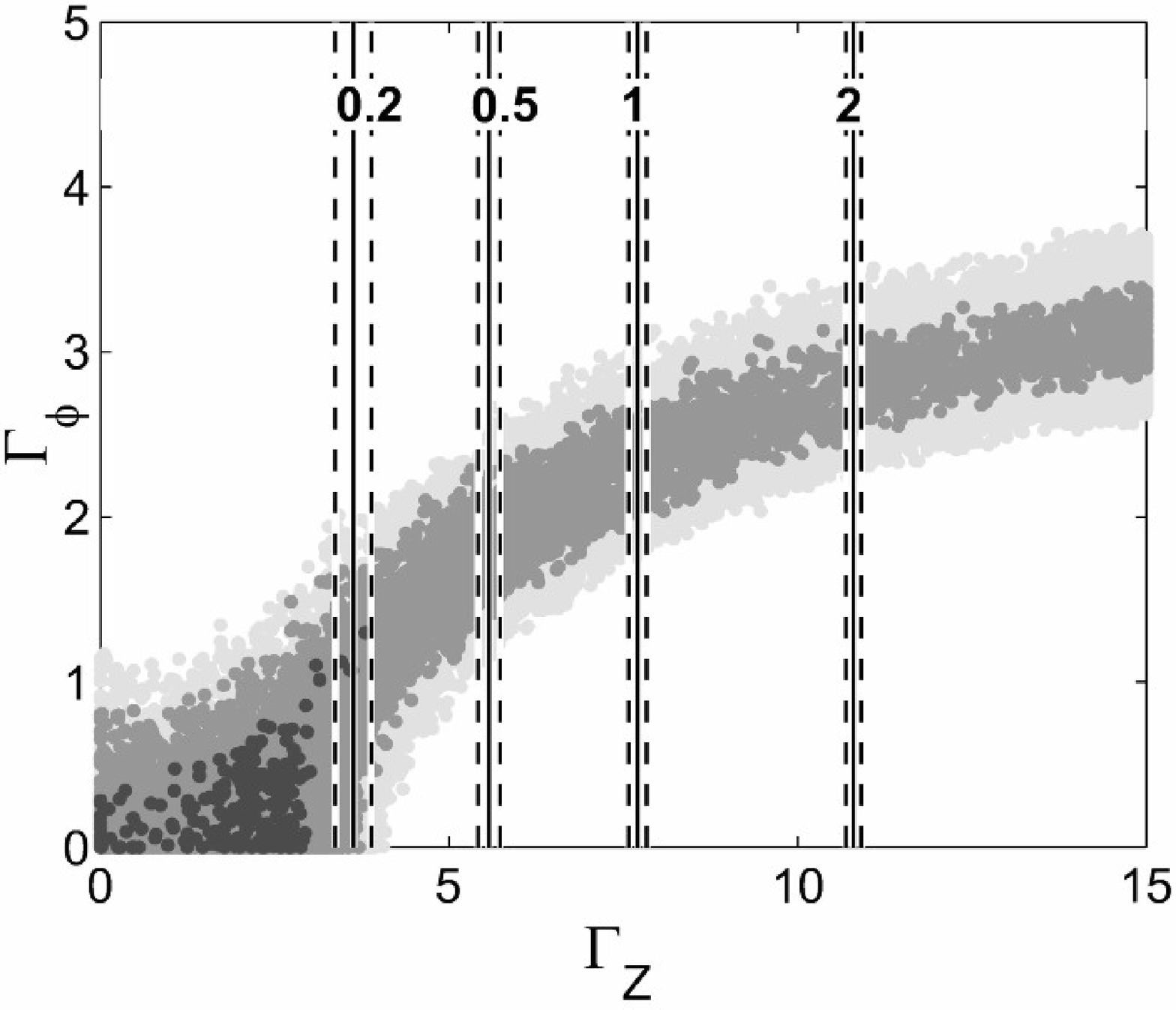}
\caption{The FP magnetic energy parameter for region 1.  (panel a., left) We show the correlation of M/K, defined by equation \ref{eq:MK}, with $\Gz$.  (panel b., right) We overlay several values of M/K on a graph of $\Gphi$ vs. $\Gz$.  The dashed lines represent the standard deviation of the value of $\Gz$ corresponding to each value of M/K.}
\label{fig:FP_F1_MK}
\end{figure}

\begin{figure}
\includegraphics[width=.49\textwidth]{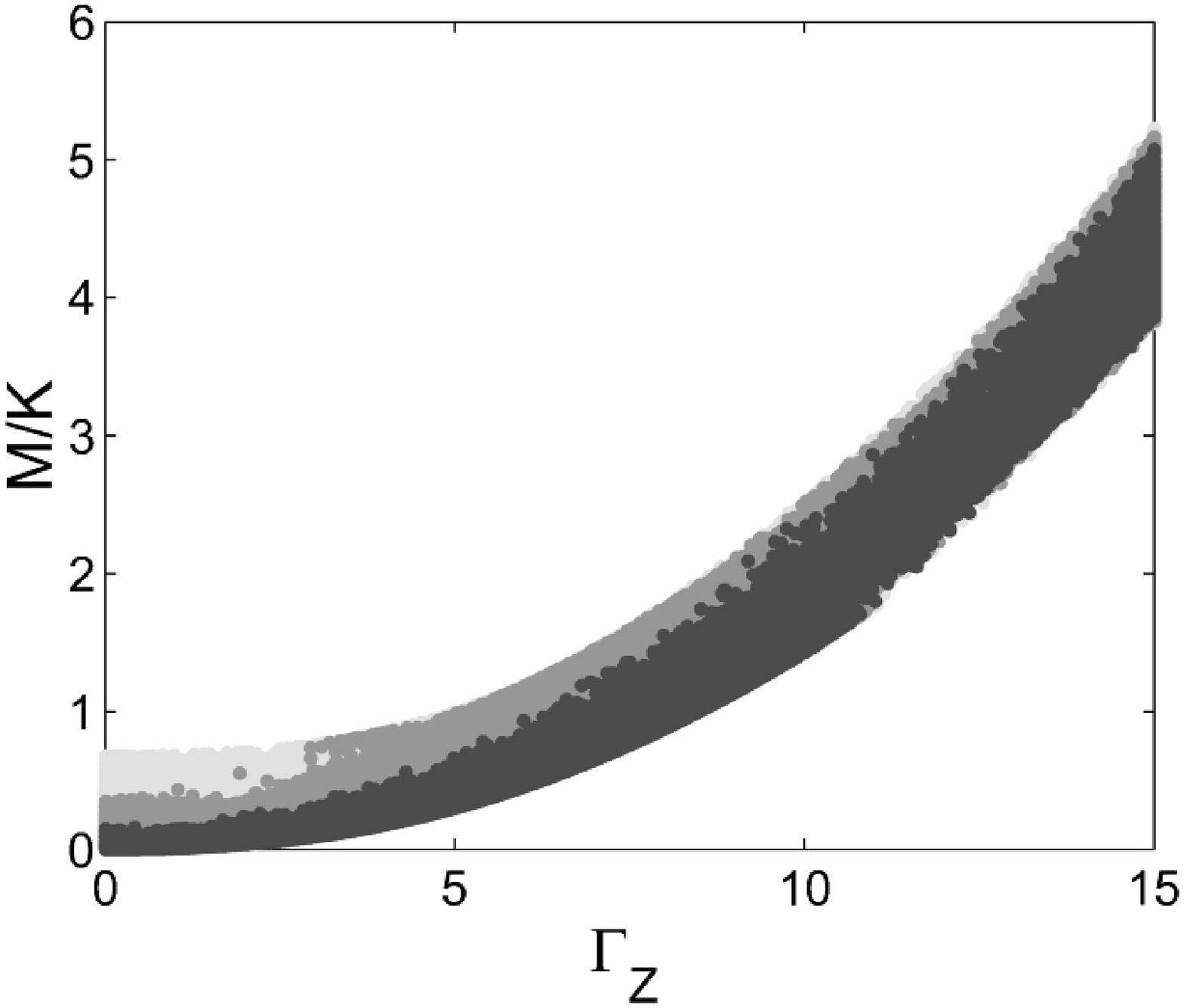}
\includegraphics[width=.49\textwidth]{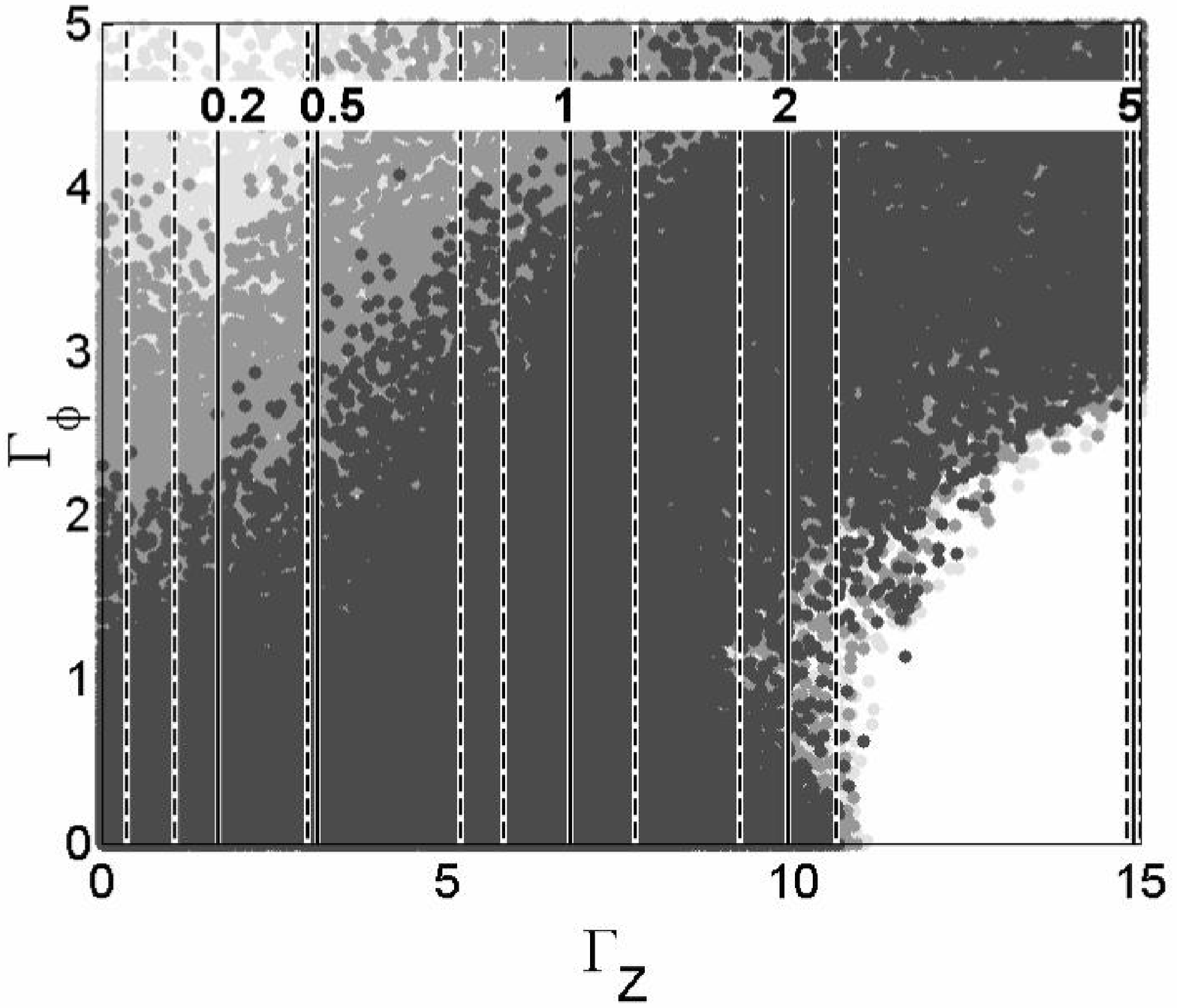}
\caption{The FP magnetic energy parameter for region 2.  See Figure \ref{fig:FP_F1_MK} for explanation.}
\label{fig:FP_F2_MK}
\end{figure}

FP defined a magnetic energy parameter M/K to determine the relative importance of the volume-averaged magnetic energy density $M$, compared with the volume-averaged turbulent energy density $K$.  These quantities are defined as follows:
\bea
M &=& \frac{\int_V (\Bz^2+\Bphi^2)dV}{8\pi V} \\
K &=& \frac{3}{2} <\rho><\sigma^2>,
\label{eq:MK}
\eea
where all quantities are defined as in $\S$\ref{sec:FP}.  FP considered $M/K \approx 5$ to be the upper limit for reasonable models, and $0.5 \appleq M/K \appleq 2$ to be the preferred range.  Figure \ref{fig:FP_F1_MK} shows the distribution of M/K for region 1.  This figure shows that M/K is almost exactly a function of $\Gz$ alone, and that solutions with $\Gz \le 15$ have $M/K \appleq 4$.  Figure \ref{fig:FP_F2_MK} is the corresponding one for region 2.  There is more scatter in this figure, but M/K remains highly correlated with $\Gz$.  Solutions with $\Gz \le 15$ are consistent with $M/K \appleq 5$.

\begin{figure}
\includegraphics[width=.49\textwidth]{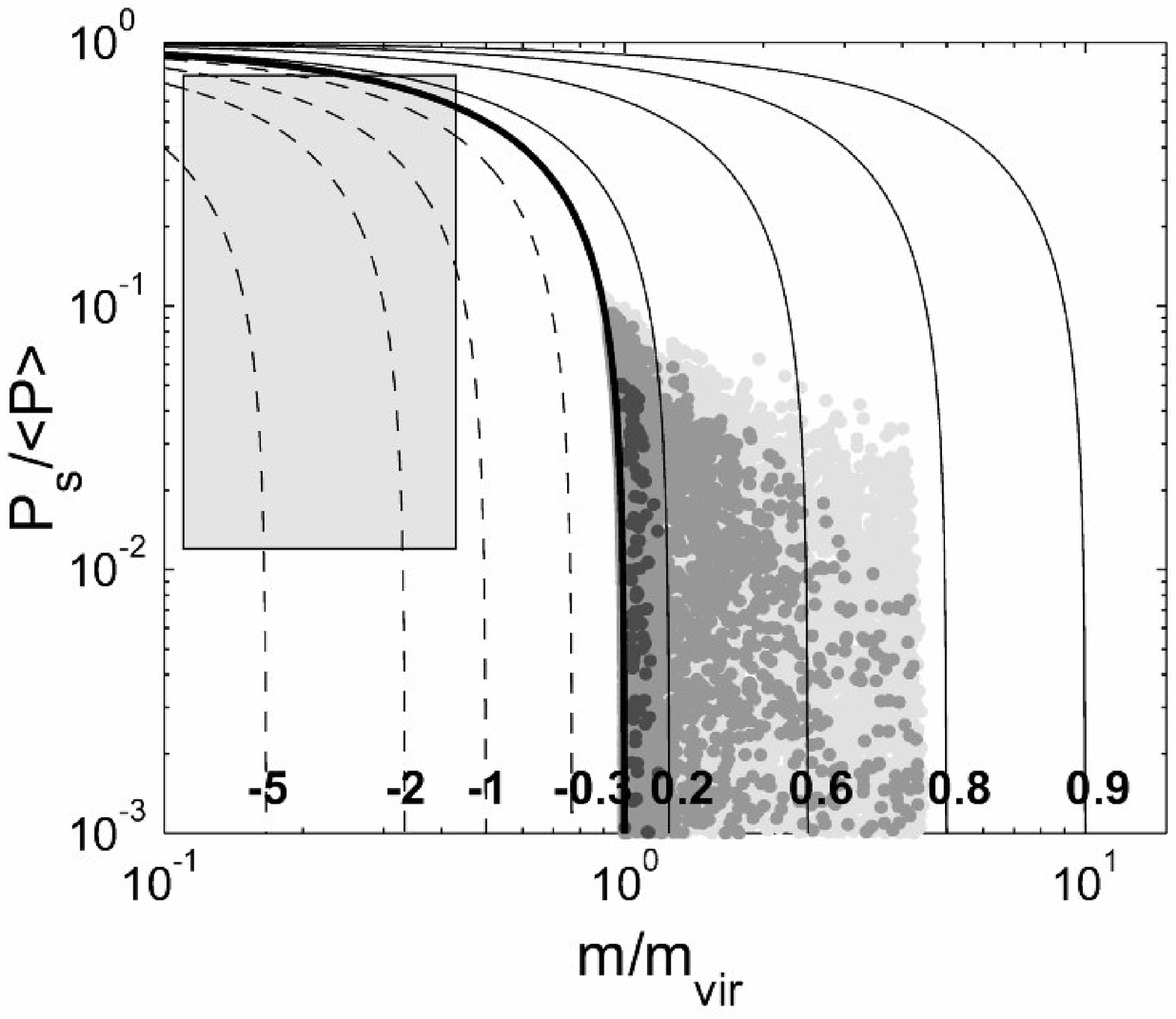}
\includegraphics[width=.49\textwidth]{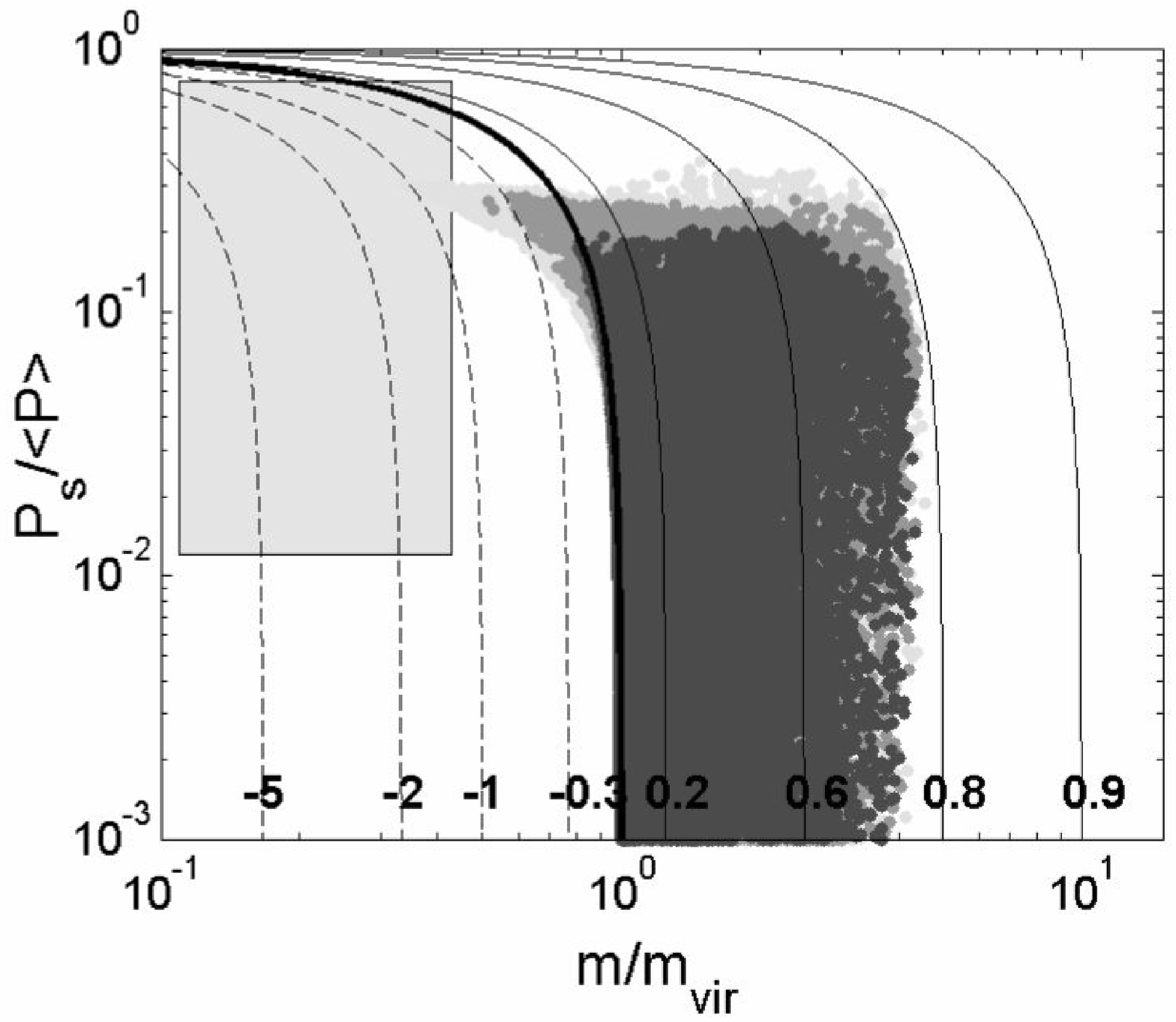}
\caption{Virial diagram for FP solutions for regions 1 (left) and 2 (right).  All quantities are as defined in Figure \ref{fig:OstVirial}.  The FP model is the only model of the three models studied that permits solutions with $\M/|\W| < 0$ (net magnetic confinement).  However, all solutions for region 1 are magnetically neutral or dominated by the poloidal field.  Most solutions for region 2 are also dominated by the poloidal field.  However, a few toroidal-field dominated solutions are permitted when the $P_S/<P>$ is in the upper range shown in the figure.}
\label{fig:FPVirial}
\end{figure}

Figure \ref{fig:FPVirial} shows the virial parameters of the FP solutions for regions 1 (left panel) and 2 (right panel).  We find that $\M/|\W| \ge 0$ for all models or Region 1 and most models of Region 2, which indicates that most models are consistent with a dominant poloidal magnetic field.  Some magnetically neutral models $\M/|\W| =0$ are allowed \footnote{Magnetically neutral models are not necessarily non-magnetic.  Filaments with $M/|\W|=0$ can be obtained by critically balancing the toroidal field and poloidal field terms in equation \ref{eq:MDef}.}, and the few toroidal field-dominated models that are allowed are characterized by rather weak fields.  This is strikingly different from the results obtained by FP1, which are indicated by the shaded box.  Toroidal-field dominated models, which were the focus of FP1, are unlikely for this filament, and our results indicate that G11.11-0.12 is an excellent candidate for a magnetically supported filament.

\section{Polarization Predictions}
\label{sec:polarimetry}
\citet{FP4} developed a technique to simulate submillimetre polarization maps of filamentary cloud models.  They considered only the FP model, but we generalize their analysis to predict the polarization structure due to all GS and FP solutions that are consistent with the radial structure of the G11.11-0.12 IRDC.  The \citet{FP4} formalism combines the uncertainties due to poorly contrained dust properties and grain alignment mechanisms into a single parameter $<\alpha>$.  This parameter determines the maximum polarization percentage that can be obtained from a region threaded by a magnetic field with the optimal geometry, which corresponds to a uniform magnetic field parallel to the plane of the sky.  We assume that $<\alpha>=0.15$, which leads to a maximum polarization percentage of 15.4\% by equation 17 in \citet{FP4}.

\citet{FP4} found that filaments aligned parallel to the plane of the sky are consistent with three general types of polarization patterns.  Type 1 patterns are characterized by polarization vectors that are aligned exactly parallel to the axis of the filament, and are consistent with toroidal-field dominated models.  Type 2 patterns have polarization vectors that are aligned exactly perpendicular to the filament, and are due to poloidal-field dominated models.  Type 3 patterns occur in models where the poloidal and toroidal fields are of comparable strength. Type 3 polarization maps are aligned perpendicular to the filament in some regions (especially near the poloidally dominated core), but are aligned parallel to the filament in other regions (i.e. a toroidally dominated envelope).    \citet{MWF} used this technique to model the `integral-shaped' filament in Orion A, which they found to be consistent with a Type 1 polarization pattern.  \citet{FP4} also showed that Type 3 patterns are most likely for FP filaments that are inclined relative to the plane of the sky.

\begin{figure}
\includegraphics[width=.49\textwidth]{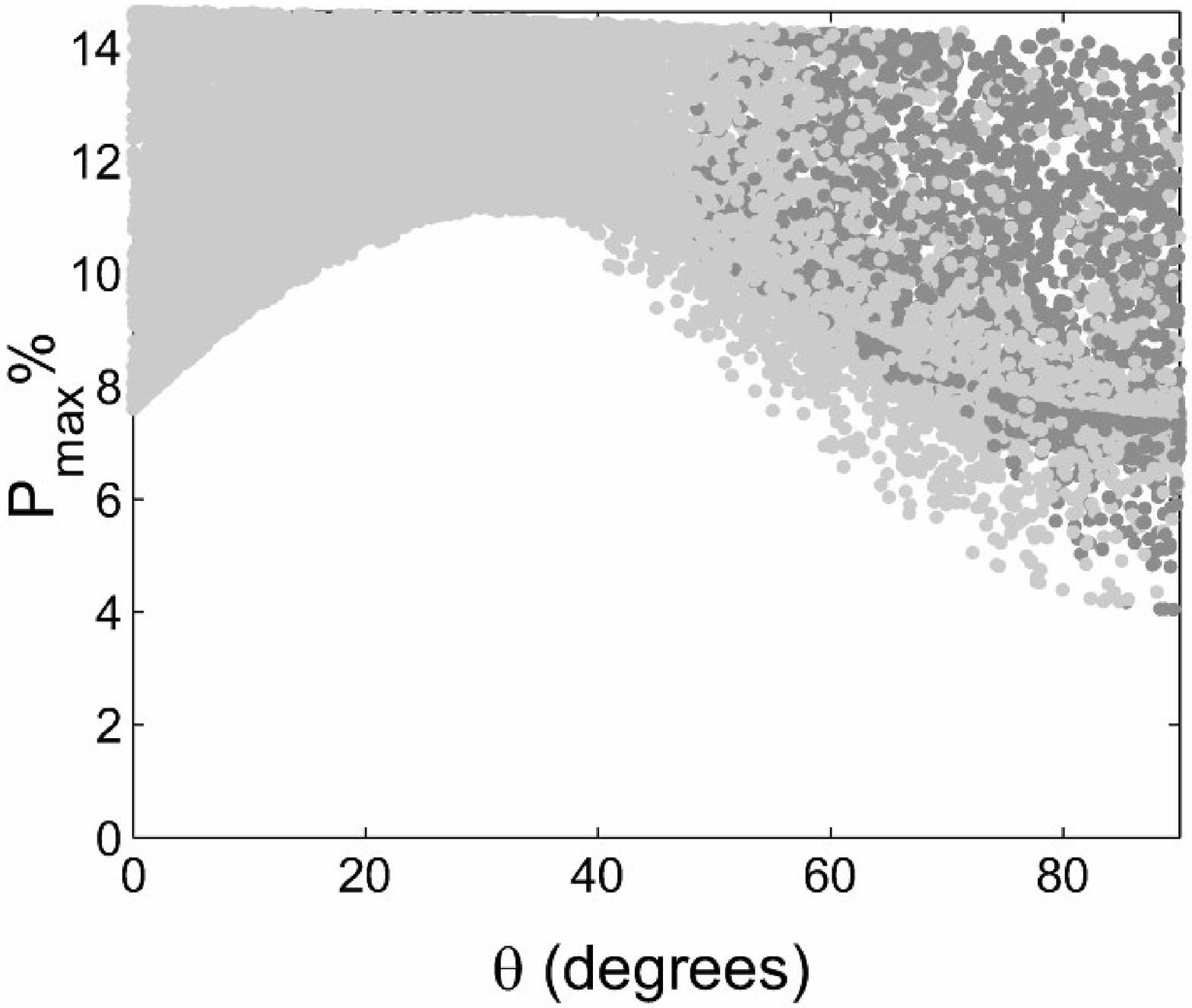}
\includegraphics[width=.49\textwidth]{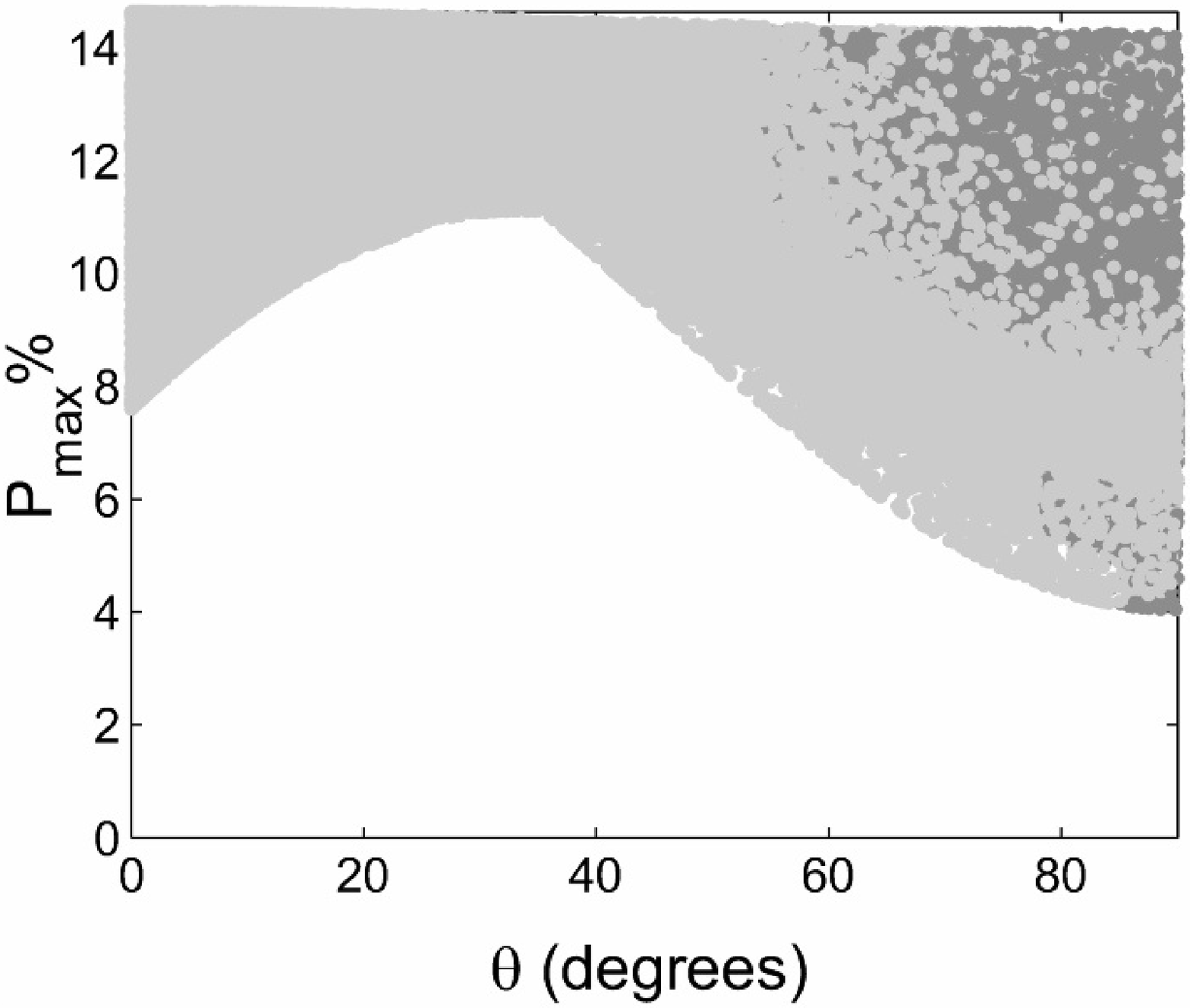}
\caption{Submillimetre polarization predictions for the GS model.  Region 1 is represented on the left, and Region 2 on the right.  We classify the polarization structure according to the classification scheme of \citet{FP4} and show the maximum predicted polarization percentage.  All solutions are Type 2 (poloidal-field dominated) and Type 3 (mixed) patterns.  The colour scheme is consistent with Figures \ref{fig:FP_F1_pol} and \ref{fig:FP_F2_pol} below, which are the corresponding figures for the FP model.  Points were layered as follows.  Successively better (lower) $\chi^2$ bins always overlay poorer bins, as in previous figures.  Within each bin, points were lain down in the order Type 1, followed by Type 2, and finally Type 3.  Some overshadowing of points was inevitible between layers.}
\label{fig:Stod_pol}
\end{figure}

We used this scheme to classify the polarization patterns of all GS and FP models computed in this paper.  All GS solutions that are consistent with the data are Type 2 or Type 3 polarization patterns.  A polarization map that shows a Type 1 pattern would rule out the GS model.  Figure \ref{fig:Stod_pol} shows that the maximum polarization percentage may range from about 4\% to 15\% for both regions.  This figure represents the three $\chi^2$ bins used elsewhere in the paper, plus the 3 possible polarization types, of which only 2 actually occur.

\begin{figure}
\includegraphics[width=.49\textwidth]{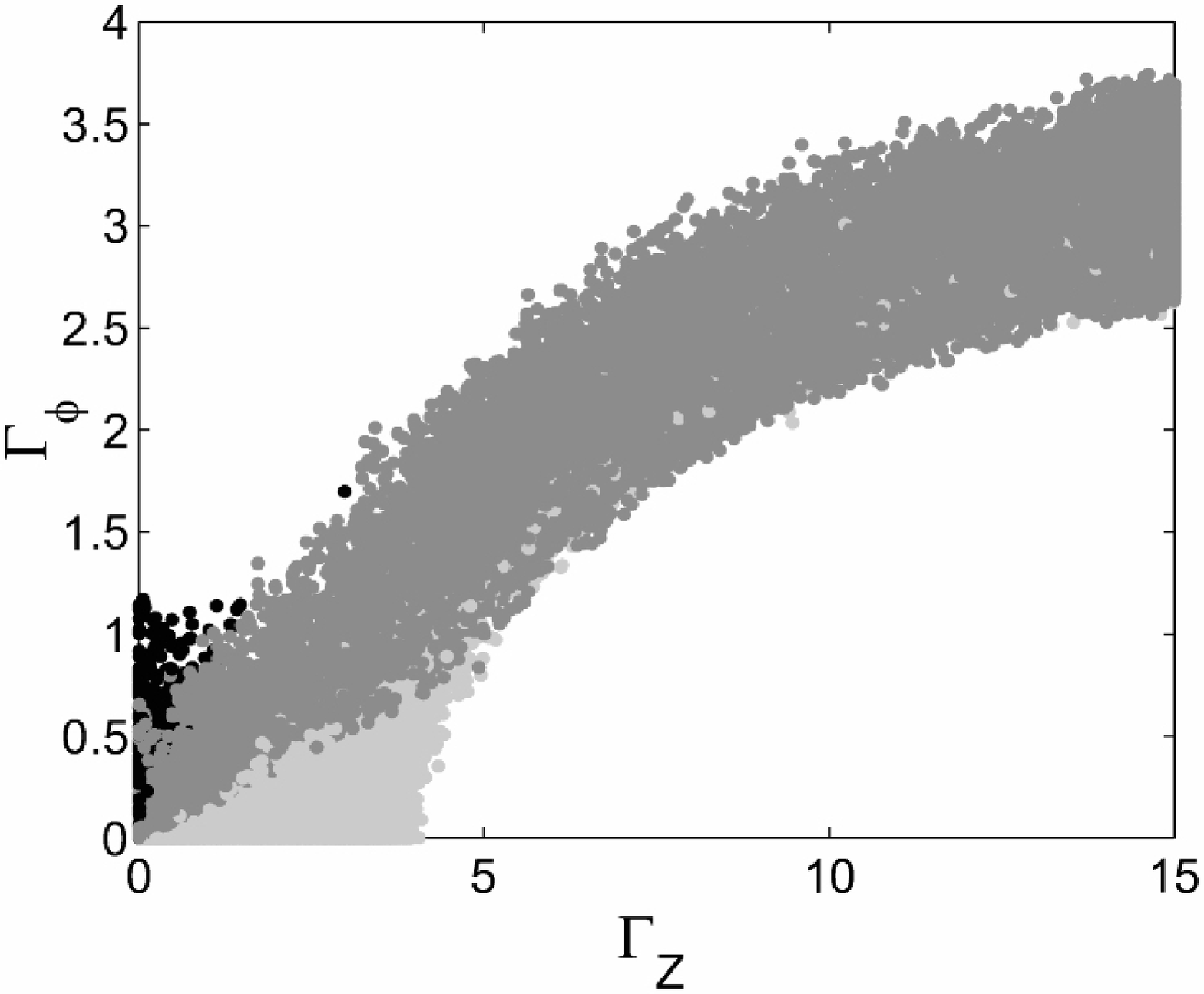}
\includegraphics[width=.49\textwidth]{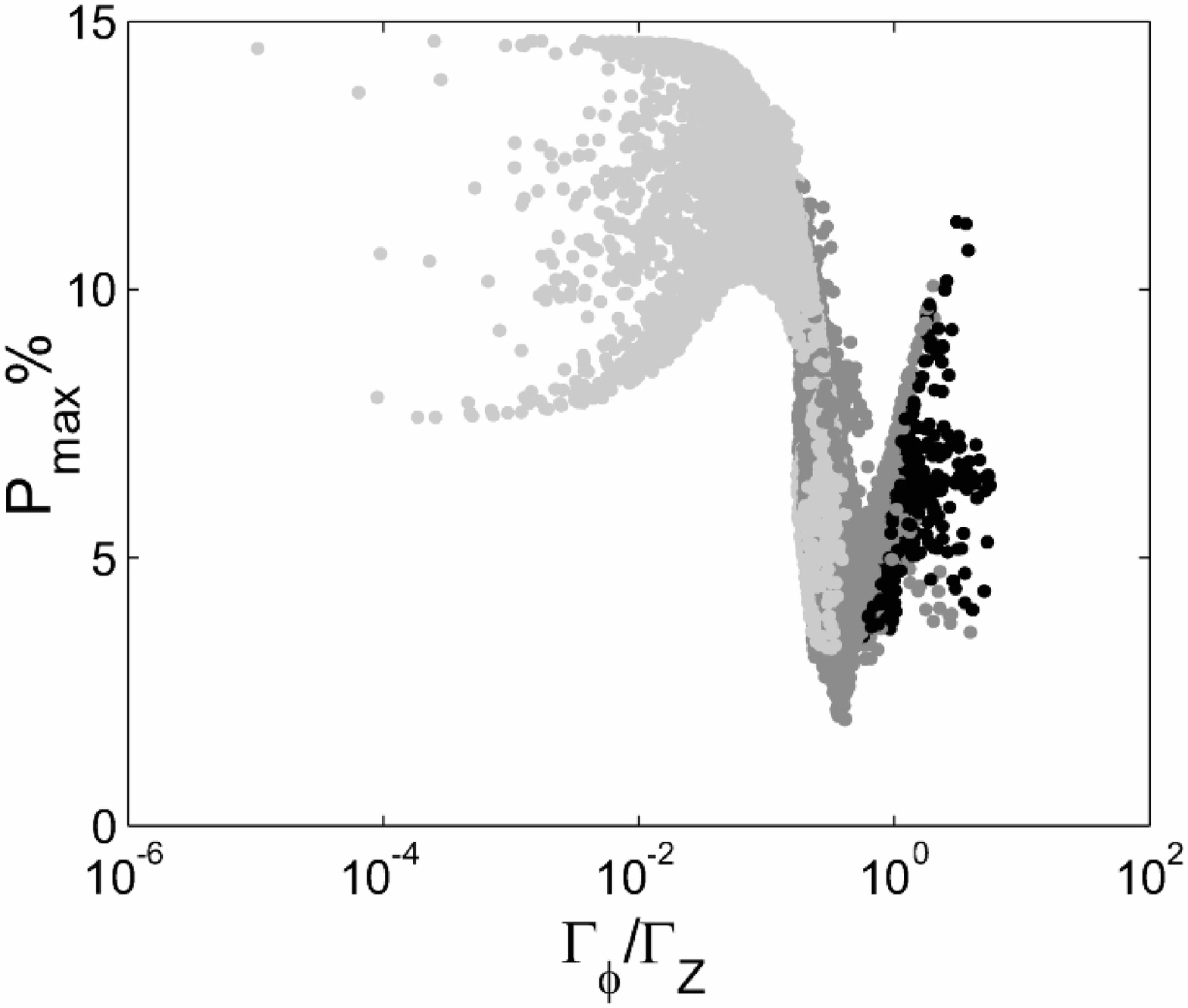}
\caption{Submillimetre polarization predictions for FP solutions for region 1.  We classify the polarization structure according to the classification scheme of \citet{FP4}: black dots indicate Type 1 patterns, medium gray dots designate Type 3 patterns and light gray is for Type 2 patterns.  These patterns are explained in the text, and the layering of points is explained in Figure \ref{fig:Stod_pol}.  [Note that this figure does not follow the grayscale convention used elsewhere in this paper, where the shade indicates the quality of the solution.]  (panel a., left) The distribution of the polarization type with respect to $\Gz$ and $\Gphi$.  (panel b., right) Polarization percentage versus $\Gphi/\Gz$.}
\label{fig:FP_F1_pol}
\end{figure}

The left panel of Figure \ref{fig:FP_F1_pol} shows the distribution of the type of pattern with respect to $\Gz$ and $\Gphi$ for the FP model, applied to region 1.  We find that Type 3 (mixed type) polarization patterns occupy the largest part of the allowed parameter space.  Type 2 (poloidal-field dominated) patterns occupy a smaller fraction of the parameter space, located at moderate $\Gz$ and low $\Gphi$.  It is interesting that we also found a few Type 1 patterns, which {\em appear} to be toroidally dominated.  These solutions are slightly dominated by their poloidal fields in a virial sense, or magnetically neutral, but would {\em appear} to be toroidally dominated in a polarization map.  This is possible because the integrals that enter the virial analysis are quite different from the Stokes integrals.  All of these solutions have very low values of $\Gz$ and $\Gphi$.

The right panel of Figure \ref{fig:FP_F1_pol} shows how the maximum polarization percentage varies with the ratio $\Gphi/\Gz$.  Maximum polarization percentages ranging from 2-15\% are consistent with the FP model for this region.  A polarization map of G11.11-0.12 that shows a Type 3 pattern with maximum polarization percentage below $\sim$5\%, or any Type 1 pattern, would be consistent with the FP model, but not the GS model.  It would be difficult to distinguish between the GS model and the FP model if a Type 2 pattern were observed, unless the peak polarization percentage was observed to be in the 2-4$\%$ range, since such low values are consistent with the FP model but not the GS model.  However, detailed fitting to the actual polarization map might reveal other differences that help to distinguish between these models.

It is clear that the type of polarization pattern, combined with the observed maximum polarization percentage provides an estimate of $\Gphi/\Gz$.  This is significant because $\Gphi/\Gz$ is the most important factor that determines the fragmentation timescale, the spacing of cores in the filament , and the masses of cores \citep{FP2}.  Therefore, polarization observations of G11.11-0.12, in combination with this work, would constrain the fragmentation properties of the cloud strongly, and determine the initial magnetohydrodynamic conditions that lead to star formation.

\begin{figure}
\includegraphics[width=.49\textwidth]{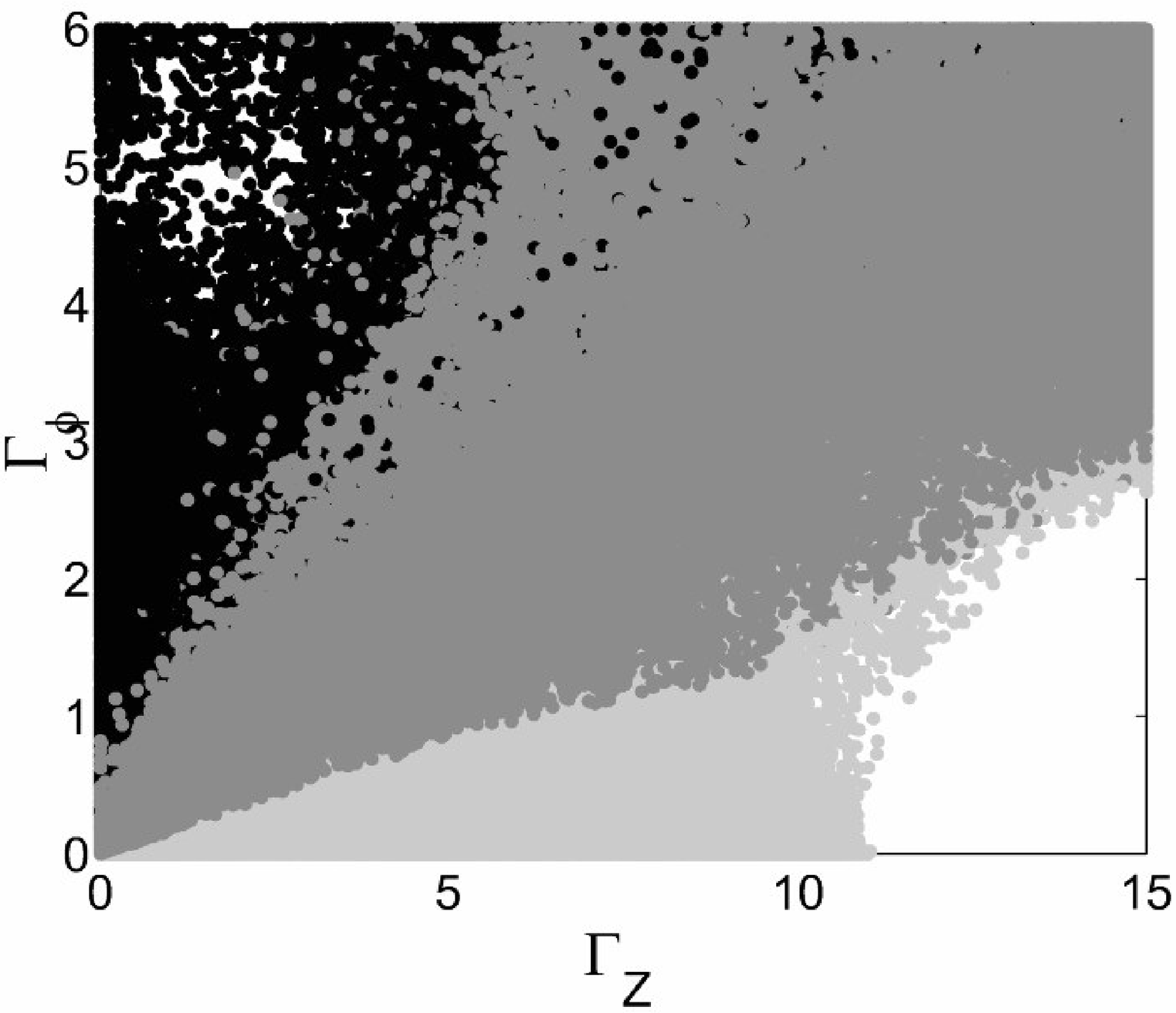}
\includegraphics[width=.49\textwidth]{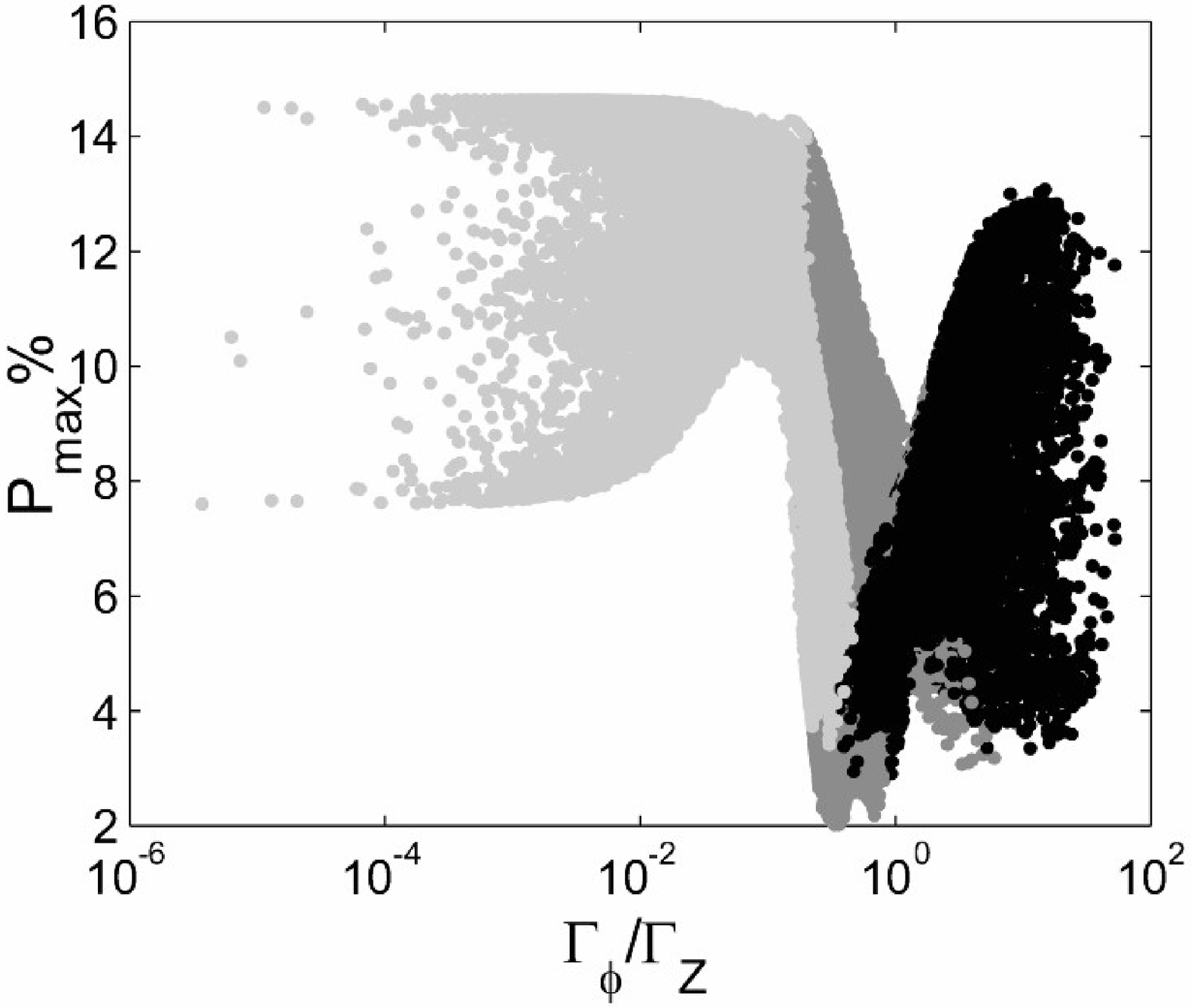}
\caption{Submillimetre polarization prediction of FP solutions for region 2.  See Figure \ref{fig:FP_F1_pol} for explanation.}
\label{fig:FP_F2_pol}
\end{figure}

Figure \ref{fig:FP_F2_pol} is the corresponding figure for region 2.  All three types of polarization patterns are possible for region 2.  The polarization percentages have a similar range of allowed values as for Region 1, although the scatter is much greater.

\section{Discussion}
\label{sec:discussion}
This paper is the first astronomical research problem attempted using a new multi-objective genetic algorithm ``Ferret'' designed by one of the authors (J.F.).  We have tested this code thoroughly, and it performs extremely well on data-fitting problems where the goal is to find the {\em class} of allowed solutions, rather than a single representative solution.  This allows the user to understand the detailed structure of the parameter space, and gives an honest representation of the degeneracies of the models being tested.  

Our previous analysis of the G11.11-0.12 infrared-dark cloud (Paper I) demonstrated that its density profile is steeper than the $~r^{-2}$ profile inferred from observations of other filamentary molecular clouds (Alves et al. 1998, Lada et al 1999, Johnstone \& Bally 1999), and that the profile can be fit reasonably well by the non-magnetic Ostriker model.  This paper provides a more detailed analysis that compares the radial structure of the G11.11-0.12 infrared-dark cloud with three models of filamentary clouds: the non-magnetic Ostriker model, the magnetic GS model, and the magnetic FP model.  We found that the density profile of G11.11-0.12 is consistent with the Ostriker model, but also with regimes of both magnetic models.  Our technique allowed us to map out the allowed regions of parameter space thoroughly for each model.

Our most significant result is that nearly all magnetic solutions that are in agreement with the data are dominated by the poloidal component of the magnetic field, or are magnetically neutral.  A few toroidal-field dominated solutions are allowed, but only for Region 2, which has a much ``noisier'' radial structure than Region 1, and even these solutions are close to magnetic neutrality.  G11.11-0.12 appears to be very different from the filaments studied by FP1, which were inferred to have a shallow $~r^{-2}$ density profile and a dominant toroidal field component.  Note that the FP model is the only model of the three models considered that can account for both toroidal-field dominated filaments and the poloidal-field dominated G11.11-0.12 filament.

We predicted the polarization patterns allowed by the GS and FP models of G11.11-0.12 that are allowed by the data.  For the GS model, all of the polarization patterns are Type 2 (poloidal-field dominated) or Type 3 (mixed) patterns \citet{FP4}.  Thus, the GS model could be ruled out for the G11.11-0.12 filament if future polarimetric observations reveal that this region is more consistent with a Type 1 (toroidal-field dominated) pattern.

For the FP model, Type 3 patterns \citet{FP4}, which are consistent with a dominant poloidal magnetic field, occupy the greatest fraction of the parameter space.  This is due primarily to the large range of inclination angles (-45$^o$ - 45$^o$) present in the solution set, since \citet{FP4} showed that type 3 patterns are the most common pattern when the filament is inclined relative to the plane of the sky.  Type 2 patterns occupy a fraction of the parameter space, and occur at moderate values of $\Gz$ and low values of $\Gphi$.  We found a few Type 1 patterns that would {\em appear} to be dominated by a toroidal field.  Most of these solutions have very weak magnetic fields that are close to magnetic neutrality or slightly dominated by the poloidal field, according to the virial equation \ref{eq:virial}.   We demonstrated that the polarization class, combined with the maximum polarization percentage, provides an estimate of the ratio of magnetic flux to mass values $\Gz/\Gphi$.  This is significant because \citet{FP2} demonstrated that this quantity is the dominant factor that controls the fragmentation timescale of a filament, the fragmentation wavelength, and ultimately the masses of clumps that form along the filament.  Thus, a polarization map of G11.11-0.12 might constrain the magnetic field sufficiently well to provide an indication of the magnetic and dynamical conditions within the filament.  Improved constraints could be obtained by detailed multi-objective modeling of the filament, which would fit the intensity structure, polarization vector orientations and polarization percentages simultaneously.

Our polarization modeling also determined that the maximum polarization percentage may vary from 2-15\% for the FP model, but is restricted to 4-15\% for the GS model.  Thus, observations of a Type 2 pattern with a low peak polarization percentage between 2-4\% would not be consistent with the GS model.

We have not considered non-axisymmetric magnetic field geometries in our analysis.  It is interesting to note that a filament threaded by a straight magnetic field perpendicular to the filament axis would generate a Type 1 polarization pattern that may be similar to that of a toroidal field-dominated filament, if {\em either} the axis of the filament or the magnetic field is parallel to the plane of the sky.  However, this is not true for the general case where neither the filament axis nor the field is parallel to the sky plane, because of projection effects.  In this case, the projection of the field in the sky plane, and hence the polarization vectors, would have a contstant angle relative to the projected axis of the filament.  The symmetry of a helical field results in polarization vectors that are always parallel or perpendicular to the projected filament axis, unless the filament is bent, in which case the polarization angle changes along the filament \citep{MWF, MFM}.  A helical field cannot produce vectors that maintain a constant angle relative to the filament axis.  Thus, it should be possible to distinguish between toroidal field-dominated helical field models and models with a non-axisymmetric orthogonal field for an ensemble of objects, although it may be difficult to distinguish between these possibilities for a single filament.  It would be interesting to include non-axisymmetric equilibrium models in future analyses.  However, no such models exist at present.

Difficult data-modeling problems are encountered frequently in astronomy and physics, which may feature large parameter spaces, noisy data, multiple constraints, and multiple modeling objectives applied to heterogenous data sets.  Our multi-objective genetic algorithm solves such problems reliably and efficiently.  We hope that our code will prove useful for a variety of problems in the physical sciences.  Our immediate goal is to build an automated system for multi-objective 2-dimensional modeling (i.e., 2 spatial map coordinates, rather than a single radial distance) of continuum and polarization maps, which would allow us to model simultaneously the intensity, polarization position angle and polarization percentage for filaments and cores.  In fact, we have already designed and tested a preliminary system that models the intensity and polarization data for one of the cores in the Orion Molecular Cloud $\int$-shaped filament, by varying the magnetic geometry, the density structure, the temperature and dust emissivity, the position angle and inclination of the core relative to the plane of the sky.  This analysis will be the subject of future work.

\section{Acknowledgements}
\label{sec:ack}
We wish to thank Sean Carey (SIRTF Science Center) for useful discussions that improved the quality of the data analyzed in this paper, and for his pioneering work in the identification and analysis of the MSX infrared-dark clouds.  We also thank Mike Peddle (HIA), who assisted us with the preparation of the figures.

\clearpage 

\end{document}